\documentclass{article}
\usepackage{graphicx}
\usepackage{booktabs}
\usepackage{amsmath}
\usepackage{multirow}
\usepackage{rotating}
\usepackage{tablefootnote}
\usepackage{mathabx}
\usepackage{tabularx}
\usepackage{fullpage}
\usepackage{parskip}
\usepackage[table,xcdraw]{xcolor}
\usepackage{pdfpages}
\usepackage{pgfplots}
\usepackage{adjustbox}
\usepackage{listings}
\usepackage{xcolor}
\usepackage[toc,page,header]{appendix}

\usepackage{minitoc}

\usepackage[sort&compress,numbers]{natbib}

\usepackage{tabularx}
\usepackage{geometry}

\usepackage{float}
\usepackage[colorlinks]{hyperref}
\hypersetup{
    colorlinks,
    linkcolor={red!50!black},
    citecolor={blue!50!black},
    urlcolor={blue!80!black}
}
\usepackage{cleveref}
\usepackage{authblk}
\usepackage{caption}
\usepackage{enumitem}

\title{Engagement, User Satisfaction, and the \\ Amplification
of Divisive Content on Social Media}
\author[a,*]{Smitha Milli}
\author[b]{Micah Carroll}
\author[c]{Yike Wang}
\author[b]{\authorcr Sashrika Pandey}
\author[b]{Sebastian Zhao}
\author[b]{Anca D. Dragan}
\affil[a]{Cornell Tech, New York, NY, USA}
\affil[b]{University of California, Berkeley, CA, USA}
\affil[c]{University of Washington, Seattle, WA, USA}
\affil[*]{Corresponding author: \texttt{smitha.milli@gmail.com}}
\date{}
\usepackage{multirow}

\begin{document}
\crefname{appendix}{SM section}{SM sections}
\Crefname{appendix}{SM section}{SM sections}
\crefname{app}{SM section}{SM sections}
\Crefname{app}{SM section}{SM sections}

\begin{titlepage}

\maketitle

\begin{center}
\textbf{Classification}: Social Sciences 
\end{center}

\vspace{1cm}

 \renewcommand{\abstractname}{Abstract / Significance Statement}
\begin{abstract}
In a pre-registered algorithmic audit, we found that, relative to a reverse-chronological baseline, Twitter's engagement-based ranking algorithm amplifies emotionally charged, out-group hostile content that users say makes them feel worse about their political out-group. Furthermore, we find that users do \emph{not} prefer the political tweets selected by the algorithm, suggesting that the engagement-based algorithm underperforms in satisfying users' stated preferences. Finally, we explore the implications of an alternative approach that ranks content based on users' stated preferences and find a reduction in angry, partisan, and out-group hostile content, but also a potential reinforcement of pro-attitudinal content. The evidence underscores the necessity for a more nuanced approach to content ranking that balances engagement and users' stated preferences.
\end{abstract}

\end{titlepage}

\section{Introduction}
Social media ranking algorithms personalize content to individual users by predicting what they will engage with~\citep{cunningham2024know,covington2016deep}. For example, in April 2023, Twitter's ranking algorithm was based on predicting whether a user would engage with a particular tweet using ten different types of engagement~\citep{Twitter_2023}. These predictions include behaviors like retweeting, replying, watching an embedded video, or lingering on a tweet for at least two minutes. Similarly, in December 2021, it was reported that TikTok's algorithm predicted how long users would watch a video and whether they would like or comment on it~\citep{smith_2021}. We refer to these kinds of ranking algorithms as \emph{engagement-based} ranking algorithms.  These algorithms can be seen as optimizing for users' \emph{revealed preferences}, i.e., they interpret a user engaging with a piece of content as an indication that the user wants to see that piece of content~\citep{ekstrand2016}.

There is concern that by focusing on maximizing user engagement---especially reactive or passive behaviors like lingering---these algorithms may amplify content that is attention-grabbing but is unaligned with the user's reflective values or with broader societal values~\citep{brady2020MAD,brady2023algorithm,kleinberg2024inversion,cunningham2024know,agan2023automating,kleinberg2024challenge}.  For example, \citet{brady2023algorithm} suggest that, ``content algorithms systematically exploit human social-learning biases because they are designed to optimize attentional capture and engagement time on the platform, and social-learning biases strongly predict what users will want to see.'' They specifically claim that ranking algorithms tend to amplify what they term `PRIME' content---prestigious, in-group, moral, and emotional content. While humans may have developed biases toward PRIME information for adaptive reasons, they suggest that in the context of modern social media, exploiting these biases can lead to heightened social misperception and conflict. Similarly, \citet{agan2023automating} suggest that since cognitive biases are more likely to be triggered during fast, intuitive thinking (``system 1''), algorithms trained on more passive or impulsive behaviors---like time spent watching a video or lingering on a tweet---are likely to replicate these biases. 

The notion that algorithms \emph{exploit} implicit human biases to drive user engagement suggests that, if users deliberately chose the content they consumed, it would differ from what the algorithm prioritizes. However, most existing research does not clearly disentangle the algorithm’s exploitation of automatic human biases from deliberate human choices. For example, many studies do not control for who the user follows. On most social media platforms, the content that is shown to a user is heavily shaped by their own explicit choices of who to follow. Many existing studies find a correlation between the engagement that a post receives and characteristics such as its emotional intensity~\citep{brady2017emotion,brady2019ideological,rathje2021out}. However, it is possible that those posts received more engagement, not due to the algorithm prioritizing them, but because users chose to follow accounts that post more emotional content. Analogously, it was often hypothesized that YouTube's recommendation algorithms led users down ``rabbit holes" to politically extreme content~\citep{tufekci2018youtube,ribeiro2023amplification,haroon2023audit,srba2023}. However, subsequent studies revealed that users often reach extreme channels through external links and subscriptions, suggesting that intentional human choice and demand might be an alternative explanation for the consumption of extreme content~\citep{chen2023,ribeiro2023amplification,hosseinmardi2021,hosseinmardi2024}.

Moreover, even controlling for who a user follows is not sufficient to disentangle deliberate choices from automatic biases. Most studies control for who a user follows by comparing posts selected by a ranking algorithm to those in a \emph{reverse-chronological} baseline which consists of recent posts from the accounts a user follows. For example, in concurrent work, \citet{bouchaud2023crowdsourced} conducted such an audit of Twitter and found that the ranking algorithm selected more emotionally valent content compared to the reverse-chronological baseline. This implies that the ranking algorithm amplifies emotional content beyond what users’ deliberate choices of who to follow alone would generate, and therefore, provide stronger evidence of the algorithm exploiting human biases. However, such evidence would be far from conclusive. The algorithm might be surfacing more emotional content because, within the accounts a user follows, the user may actually prefer more emotional posts, even if they lack a direct mechanism to filter content at such a granular level.  In other words, the algorithm could still be optimizing for what users would have deliberately selected on their own, given finer control over their content.

Disentangling deliberate human choice from algorithmic exploitation of automatic biases is important not only for scientific understanding but also for designing better systems. If as \citet{brady2023algorithm} suggest, the algorithmic exploitation of social biases leads to greater social misperception and conflict, then redesigning algorithms to avoid such exploitation becomes imperative. In fact, a growing body of research in computer science is working toward this goal. For example, \citet{milli2021optimizing} developed a method that optimizes for passive engagement only to the extent that they are consistent with and predictive of users' stated content preferences. \citet{agarwal2024system} utilize generative temporal point processes to create a so-called `System 2 recommender system' which models when users return to the platform to assess whether recommendations align with their deeper, reflective preferences. \citet{kleinberg2024challenge} provide a more theoretical analysis but suggest ways that platform can understand whether they are over-optimizing for System 1 while failing to provide reflective value. 
These approaches all operate under the assumption that algorithms exploit automatic biases and that correcting for this could provide greater value to individuals and society. To provide a stronger grounding and justification to these methods, it is crucial to empirically test this hypothesis, which is the aim of our work.

We executed a pre-registered algorithmic audit of Twitter\footnote{Twitter rebranded to ``X'' in July 2023. Our study took place before the rebranding, so we refer to the platform as ``Twitter''.} that was designed to differentiate deliberate human choices from algorithmic exploitation of biases. Over two weeks in February 2023, we recruited a group of Twitter users (N=806) and collected (1) the first ten tweets the \emph{engagement-based} personalized ranking algorithm would have shown them, and (2) the ten most recent tweets they would have seen from accounts that they follow, i.e., a \emph{reverse-chronological} baseline.\footnote{On Twitter, users can opt out of the engagement-based timeline and use the reverse-chronological baseline instead, making it an especially relevant baseline to test against}. By comparing to the reverse-chronological baseline, we are able to understand the effects of the engagement-based timeline, beyond users' own deliberate decisions of who to follow.

Moreover, beyond simply controlling for users' decisions of who to follow, we also explicitly surveyed users regarding their preferences for the tweets shown in both their engagement-based and reverse-chronological timelines. In exploratory analysis, we used these survey responses to construct a counterfactual \emph{stated preference timeline} consisting of the collected tweets that the user said they valued the most. The stated preference timeline simulates how a timeline might look if algorithms were more aligned with users' stated preferences, i.e., preferences that are derived from what users say they want, rather than users' revealed preferences which are based on which items they engage with. We then compared this stated preference timeline to that of the engagement-based and reverse-chronological timeline. We then compared this stated preference timeline with the engagement-based and reverse-chronological timelines, allowing us to assess whether the effects of the engagement-based algorithm were still more pronounced than what might occur if users could curate their content at the tweet level (rather than merely by whom they follow).

\subsection{Hypotheses and research questions}
We pre-registered five primary hypotheses and four secondary research questions. All hypotheses and research questions are stated as expectations about the effects of the engagement-based timeline relative to their reverse-chronological timeline. In exploratory analysis, we also analyzed the same outcomes for the stated preference timeline that we constructed using users' explicit stated value for individual tweets.

Our first three hypotheses focused on the differences in the content selected by the engagement-based timeline and the reverse-chronological timeline. In particular, we hypothesized that, compared to the reverse-chronological timeline, the engagement-based timeline would show tweets (i) that were more emotional, (ii) had a stronger ideological leaning, and (iii) contained more expressions of animosity towards participants' political out-group.

\begin{enumerate}[label=\textbf{Hypothesis \arabic*.},leftmargin=*,start=1]
    \item The engagement-based timeline will show tweets that are more emotional (along four dimensions of anger, sadness, anxiety and happiness), compared to the reverse-chronological timeline.
\end{enumerate}

\begin{enumerate}[label=\textbf{Hypothesis \arabic*.},leftmargin=*,start=2]
    \item The engagement-based timeline will show tweets with a stronger ideological leaning, compared to the reverse-chronological timeline.
\end{enumerate}

\begin{enumerate}[label=\textbf{Hypothesis \arabic*.},leftmargin=*,start=3]
    \item The engagement-based timeline will show tweets containing greater out-group animosity, compared to the reverse-chronological timeline.
\end{enumerate}

These three hypotheses are consistent with both prior theoretical and empirical work. They are consistent with the theoretical framework put forth by \citet{brady2023algorithm}, who suggest that social media ranking algorithms exploit humans' social biases towards emotional, moralized, and in-group content. Moreover, previous observational studies have found that tweets with more emotion and out-group animosity tend to receive higher engagement~\citep{rathje2021out,brady2017emotion}. However, as previously noted, these studies do not account for users’ choices in whom they follow, making it challenging to assess the specific influence of the ranking algorithm in amplifying emotional and out-group antagonistic content. Our hypotheses posit that engagement-driven algorithms do indeed amplify such emotionally charged, politically divisive content beyond the impact of users’ own follow choices.

We also had two hypotheses about how participants would feel after reading the tweets in their engagement-based timeline. The first was that, by selecting tweets with greater ideological leaning and out-group animosity, we expected readers to feel worse about their political out-group after reading tweets from their engagement-based timeline.
\begin{enumerate}[label=\textbf{Hypothesis \arabic*.},leftmargin=*,start=4]
    \item Participants will feel worse about their political out-group after reading tweets from their engagement-based timeline, compared to the reverse-chronological timeline.
\end{enumerate}

Our second reader-focused hypothesis was that participants would feel happier but less angry, sad, or anxious reading tweets in their engagement-based timeline.  This contrasted with our expectation that the engagement-based timeline would increase the frequency of expressions of all four emotions (Hypothesis 1). The rationale for this difference is that platforms test ranking algorithms in A/B tests to identify those that optimize user retention~\citep{cunningham2024know}; thus, we assumed that fostering positive emotions would encourage users to return to the platform. (Ultimately this turns out to be the only hypothesis we did not find supporting evidence for --- readers reported feeling higher levels of all four emotions on the engagement-based timeline.)
\begin{enumerate}[label=\textbf{Hypothesis \arabic*.},leftmargin=*,start=5]
    \item Participants will feel happier and less angry, anxious, or sad reading tweets in their personalized timeline, compared to the reverse-chronological timeline.
\end{enumerate}

We also formulated four secondary pre-registered research questions. The first two explored the emotional impact of the engagement-based timeline on both the content it curates (as in Hypothesis 1) and the emotions experienced by readers (as in Hypothesis 4), but focused specifically on political tweets. While we expected that the engagement-based timeline would increase the expression of all four measured emotions—anger, happiness, sadness, and anxiety—across tweets overall (Hypothesis 1), we were uncertain about the effects when considering only political tweets. It is possible that political tweets might primarily amplify negative emotions rather than emotional content more broadly, with a similar pattern reflected in the emotions experienced by readers.

\begin{enumerate}[label=\textbf{Question \arabic*.},leftmargin=*,start=1]
    \item How angry, happy, sad, or anxious are the political tweets selected by the engagement-based timeline, compared to the reverse-chronological timeline?
\end{enumerate}

\begin{enumerate}[label=\textbf{Question \arabic*.},leftmargin=*,start=2]
    \item How happy angry, happy, sad, or anxious do readers feel after reading political tweets selected by the engagement-based timeline, compared to the reverse-chronological timeline?
\end{enumerate}

Our next question related to the reader's change in perception of their political in-group. We hypothesized that the engagement-based timeline would make the reader feel worse about their out-group (Hypothesis 4) because of increased exposure to tweets expressing animosity to the out-group. However, we were uncertain whether, conversely, the reader’s perception of their in-group would improve when exposed to tweets from the engagement-based timeline, especially because previous research has found that users are more likely to share tweets that are negative towards their out-group than those that are positive towards their in-group~\citep{yu2023partisanship}. 

\begin{enumerate}[label=\textbf{Question \arabic*.},leftmargin=*,start=3]
    \item How do readers' perceptions of their  political in-group change after reading tweets in their engagement-based timeline, compared to their reverse-chronological timeline?
\end{enumerate}

Our final question investigated whether the engagement-based timeline aligns with users' stated preferences—specifically, whether it shows them tweets they say they value. Users' revealed preferences appear to favor the engagement-based timeline over the reverse-chronological one, as randomized experiments at Twitter have shown that it increases the amount of time users spend on the platform compared to the reverse-chronological timeline~\citep{milli2022causal}. However, theories suggesting that ranking algorithms exploit social biases~\citep{brady2023algorithm, agan2023automating} suggest that users’ revealed preferences, which are given in an automatic state, may significantly differ from their stated preferences, which are obtained in a more reflective way, e.g. through user surveys.
\begin{enumerate}[label=\textbf{Question \arabic*.},leftmargin=*,start=4]
    \item Do users see more tweets they (say they) value in the engagement-based timeline, compared to the reverse-chronological timeline?
\end{enumerate}

\section{Results} \label{sec:results}
From February 11 to February 27, 2023, we conducted our study on CloudResearch Connect, an online crowd-working platform. The study period was broken into four waves\footnote{The time periods for the waves (inclusive) were 02/11-02/14, 02/16-02/19, 02/21-02/23, and 02/25-02/27.} and participants could complete the study once during each wave. Every day, we recruited up to 150 eligible participants who lived in the United States, were at least 18 years old, and used Google Chrome. Furthermore, participants were required to use Twitter at least a few times a week and follow at least 50 people on Twitter (both gauged through self-reports). To collect data, participants were directed to download a Chrome extension that we developed which scraped their Twitter homepage to collect the top tweets from their personalized, engagement-based timeline. While scraping, the Chrome extension added an overlay to the homepage that prevented the user from seeing the tweets during collection. At the same time that the engagement-based timeline was collected, we queried the Twitter API to get the top tweets from the reverse-chronological timeline. Only public tweets were collected and no promoted tweets (advertisements) were collected.

After collecting both sets of tweets, participants were directed to complete a survey on Qualtrics that asked questions about each of the top ten tweets from their engagement-based and reverse-chronological timeline. All tweets were displayed in a randomized order (thus, tweets from both timelines were typically interwoven rather than, for example, first showing all the engagement-based tweets and then all reverse-chronological tweets). If the same tweet was present in both the engagement-based and reverse-chronological timeline, then participants were only shown it once. If a tweet was a reply to another tweet, the user was shown both the replied tweet and the main tweet, and asked to answer the questions for both tweets. Similarly, if a tweet was a quote tweet, then the user was asked to answer the questions for both the quoted tweet and the main tweet. 



For each tweet, we asked the user to assess whether it was about a political or social issue.  If they marked the tweet as political, we asked them about the tweet's ideological leaning,  their own perception of their political in-group and out-group after reading the tweet, and whether the tweet author was expressing out-group animosity. For all tweets (including non-political tweets), we asked users to assess the author's emotions as well as the emotions the tweet made them feel, along four dimensions: happiness, anger, sadness, and anxiety. Finally, we also asked the reader for their stated preference about the tweet, i.e., whether they wanted to see tweets like it when they used Twitter. The full survey is provided in \Cref{app:survey}.

For our analysis of outcomes that are about the tweet itself (the tweet's ideological leaning, the emotions expressed by the author, and whether the tweet expresses out-group animosity), as opposed to the reader's emotions, we measure these outcomes in two ways. First, as just described, we collect readers' perceptions of these outcomes. Second, we ask the same questions to GPT-4~\cite{openai2023gpt}, a large language model (LLM). The main text presents the results based on readers' judgments, and \Cref{appendix:gpt-effects} includes results using GPT-generated labels. In the main text, we present the results based on readers' judgments, while \Cref{appendix:gpt-effects} contains results using GPT-provided labels. The findings indicate that both sets of results align qualitatively, providing reassurance given the limitations of each approach. Reader-given labels have the disadvantage that the reader's own background, e.g. political leaning, education, etc, might cause them to systematically misinterpret certain outcomes such as the author’s emotions~\citep{brady2023overperception}. Conversely, while humans can grasp broader context beyond the tweet's text, LLMs are limited to textual input and cannot process associated images or linked content. Moreover, large language models have their own biases in responses that may not be reflective of the human population~\citep{santurkar2023whose,dominguez2023questioning}. Nevertheless, the consistency between the two methods, each with their advantages and disadvantages, supports the robustness of our results.

A full description of our study procedure can be found in \Cref{app:exp-design}. In \Cref{app:metadata}, we provide descriptive statistics of the metadata in both timelines, e.g., the number of likes, retweets, links, photos, etc in each tweet, as well as analysis on the amplification of individual user accounts.  

\subsection{Effects of engagement-based ranking}

\begin{figure}[hbtp]
    \centering
    \includegraphics[width=0.9\columnwidth]{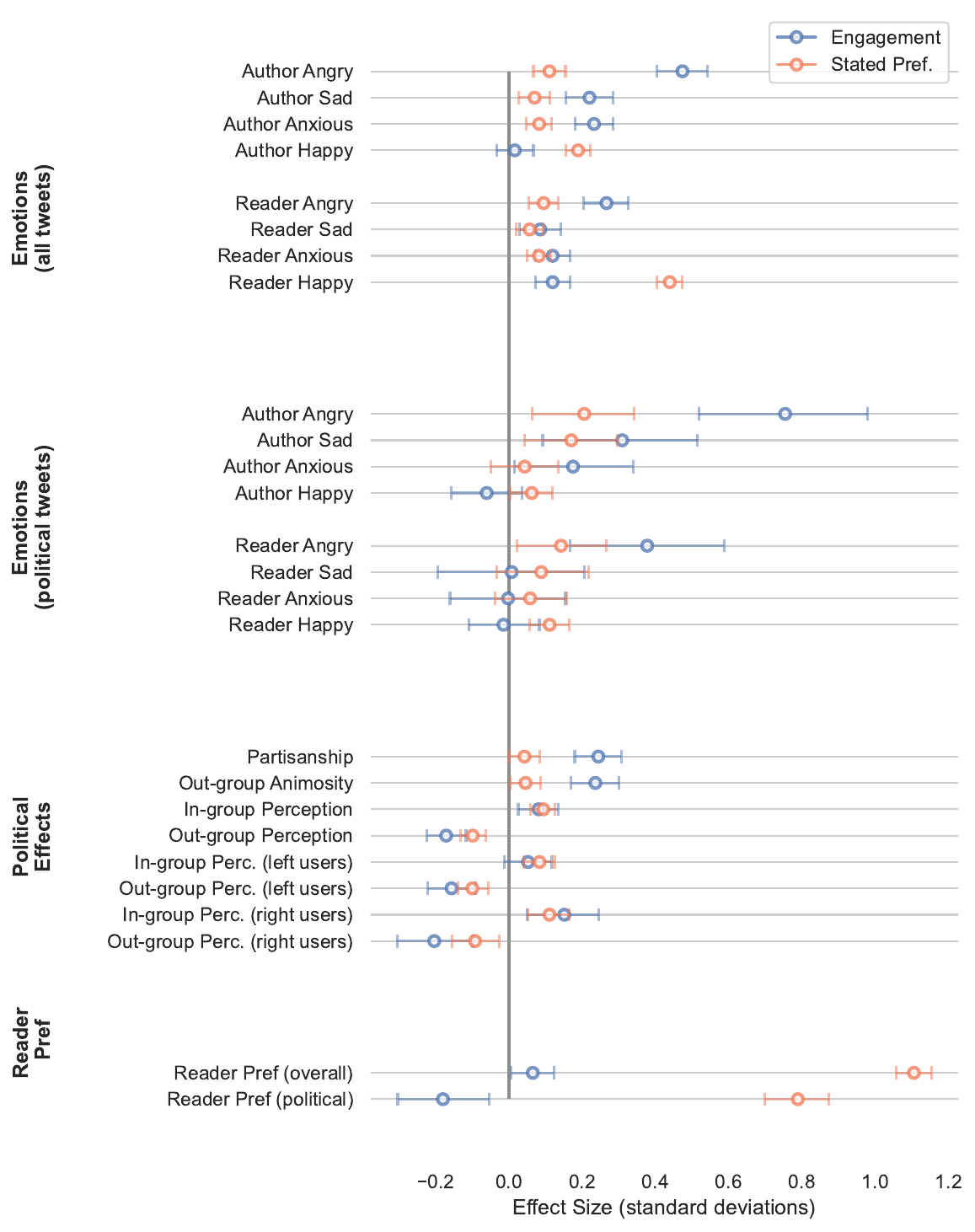}
    \caption{\small \textbf{Average treatment effects for all outcomes.} ATEs are shown with 95\% Bootstrap confidence intervals (unadjusted for multiple testing). The treatment effects of two different timelines are shown, relative to the reverse-chronological timeline: (1) Twitter's own engagement-based timeline, (2) our exploratory timeline that ranks based on users' stated preferences. The outcomes shown here are based on reader judgments, the analogous effects for GPT-4 based labels can be found in \Cref{appendix:gpt-effects}. The effect sizes for both timelines are relative to the reverse-chronological timeline (the zero line). Average treatment effects are standardized using the standard deviation of outcomes in the reverse-chronological timeline (see \Cref{app:ate-estimation} for details).}
    \label{fig:effects}
\end{figure}

First, we state our findings on the effects of Twitter's engagement-based algorithm. All tested outcomes and our analysis plan were pre-registered at \url{https://osf.io/upw9a}. \Cref{fig:effects} shows a summary of the average treatment effect for each outcome. As specified in our pre-analysis plan, the average treatment effect (of the engagement-based algorithm, relative to the control of the reverse-chronological ranking) is estimated through a difference in means (see \Cref{app:ate-estimation}), and two-sided $p$-values are estimated by paired permutation tests. In total, we tested 26 outcomes, and all results that are significant (at a $p$-value threshold of $0.05$) remain significant at a false discovery rate (FDR) of $0.01$. Thus, in expectation, none of our discoveries are false discoveries. The full table of standardized and unstandardized effect sizes, $p$-values, and FDR-adjusted $p$-values can be found in \Cref{appendix:full-effects}.

\textbf{Political outcomes}. We defined the \emph{partisanship} of a tweet as the absolute value of the reader-provided ideological leaning of the tweet (ranging from -2 = ``Far left'' to +2 = ``Far right''). Relative to the reverse-chronological baseline, we found that the engagement-based algorithm amplified tweets that exhibited greater partisanship ($0.24$ SD, $p<0.001$) and expressed more out-group animosity ($0.24$ SD, $p<0.001$). Furthermore, tweets from the engagement-based algorithm made users feel significantly worse about their political out-group ($-0.17$ SD, $p<0.001$) and better about their in-group ($0.08$ SD, $p=0.0014$). These effects remained significant when considering left and right-leaning users specifically, except for in-group perception in left-leaning users, where we found no significant effect.

\textbf{Amplified emotionality.} The engagement-based algorithm significantly amplified tweets that expressed negative emotions---anger ($0.47$ SD, $p<0.001$), sadness ($0.22$ SD, $p<0.001$), and anxiety ($0.23$ SD, $p<0.001$). It also led readers to feel more of all four emotions---anger ($0.27$ SD, $p<0.001$), sadness ($0.09$ SD, $p=0.003$), anxiety ($0.12$ SD, $p<0.001$), and happiness ($0.12$ SD, $p<0.001$). When considering only political tweets, we found that anger was by far the predominant emotion amplified by the engagement-based algorithm, both in terms of the emotions expressed by authors ($0.75$ SD, $p<0.001$) and the emotions felt by readers ($0.37$ SD, $p<0.001$).

\textbf{User's stated preference.} For each tweet, we also asked users whether they wanted to see tweets like it when they used Twitter. We found that overall, tweets shown by the engagement-based algorithm are rated slightly higher ($0.06$ SD, $p=0.022$). Interestingly, however, the political tweets recommended by the engagement-based algorithm led to significantly lower user value than the political tweets in the reverse-chronological timeline ($-0.18$ SD, $p=0.005$).

\begin{figure}[t]
    \centering
    \includegraphics[width=0.98\columnwidth]{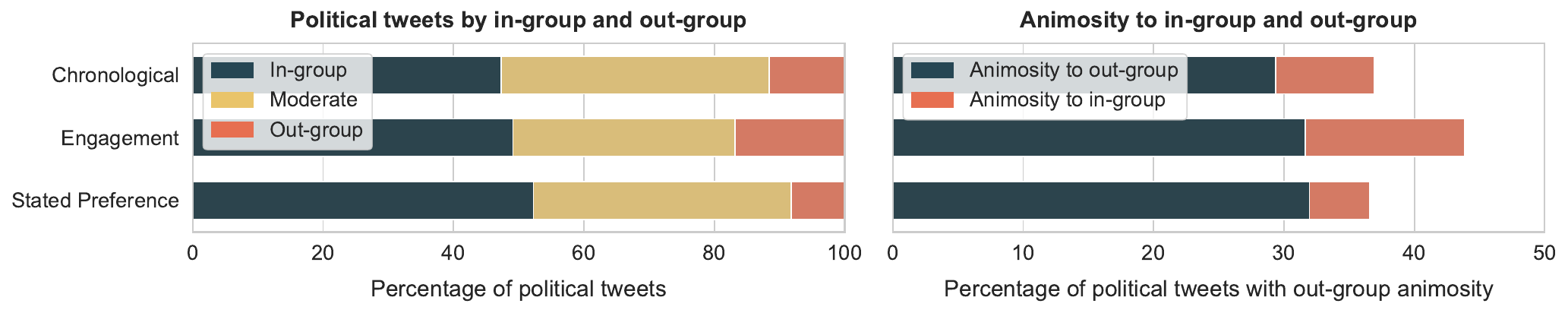}
    \caption{\small \textbf{The distribution of political tweets and out-group animosity.} The graph on the left shows the distribution of political tweets in each timeline, categorized by whether they align with the reader's in-group, out-group, or are moderate. Meanwhile, the graph on the right delineates the proportion of political tweets that express out-group animosity, broken down by whether they target the reader's in-group or out-group. The stated preference timeline has a lower percentage of tweets with animosity than the engagement timeline. However, this decrease is mainly due to a decrease in animosity towards the reader's \emph{in-group}; the percentage of tweets with animosity towards the reader's \emph{out-group} stays roughly the same.}
    \label{fig:oga_breakdown}
\end{figure}

\subsection{Ranking by stated preferences}
Next, we conducted an exploratory analysis in which we simulated an alternative ranking of tweets that is based on users' stated preferences. In particular, we took the approximately twenty unique tweets that each user was surveyed about and re-ranked them by the users' stated preference for the tweet. Each tweet received a score of $1$ = ``Yes'', $0$ = ``Indifferent'', or $-1$ = ``No'', depending on the users' stated preference for the tweet.\footnote{If a tweet was a quote tweet or a reply to another tweet, then we combined the two tweets into one ``tweet object'' and score it by the average of the users' stated preferences for both tweets.} The ``Stated Preference'' (SP) timeline consisted of the ten tweets (out of the approximately twenty unique tweets) that scored highest according to the users' stated preferences (ties are broken at random). We chose the top ten tweets to maintain consistency with the engagement-based and reverse-chronological timelines (where we also considered only the top ten tweets).
 
 As shown in Figure \ref{fig:effects}, relative to the engagement timeline, the SP timeline reduced negativity (anger, sadness, anxiety) and increased happiness, both in terms of the emotions expressed by authors and the emotions felt by readers. Moreover, the content shown in the SP timeline is less partisan and less likely to contain expressions of out-group animosity. However, further analysis shown in \Cref{fig:oga_breakdown} reveals that the reduction in partisanship and animosity in the SP timeline is due almost entirely to reducing the number of tweets from the reader's out-group, and by reducing animosity directed toward the reader's in-group (but not animosity towards the reader's out-group). 

\section{Discussion} \label{sec:disc}

\paragraph{The engagement-based algorithm does not cater to users' stated preferences.} For tweets overall, we found that users had only a slight preference for tweets in their engagement-based timeline compared to tweets in their reverse chronological timeline. Moreover, users were \emph{less} likely to prefer the political tweets that the engagement-based algorithm selected. This is notable because, as measured by internal A/B tests at Twitter, the engagement-based algorithm causes users to spend more time on the platform, relative to the reverse-chronological timeline~\cite{milli2022causal,bandy2023exposure}. Thus, the engagement-based algorithm seems to be catering to users' \emph{revealed preferences} (in terms of engagement and usage patterns) but not to their \emph{stated preferences}, particularly when it comes to political content. \citet{rathje2022people} found that U.S. adults believe social media platforms amplify negative, emotionally-charged, and out-group hostile content but should not. This perspective may help explain why users were dissatisfied with the political tweets selected by the engagement-based algorithm, which tended to be angrier and more out-group hostile. The fact that the engagement-based algorithm is not aligned with users' stated preference for tweets also suggests that its effects cannot entirely be explained by users' deliberative choices.

\paragraph{The engagement-based algorithm amplifies partisan, emotional, out-group hostile tweets.} We found that the engagement-based algorithm tends to select more emotionally-charged, partisan, out-group hostile content than both the reverse-chronological timeline and the stated preference timeline. This finding helps clarify inconsistencies in previous research. Earlier studies relied on observational methods that were limited to specific subsets of content on the platform~\cite{brady2017emotion,rathje2021out} and replications produced mixed  results~\cite{burton2021reconsidering,brady2021estimating}. These studies found a correlation between emotion and engagement but did not control for who users followed, leaving open the possibility that the observed correlation was due to users' choices of who to follow. Our study provides insight into this debate by providing evidence that the engagement-based algorithm amplifies emotionally charged content, beyond users' choices of who to follow, and even beyond what users' stated preferences at the tweet-level would indicate. It also provides evidence for the theory of \citet{brady2023algorithm} who suggest that ranking algorithms exploit humans' natural social biases towards moralized, emotional content.

 \paragraph{The engagement-based timeline may be more polarizing than the reverse-chronological timeline}  We also asked users how they felt about their political in-group and out-group (from ``Much worse'' to ``Much better'') after each individual piece of content they saw from either the engagement-based or reverse-chronological timeline. We found that, after reading tweets selected by the engagement-based algorithm, users tended to have more positive perceptions of their in-group and more negative perceptions of their out-group, compared to the reverse-chronological timeline. Our results suggest that the engagement-based ranking algorithm selects more polarizing content compared to what would be expected from users' following choices alone.
 
 However, it is unclear whether ranking algorithms have lasting effects on users' attitudes and affective polarization. In contrast to our content-specific surveying, Guess et al. (2023) measured the effect of the Facebook and Instagram ranking algorithms on users' affective polarization, in general, between two weeks to three months after the start of their experiment~\cite{guess2023social}. They found no significant impact on users' affective polarization. Crucially, however, the scope of Guess et al. (2023)'s research does not encompass ``general equilibrium" effects, such as the possibility for ranking algorithms to indirectly shape user attitudes by incentivizing and changing the type of content that users produce in the first place.

\paragraph{Ranking by stated preferences may reinforce in-group bias.} We also found that while the stated preference timeline had fewer partisan tweets than the engagement-based timeline, this was primarily due to a reduction in tweets from the reader's political out-group. Similarly, the lower out-group animosity in the stated preference timeline (relative to the engagement-based timeline) was primarily due to reducing hostility towards the readers' in-group (but not towards their out-group). And we did not find a significant difference between the engagement-based timeline and the stated preference timeline, when it came to effects on readers' perceptions of their in-group and out-group. Overall, these results suggests that it is possible that ranking by stated preferences could \emph{heighten} users' exposure to content that reinforces their pre-existing beliefs, relative to the engagement-based algorithm. This is in line with prior research suggesting that users' preference for pro-attitudinal political content and sources is a key factor contributing to greater exposure to in-group content compared to out-group content~\cite{bakshy2015exposure,gonzalez2023asymmetric,wojcieszak2022most}. 

Our findings contrast with that of \citet{agan2023automating}. They posit that algorithms that learn from users' reflexive engagement tend to replicate automatic ``system 1" biases. As supporting evidence, they show that, on Facebook, ranking content based on engagement increases exposure to in-group content (where in-groups are defined by race or religion), whereas ranking by stated preferences reduces this exposure. However, our results reveal an opposite trend when defining in-groups and out-groups by political affiliation. This might indicate that political in-group preferences are not solely “system 1” biases but also involve more deliberate “system 2” preferences. This distinction between political out-groups and other societal out-groups may also be an important distinction for the theory proposed by \citet{brady2023algorithm}, which also posits that ranking algorithms capitalize on humans' social biases towards in-group favoritism. 

An important caveat in interpreting our results is that we only had a pool of approximately twenty tweets---the top ten in the users' chronological and engagement timelines---to choose from when ranking by stated preferences. The platform has access to a much larger pool of tweets, and more research is needed to understand the impact of ranking by stated preferences in such a context. For example, users might be open to seeing respectful and civil content from their political out-group, but those tweets may not have been common in our limited pool of tweets---especially since about half were explicitly selected using engagement metrics.

It also may be that variants of the stated preference timeline can avoid reinforcing in-group bias. In \Cref{app:sp-oa}, we investigate a slight modification to the stated preference timeline, one which involves tie-breaking content ranking based on the presence of out-group animosity, rather than random tie-breaking. This approach is similar to that of \citet{jia2024embedding} who suggest ranking based on ``societal objective functions'', e.g., an objective that prioritizes content with pro-democratic attitudes. Compared to the engagement-based and reverse-chronological timeline, we find that this modified SP timeline greatly reduces the amplification of out-group hostile content while avoiding heightened exposure to in-group content.

\paragraph{Implications on algorithm design.} On real-world platforms, users' stated preferences (in the form of user surveys or user controls like a ``Do not recommend'' button) are rarely observed, making engagement-based signals still the predominant target for ranking (see \cite{cunningham2024know} for a review of the use of non-engagement-based signals for ranking). Nonetheless, recent research in computer science has explored more effective methods for incorporating stated preference signals~\citep{milli2021optimizing,agarwal2024system,kleinberg2024challenge}, making it increasingly feasible to optimize for these preferences. A fundamental, yet largely untested, assumption in this line of work is that aligning algorithms with users’ stated preferences can lead to improved outcomes for both individuals and society. Our study contributes to addressing this gap by showing that optimizing for stated preferences could decrease the prominence of emotionally charged, divisive content, which others have argued can lead to greater social misperception and conflict~\citep{brady2023algorithm}. However, there is also a possibility that ranking by stated preferences could increase exposure to pro-attitudinal content, which warrants further investigation.

\subsection{Limitations}
There are several limitations to our study that should be regarded when interpreting our results. First, our study analyzed the difference in users' engagement-based and reverse-chronological timelines during only one point in time. Such an approach does not capture the engagement-based algorithm's long-term effects. For example, by incentivizing some types of content over others, the algorithm changes what type of content is produced in the first place, which, in turn, affects what is available in the reverse-chronological timeline. If users have learned to produce more of the content that the engagement-based algorithm incentivizes~\cite{brady2023algorithm,cen2023user}, then its long-term effects on the content-based outcomes we measure (emotions, partisanship, and out-group animosity in tweets) may be even greater. On the other hand, other research has failed to find long-term changes in user beliefs due to social media ranking algorithms~\citep{guess2023social,guess2023reshares,nyhan2023like,haroon2023nudging}, and the short-term changes in users' perceptions of their in-group and out-group that we found may not persist. Rigorous experimentation needs to be done to establish how all the effects found in our study change over time in response to feedback loops~\cite{thorburn_2023}.

Second, we required participants to install a Chrome extension and relied on an online crowd-working platform for participant recruitment, and this may have impacted the types of participants we could attract. In particular, compared to Twitter users in the 2020 ANES study~\cite{ANES2020}, our population tended to be younger (53\% of our study were aged 18-34 years old, compared to 33\% in the ANES study) and more likely to affiliate with the Democratic Party (56\% Democrat in our study versus 43\% in the ANES study) (see \Cref{appendix:user-demographics} for full demographic statistics). We report on heterogenous effects by different demographic groups in \Cref{appendix:heterogenous-effects}, however, more research may be warranted before generalizing to the full Twitter population.

Another concern pertains to the ecological validity of our study. We surveyed participants about each of the first ten tweets in each timeline, a method that deviates from real-world Twitter usage where users can scroll past various tweets. Looking at the first ten tweets might be a reasonable approximation though---Bouchaud et al. (2023) found that users viewed an average of 30 tweets in one session of using Twitter~\cite{bouchaud2023crowdsourced}. In \Cref{appendix:robustness_to_tweet_threshold}, we show that our results are robust to using a range of thresholds, spanning from the first five tweets to the first ten tweets. Still, the effects of real-world social media use may not be accurately captured by averaging the impact of tweets shown to users. For instance, users may only be influenced by a few particularly notable tweets as they scroll through their timeline.

There is also the possibility that respondents experienced survey fatigue, as we repeatedly asked them about twenty tweets. To address this, we included attention checks in the survey and excluded participants who failed these checks. Nonetheless, we cannot completely rule out the effect of fatigue on the study's results. Additionally, when evaluating certain tweet-related outcomes (e.g., perceived partisanship or emotional tone) based on reader perception, participants’ own backgrounds is likely to influence their preceptions~\citep{brady2023overperception}. Although we conducted parallel analyses using GPT-4 to label these outcomes (see \Cref{appendix:gpt-effects}), large language models have their own biases~\citep{santurkar2023whose,dominguez2023questioning} and are limited to analyzing only the textual content of the tweets, excluding images, videos, and linked websites. The similar results from both GPT-4 and reader assessments support the robustness of our findings. Still, it is essential to consider the complementary biases from both readers and GPT-4 when interpreting both sets of results.

Finally, throughout this paper, we frequently discuss disentangling the effects of ranking algorithms from intentional human decisions such as which accounts to follow. However, over a longer period, ranking algorithms also shape these choices since users are more inclined to follow accounts that appear in their timelines. Consequently, even when using an engagement-based timeline, users’ follow decisions—shaped by the engagement-based algorithm’s recommendations—can gradually influence their reverse-chronological timeline, which is based on their followed accounts. This means that, over time, the engagement-based algorithm indirectly affects the content of the reverse-chronological timeline. Our study does not account for the potential long-term impact of the engagement-based algorithm on users’ choices of who to follow.

\clearpage
\subsection*{Acknowledgements}
We thank Katy Glenn Bass, Fabian Baumann, William Brady, Molly Crockett, Tom Cunningham, Ravi Iyer, Philipp Lorenz-Spreen, Solomon Messing, Aviv Ovadya, Sara Johansen, Arvind Narayanan, Jonathan Stray, Luke Thorburn, and Nina Vasan for their discussion and feedback.

\subsection*{Funding.} This study was financially supported by the UC Berkeley Center for Human-Compatible AI. MC was supported by the National Science Foundation Graduate Research Fellowship Program. Authors declare no competing interests.

\subsection*{Author contributions.} 

Conceptualization: SM

Methodology: SM, MC

Software: SM, SP, MC, YW, SZ

Formal analysis: SM, MC, YW, SP, SZ

Investigation: SM, MC, YW, SP

Data curation: YW, SP, SM, MC, SZ

Writing - original draft: SM, MC, YW, SP, SZ

Writing - review \& editing: SM, MC, YW, SP, SZ, AD

Supervision: SM, AD

Project administration: SM, AD

Funding acquisition: SM, AD

\subsection*{Data availability.} The data and code needed to reproduce our results are available to the research community at \url{http://github.com/smilli/twitter}.


\clearpage

\bibliography{refs}
\bibliographystyle{plainnat}

\clearpage
\begin{appendix}

\renewcommand{\thesection}{S\arabic{section}}
\renewcommand{\thesubsection}{\thesection.\arabic{subsection}}

\doparttoc
\faketableofcontents
\part{}
\part{Supplementary Materials} 
\parttoc 
\newpage
\section{Materials and methods} \label{app:exp-design}

\subsection{Study procedure} 

We conducted our study between February 11 to February 27, 2023 on CloudResearch Connect\footnote{Prior academic research has found that participants recruited on CloudResearch or Prolific tend to provide higher-quality data than those recruited from Mechanical Turk or Qualtrics~\citep{douglas2023,eyal2021data}. We selected CloudResearch over Prolific because of its functionality that enables requesting that participants download a Chrome extension. CloudResearch enforces one account per participant, verifies that each IP address aligns with the reported location, and ensures that the bank or PayPal accounts for cashing out are unique to each participant. Participants are only allowed to sign up for the platform if they pass an onboarding process that checks for properties such as attention, language comprehension, honesty, and their ability to follow instructions. To maintain high data quality, CloudResearch continually monitors participants, using random attention checks and investigating those who are frequently rejected or flagged by researchers~\citep{hartman2023introducing}.}, a crowd-working platform. Informed consent was obtained from all participants. The study was approved by UC Berkeley's IRB under the CPHS protocol ID number 2021-09-14618 and complies with all ethical regulations. Moreover, at the end of the study period, our rating (given by study participants) on CloudResearch was 4.9 stars, higher than 99 percent of other researchers on the platform. 

The study period was broken into four waves and participants could complete the study once during each wave.\footnote{To prevent repeat participation, we ensured that each Twitter user ID could only be used once per wave. For a participant to participate multiple times in a single wave, they would have needed to create multiple Twitter accounts and bypass CloudResearch's safeguards to register multiple CloudResearch Connect accounts.} The time periods for the waves (inclusive) were 02/11-02/14, 02/16-02/19, 02/21-02/23, and 02/25-02/27. Every day, we recruited up to 150 eligible participants who lived in the United States, were at least 18 years old, and used Google Chrome. Furthermore, participants were required to use Twitter at least a few times a week and follow at least 50 people on Twitter (both gauged through self-reports). To collect data, participants were directed to download a Chrome extension that we developed which scraped their Twitter homepage to collect the top tweets from their personalized timeline. While scraping, the Chrome extension added an overlay to the homepage that prevented the user from seeing the tweets during collection. At the same time that the personalized tweets were collected, we queried the Twitter API to get the top tweets from the chronological timeline. Only public tweets were collected and no promoted tweets (advertisements) were collected.

\begin{figure}
    \centering
    \includegraphics[width=0.95\textwidth]{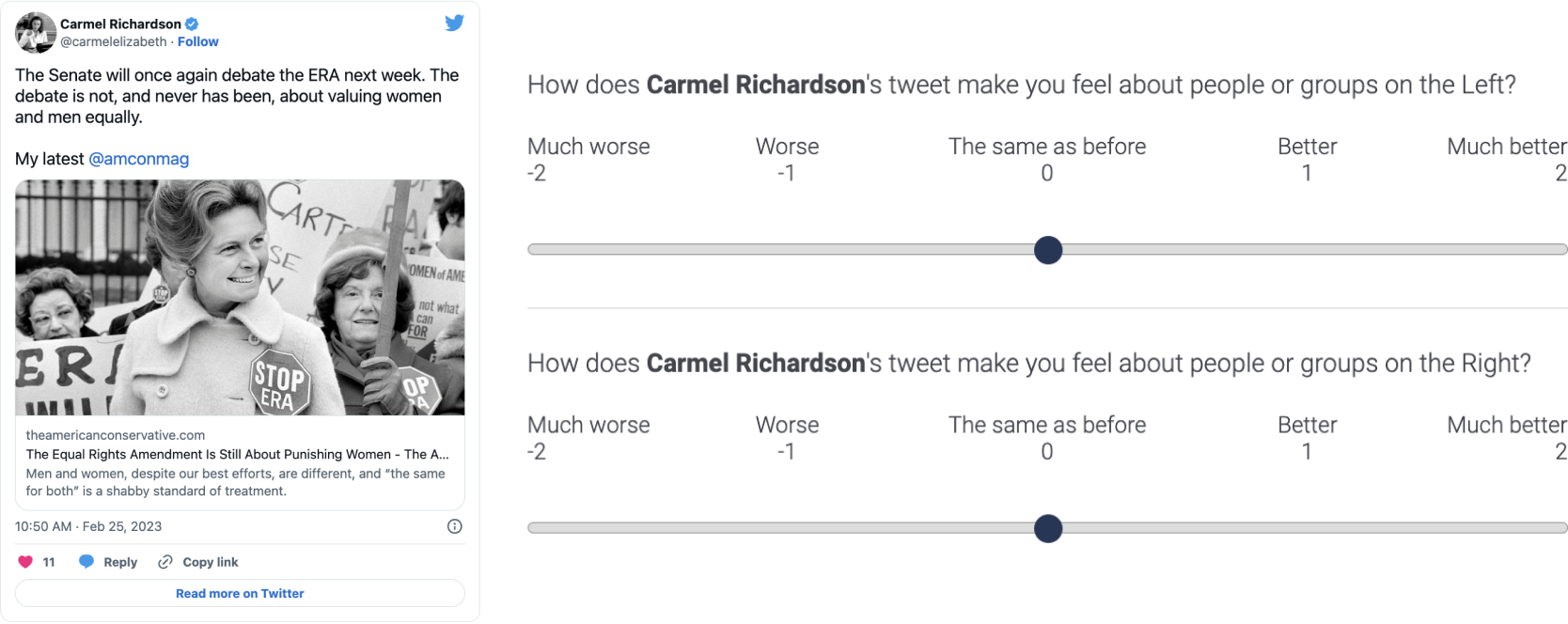}
    \caption{Sample view of the survey: users saw tweets embedded alongside each question for reference.}
    \label{fig:embedded-tweet}
\end{figure}

After collecting both sets of tweets, participants were directed to complete a survey on Qualtrics that asked questions about each of the top ten tweets from their personalized and chronological timeline. All tweets are displayed in a randomized order (thus, tweets from both timelines are typically interwoven rather than, for example, first showing all the personalized and then all chronological tweets). If the same tweet was present in both the personalized and chronological timeline, then participants were only shown it once. Out of the 10 tweets in both timelines, on average, 2.35 of these tweets were common to both of the timelines. On each question, the tweet is displayed for reference as an embedded tweet (see \Cref{fig:embedded-tweet}), so it looks as similar as possible to the way it would on Twitter. If a tweet is a reply to another tweet, the user is shown both the replied tweet and the main tweet, and asked to answer the questions for both tweets. Similarly, if a tweet is a quote tweet, then the user is asked to answer the questions for both the quoted tweet and the main tweet.

The outcomes we measured concerned the emotions expressed by the author (on four axes: anger, sadness, anxiety, and happiness), the reader emotions (on four axes: anger, sadness, anxiety, and happiness), the author's expression of out-group animosity, the partisan leaning of the tweet, the readers' in-group and out-group perception after reading the tweet, and readers' explicit preference for the tweet. The full survey is provided in \Cref{app:survey}, but the question phrasings and their possible answers are provided below for quicker reference:

\begin{itemize}
    \item \textbf{Author emotions questions.} Participants are asked: ``How is [@author-handle] feeling in their tweet?'' For each emotion (angry, anxious, happy, sad), the participant responds on a Likert scale of “Not at all,” “Slightly,” “Somewhat,” “Moderately,” or “Extremely.”
    \item \textbf{Reader emotions.} Participants are asked: ``How did [@author-handle]'s tweet make you feel?'' For each emotion (angry, anxious, happy, sad), the participant responds on a Likert scale of “Not at all,” “Slightly,” “Somewhat,” “Moderately,” or “Extremely.”
    \item \textbf{Is tweet political.} Participants are asked: ``Is [@author-handle]'s tweet about a political or social issue?'' with a binary response of ``Yes'' or ``No''
    \item \textbf{Political leaning.} Participants are asked: ``How does [@author-handle]'s tweet lean politically?'' Participants respond on a Likert scale of “Far left,” “Left,” “Moderate,” “Right,” or “Far right.”
    \item \textbf{Political affect towards Left.} Participants are asked: ``How does [@author-handle]'s tweet make you feel about people or groups on the Left?'' Participants respond on a Likert scale of ``Much worse,'' ``Worse,'' ``The same as before,'' ``Better,'', or ``Much better.''
    \item \textbf{Political affect towards Right.} Participants are asked: ``How does [@author-handle]'s tweet make you feel about people or groups on the Right?'' Participants respond on a Likert scale of ``Much worse,'' ``Worse,'' ``The same as before,'', ``Better,'' or ``Much better,''
    \item \textbf{Out-group animosity towards the Left.} Participants are asked: ``Is [@author-handle]'s tweet expressing anger, frustration, or hostility towards a person or group on the Left?'' Participants choose between a response of ``Yes'' or ``No'' This question is only asked if political leaning is “Right” or “Far right.”
    \item \textbf{Out-group animosity towards the Right.} Participants are asked: ``Is [@author-handle]'s tweet expressing anger, frustration, or hostility towards a person or group on the Right?'' Participants choose between a response of ``Yes'' or ``No.'' This question is only asked if political leaning is “Left” or “Far left.”
    \item \textbf{Value.} Participants are asked: ``When you use Twitter, do you want to be shown tweets like [@author-handle]'s tweet?'' Participants choose between “No,” “Indifferent,” “Yes.”
\end{itemize}

\textbf{Data exclusions.} Those who passed the pre-screen but did not complete data collection with the Chrome extension were paid \$0.75, while those who completed data collection and the full survey (which takes approximately 30 minutes) were paid \$10 (a rate of \$20/hr). Overall, 18 percent of participants who consented to the experiment did not complete data collection with the Chrome extension. This was either because they chose to not use the Chrome extension or ran into an error during data collection. Since the Chrome extension relies on scraping Twitter's homepage to retrieve tweets it may not work if, for example, the user has conflicting Chrome extensions or is in a Twitter A/B test that changes the UI of the Twitter homepage. Furthermore, we also surveyed participants about two attention-check tweets (that changed during each of the four waves) that were unambiguously either left-leaning or right-leaning. If a participant's survey response did not pass the attention checks, it was excluded from the final data set used for analysis. Furthermore, if a participant did not successfully complete the survey (e.g. did not complete data collection or pass the attention checks) in one wave, then they were excluded from participating in future waves. Our final data set consisted of 1730 responses from 806 unique participants. \Cref{tab:waves} shows statistics on the attrition of participants from consenting to passing the attention checks.

\subsection{Pre-registration and deviations} 
All our hypotheses, outcome measures, and statistical analyses were pre-registered on Open Science Framework (OSF) at \url{https://osf.io/upw9a}. However, we made the following deviations from the pre-registration. First, we opted to use CloudResearch Connect instead of CloudResearch's Mechanical Turk toolbelt for participant recruitment because CloudResearch Connect had features better suited for our study, such as the ability to flag our study as requiring a software download. Second, due to limitations in the number of unique Twitter users we could recruit on CloudResearch Connect, we divided our data collection period into four intervals and allowed participants to complete the study once in each interval. Third, we underestimated the study's costs, resulting in us stopping data collection when we reached our budget (at around 1700 pairs of timelines) instead of our initial target of 2000 pairs of timelines. And finally, we asked participants to rate out-group animosity on a five-point Likert scale from ``Not at all'' to ``Extremely'' for main tweets but accidentally left the response options for quoted and replied tweets as binary ``Yes'' or ``No'' options. Thus, for analysis, we collapse the Likert responses to a binary scale where ``Not at all'' corresponds to ``No'' and anything higher corresponds to ``Yes.''

\begin{center}
\begin{table}[t]
\begin{tabularx}{\textwidth}{@{} ll @{}}
\toprule
\textbf{Times Participant Completed Survey} & \textbf{Number of Unique Participants} \\
\midrule
1 & 359 \\
2 & 139 \\
3 & 139 \\
4 & 169 \\
\midrule
\textbf{Total} & \textbf{806} \\
\bottomrule
\end{tabularx}
\caption{The table displays the distribution of the number of times that each participant took part in the study. Most participants only participated once, but many participated two or more times.}
\end{table}
\end{center}

\begin{center}
\begin{table}[t]
\begin{tabularx}{\textwidth}{@{} lllll @{}}
\toprule
\textbf{Wave}  & \textbf{Consented} & \textbf{Collected Data} & \textbf{Completed Survey} & \textbf{Passed  Attention Check}\\
\midrule
\textbf{1}  & 816 & 599 & 586 & 508 \\
\textbf{2}& 688 & 585 & 574 & 454 \\
\textbf{3}  & 514 & 440 & 434 & 405 \\
\textbf{4} & 486 & 441 & 432 & 363 \\
\midrule
\textbf{Total} & \textbf{2504} & \textbf{2065} & \textbf{2026} & \textbf{1730} \\
\bottomrule
\end{tabularx}
\caption{The table displays, for each study period, the number of participants who consented to the study, successfully completed the data collection through the Chrome extension, completed the survey, and passed the attention check.}
\label{tab:waves}
\end{table}
\end{center}

\newpage
\subsection{Estimation of average treatment effects} \label{app:ate-estimation}
 As specified in our pre-registered analysis plan, our main estimation of the average treatment effect is through a simple difference in mean outcomes between the personalized and chronological timelines. Let $\mathcal{U} = \{1, 2, \dots, n\}$ be the set of participants and $\mathcal{W}_u \subseteq \{1, 2, 3, 4\}$ be the set of study waves that a user $u \in \mathcal{U}$ participates in. During study wave $w \in \mathcal{W}_u$, participant $u$ rates both a set of tweets $\mathcal{T}_{w,u}(1)$ from their engagement-based timeline (denoted as timeline $1$) and a set of tweets $\mathcal{T}_{w,u}(0)$ from their chronological timeline (denoted as timeline $0$). The number of tweets that each participant rates is approximately ten from each timeline, i.e., $|\mathcal{T}_{w,u}(1)| \approx 10$ and $|\mathcal{T}_{w,u}(0)| \approx 10$. However, this number can be higher if a timeline contains quote tweets or replies. For both quote tweets and replies, participants rate two tweets: the quote tweet and the tweet being quoted or the reply and the tweet being replied to.  

Let $Y_{w, u, t, q}(x)$ be the response of participant $u$ to question $q$ on tweet $t \in \mathcal{T}_{w,u}(x)$ from timeline $x \in \{0, 1\}$ in study wave $w$.  Certain questions (partisanship, out-group animosity, in-group and out-group perception) are only asked if a participant labeled the tweet as being a political tweet. If the participant did not label the tweet as being political, then we let $Y_{w, u, t, q}(x) = 0$ for those questions. This corresponds to assuming that the tweet does not have a partisan leaning, does not contain out-group animosity, and did not impact the readers' perception of their in-group or out-group.

Then, let 
\begin{align}
\widebar{Y}_{u, q}(x) = \sum_{w \in \mathcal{W}_u} \frac{1}{|\mathcal{W}_u|} \sum_{t \in \mathcal{T}_{w,u}(x)} \frac{Y_{w, u, t, q}(x)}{|\mathcal{T}_{w,u}(x)|}
\end{align}
be the mean outcome on question $q$ across the tweets in timeline $x$ for individual $u$. The individual effect for participant $u$ on question $q$ is estimated as $\widebar{Y}_{u, q}(1) - \widebar{Y}_{u, q}(0)$. Finally, the average treatment effect on question $q$ across participants is estimated as 
\begin{align}
    \widehat{ATE}(q) = \frac{\sum_{u} \widebar{Y}_{u, q}(1) - \widebar{Y}_{u, q}(0)}{|\mathcal{U}|} \,.
\end{align}

In addition to considering tweets overall, we also look at the effect on authors' emotions, readers' emotions, and the users' stated preference for political tweets only. Let $\mathcal{T}^{\text{pol}}_{w, u}(x)$ be the tweets that participant $u$ labeled as being political in study wave $w$ and timeline $x$. Furthermore, let $\mathcal{W}_u^{\text{pol}}$ be the set of waves in which participant $u$ had at least one political tweet in both the chronological and engagement timeline. Then, let
\begin{align}
\widebar{Y}^{\text{pol}}_{u, q}(x) = \sum_{w \in \mathcal{W}_u^{\text{pol}}} \frac{1}{|\mathcal{W}_u^{\text{pol}}|} \sum_{t\in \mathcal{T}^{\text{pol}}_{w, u}(x)} \frac{Y_{w, u, t, q}(x)}{|\mathcal{T}^{\text{pol}}_{w, u}(x)|}
\end{align}
be the participant's mean outcome for question $q$ when considering only political tweets. Then, the average treatment effect for political tweets only is
\begin{align}
    \widehat{ATE}_{\text{pol}}(q) = \frac{\sum_{u \in \mathcal{U}_{\text{pol}}} \widebar{Y}^{\text{pol}}_{u, q}(1) - \widebar{Y}^{\text{pol}}_{u, q}(0)}{|\mathcal{U}_{\text{pol}}|} \,,
\end{align}
where $\mathcal{U}_{\text{pol}}$ is the set of participants with at least one political tweet in each timeline.

\textbf{Significance and multiple testing.} To compute $p$-values, we conducted two-tailed paired permutation tests. A participant’s responses to their personalized tweets and responses to their chronological tweets are considered paired data, i.e., $\widebar{Y}_{u, q}(1)$ and $\widebar{Y}_{u, q}(0)$, or $\widebar{Y}^{\text{pol}}_{u, q}(1)$ and $\widebar{Y}^{\text{pol}}_{u, q}(0)$ are considered paired data. We used bootstrap sampling to calculate the 95\% confidence intervals displayed in \Cref{fig:effects}. To adjust for the possibility of chance findings when conducting multiple tests, we used the Benjamini-Krieger-Yekutieli two-stage method~\cite{benjamini2006adaptive}, as described by~\cite{anderson2008multiple}, to compute false discovery rate (FDR) adjusted $p$-values. The FDR-adjusted $p$-value (sometimes called a \emph{$q$-value}) for each test is the lowest false discovery rate such that the test would still be rejected at a  significance level of $0.05$. We test 26 outcomes in total, and all significant results (significant at a $p$-value threshold of $0.05$) remain significant at a false discovery rate of $0.01$. Thus, on expectation, none of our discoveries are false discoveries. The full table of standardized and unstandardized effect sizes, $p$-values, and FDR-adjusted $p$-values can be found in \Cref{appendix:full-effects}.

\textbf{Effect size standardization.} When displaying the effect sizes for all outcomes in \Cref{fig:effects}, we standardize using the standard deviation of the outcome in the chronological timeline, i.e., $\sigma(\widebar{Y}_{u, q}(0))$. Note that we do not use the pooled standard deviation between the chronological and engagement-based timeline as would be done in Cohen's $d$. The standardized effect size we present is also known as Glass's $\Delta$ and uses the standard deviation of the control group in order to allow comparison between multiple treatments, i.e., not just Twitter's engagement-based algorithm but also our alternative ranking algorithm based on stated preferences. One exception is that when calculating the standardized effect sizes for author emotion, reader emotion, and explicit value for \emph{political} tweets only, we standardize the effect using the standard deviation of that outcome across \emph{all} tweets. This was done to ensure that the effect sizes found when restricting to political tweets could be compared with the effect sizes observed for all tweets, i.e., we use $\sigma(\widebar{Y}_{u, q}(0))$ instead of $\sigma(\widebar{Y}^{\text{pol}}_{u, q}(0))$. 
\newpage
\section{User-level survey questions and demographics}\label{appendix:user-demographics}
In this section, we report on the demographic distribution of our participants and compare our distribution to that of Twitter users from the 2020 ANES study~\cite{ANES2020}. The primary differences are that our population is younger (53 percent of our study are aged 18-34 years old, compared to 33\ percent in the ANES study) and more likely to affiliate with the Democratic Party (56 percent Democrat in our study versus 43 percent in the ANES study).

\subsection{Demographics of our study}
Participants in our study had the opportunity to take part in up to four waves, with demographic information collected at each wave. In total, 806 unique users participated in our study 1730 times across waves. The following tables report demographics across all users. For users that participated in multiple waves, we report on their demographic information from their first wave. In addition to demographic questions, we also asked two user-level survey questions about the primary reason participants use Twitter and the primary type of content they saw in the tweets we showed them (\Cref{tab:primary-content-category}).

\begin{center}
\begin{table}[h]
\begin{tabularx}{\textwidth}{@{} p{10cm}cc @{}}
\toprule
\textbf{Race (Our Study)} & $\mathbf{n}$ & \textbf{Percent} \\
\midrule
White & 537 & 66.63 \\
Black or African American & 89 & 11.04 \\
Hispanic & 81 & 10.05 \\
Asian or Native Hawaiian/other Pacific Islander & 63 & 7.82 \\
American Indian/Alaska Native or Other & 8 & 0.99 \\
Multiple races, non-Hispanic & 28 & 3.47 \\
\textbf{Total} & \textbf{806} & \textbf{100} \\
\bottomrule
\end{tabularx}
\caption{We ask two separate questions about race and ethnicity to participants, and for comparison, we combine them together in the same way that the 2020 ANES study does.}
\end{table}
\end{center}

\begin{center}
\begin{tabularx}{\textwidth}{@{} p{10cm}cc @{}}
\toprule
\textbf{Gender (Our Study)} & $\mathbf{n}$ & \textbf{Percent} \\
\midrule
Man & 410 & 50.87 \\
Woman & 371 & 46.03 \\
Non-binary & 21 & 2.61 \\
Other & 4 & 0.50 \\
\textbf{Total} & \textbf{806} & \textbf{100} \\
\bottomrule
\end{tabularx}
\end{center}

\begin{center}
\begin{tabularx}{\textwidth}{@{} p{10cm}cc @{}}
\toprule
\textbf{Ideological Leaning (Our Study)} & $\mathbf{n}$ & \textbf{Percent} \\
\midrule
Far left & 170 & 21.09 \\
Left & 322 & 39.95 \\
Moderate & 196 & 24.32 \\
Right & 86 & 10.67 \\
Far right & 25 & 3.10 \\
Other & 7 & 0.87 \\
\textbf{Total} & \textbf{806} & \textbf{100} \\
\bottomrule
\end{tabularx}
\end{center}

\begin{center}
\begin{table}[h]
\centering
\begin{tabularx}{\textwidth}{@{} p{10cm}cc @{}}
\toprule
\textbf{Political Leaning Further (Our Study)} & $\mathbf{n}$ & \textbf{Percent} \\
\midrule
Towards the Left & 136 & 62.67 \\
Towards the Right & 81 & 37.33 \\
\textbf{Total} & \textbf{217} & \textbf{100} \\
\bottomrule
\end{tabularx}
\caption{Those who identified as being moderate were asked whether they leaned more toward the Left/Right.}
\end{table}
\end{center}

\begin{center}
\begin{table}[h]
\centering
\begin{tabularx}{\textwidth}{@{} p{10cm}cc @{}}
\toprule
\textbf{Summary Leaning (Our Study)} & $\mathbf{n}$ & \textbf{Percent} \\
\midrule
Left-leaning & 618 & 76.67 \\
Right-leaning & 188 & 23.33 \\
\textbf{Total} & \textbf{806} & \textbf{100} \\
\bottomrule
\end{tabularx}
\caption{Participants who selected ``Moderate'' or ``Other'' in response to the ideological leaning question were asked a follow-up question about whether they lean towards left or right more as of today. The table shows aggregate counts of those who identified as being on the Left/Right or that they leaned more towards the Left/Right.}
\end{table}
\end{center}

\begin{center}
\begin{tabularx}{\textwidth}{@{} p{10cm}cc @{}}
\toprule
\textbf{Political Party (Our Study)} & $\mathbf{n}$ & \textbf{Percent} \\
\midrule
Democrat & 451 & 55.96 \\ 
Republican & 111 & 13.77 \\
Independent & 205 & 25.43 \\
Something else & 39 & 4.84 \\
\textbf{Total} & \textbf{806} & \textbf{100} \\
\bottomrule
\end{tabularx}
\end{center}

\begin{center}
\begin{table}[h]
\centering
\begin{tabularx}{\textwidth}{@{} p{10cm}cc @{}}
\toprule
\textbf{Political Party Further (Our Study)} & $\mathbf{n}$ & \textbf{Percent} \\
\midrule
Republican & 72 & 29.27 \\
Democrat & 174 & 70.73 \\  
\textbf{Total} & \textbf{246} & \textbf{100} \\
\bottomrule
\end{tabularx}
\caption{Participants who selected ``Independent'' or ``Something else'' in response to the political party question were asked a follow-up question about whether they lean towards the Democrat or Republican party more as of today.}
\end{table}
\end{center}

\begin{center}
\begin{table}[h]
\centering
\begin{tabularx}{\textwidth}{@{} p{10cm}cc @{}}
\toprule
\textbf{Summary Party (Our Study)} & $\mathbf{n}$ & \textbf{Percent} \\
\midrule
Republican & 182 & 22.58 \\
Democrat & 624 & 77.42 \\  
\textbf{Total} & \textbf{806} & \textbf{100} \\
\bottomrule
\end{tabularx}
\caption{Participants who selected ``Independent'' or ``Something else'' in response to the political party question were asked a follow-up question about whether they lean towards the Democrat or Republican party more as of today. The table shows aggregate counts of those who said they were Democrat/Republican or lean Democrat/Republican.}
\end{table}
\end{center}

\begin{center}
\begin{tabularx}{\textwidth}{@{} p{10cm}cc @{}}
\toprule
\textbf{Education Levels (Our Study)} & $\mathbf{n}$ & \textbf{Percent} \\
\midrule
Some high school & 5 & 0.62 \\
High school graduate & 242 & 30.02 \\
Associate degree & 102 & 12.66 \\
Bachelor's degree & 310 & 38.46 \\
Master's degree or above & 128 & 15.88 \\
Other & 17 & 2.11 \\
Prefer not to answer & 2 & 0.25 \\
\textbf{Total} & \textbf{806} & \textbf{100} \\
\bottomrule
\end{tabularx}
\end{center}

\begin{center}
\begin{tabularx}{\textwidth}{@{} p{10cm}cc @{}}
\toprule
\textbf{Age Group (Our Study)} & $\mathbf{n}$ & \textbf{Percent} \\
\midrule
18-24 years old & 120 & 14.89 \\
25-34 years old & 307 & 38.09 \\
35-44 years old & 211 & 26.18 \\
45-54 years old & 93 & 11.54 \\
55-64 years old & 51 & 6.33 \\
65-74 years old & 23 & 2.85 \\
75 years or older & 1 & 0.12 \\
\textbf{Total} & \textbf{806} & \textbf{100} \\
\bottomrule
\end{tabularx}
\end{center}

\begin{center}
\begin{tabularx}{\textwidth}{@{} p{10cm}cc @{}}
\toprule
\textbf{Annual Household Income (Our Study)} & $\mathbf{n}$ & \textbf{Percent} \\
\midrule
Less than \$25,000 & 118 & 14.64 \\
\$25,000-\$50,000 & 188 & 23.33 \\ 
\$50,000-\$100,000 & 300 & 37.22 \\ 
\$100,000-\$200,000 & 149 & 18.49 \\
More than \$200,000 & 35 & 4.34 \\
Prefer not to say & 16 & 1.99 \\ 
\textbf{Total} & \textbf{806} & \textbf{100} \\
\bottomrule
\end{tabularx}
\end{center}

\begin{center}
\begin{table}[h]
\begin{tabularx}{\textwidth}{@{} p{10cm}cc @{}}
\toprule
\textbf{Primary Category of Content (Our Study)} & $\mathbf{n}$ & \textbf{Percent} \\
\midrule
Entertainment & 416 & 28.75 \\
Other & 108 & 7.46 \\
Politics & 483 & 33.38 \\
News & 272 & 18.80 \\
Hobbies & 152 & 10.50 \\
Work & 16 & 1.11 \\
\textbf{Total} & \textbf{1447} & \textbf{100} \\
\bottomrule
\end{tabularx}
\caption{At the end of the study, we ask participants about the content shown to them: "What were the tweets we showed you today predominantly about? Select a maximum of two."}
\label{tab:primary-content-category}
\end{table}
\end{center}

\begin{center}
\begin{table}[h]
\begin{tabularx}{\textwidth}{@{} p{10cm}cc @{}}
\toprule
\textbf{Primary Reason for Using Twitter (Our Study)} & $\mathbf{n}$ & \textbf{Percent} \\
\midrule
A way to stay informed & 293 & 36.35 \\
Entertainment & 393 & 48.76 \\
Keeping me connected to other people & 53 & 6.58 \\
It's useful for my job or school & 28 & 3.47 \\
Lets me see different points of view & 18 & 2.23 \\
A way to express my opinions & 21 & 2.61 \\
\textbf{Total} & \textbf{806} & \textbf{100} \\
\bottomrule
\end{tabularx}
\caption{We ask participants about the primary reason they use Twitter: "What would you say is the main reason you use Twitter?"}
\end{table}
\end{center}

\subsection{Demographics of Twitter users in ANES 2020 study}
In this subsection, we present demographic distributions for the 1030 participants in the 2020 ANES study~\cite{ANES2020} who reported using Twitter at least a few times a week (only those who reported using Twitter at least a few times a week were allowed to participate in our study). 
For certain demographic questions, participants were not given the option to opt out in our study, but they were in the ANES study. Thus, for those cases, we do not include the ANES participants who opted out of the question.

Comparing the demographics of users in the ANES 2020 study to those of users in our study, the largest differences are that our population is younger (53 percent of our study are aged 18-34 years old, compared to 33 percent in the ANES study) and more likely to affiliate with the Democratic Party (56 percent Democrat in our study versus 43 percent in the ANES study). That said, all distributions over demographic attributes are significantly different (when using a chi-squared test), apart from sex/gender.

\begin{table}[H]
\begin{center}
\begin{tabularx}{\textwidth}{@{} p{8cm}cccc @{}}
\toprule
\textbf{Race/Ethnicity} & \multicolumn{2}{c}{\textbf{ANES}} & \multicolumn{2}{c}{\textbf{Our Study}} \\
\cmidrule(lr){2-3} \cmidrule(lr){4-5}
& $n$ & Percent & $n$ & Percent \\
\midrule
White                                           & 729 & 71.19 & 537 & 66.63 \\
Black or African American                       &  71 &  6.93 &  89 & 11.04 \\
Hispanic                                        & 113 & 11.04 &  81 & 10.05 \\
Asian or Native Hawaiian/other Pacific Islander &  49 &  4.79 &  63 &  7.82 \\
American Indian/Alaska Native or Other          &  25 &  2.44 &   8 &  0.99 \\
Multiple races, non-Hispanic                    &  37 &  3.61 &  28 &  3.47 \\
\midrule
\textbf{Total} & \textbf{1024} & \textbf{100.00} & \textbf{806} & \textbf{100.00} \\
\bottomrule
\end{tabularx}
\end{center}
\caption{Distribution of race and ethnicity comparing 2020 ANES Twitter population with our study population ($\chi^2(5) = 22.53$, $p < .001$).}
\end{table}

\begin{table}[H]
\begin{center}
\begin{tabularx}{\textwidth}{@{} p{8cm}cccc @{}}
\toprule
\textbf{Sex/Gender} & \multicolumn{2}{c}{\textbf{ANES}} & \multicolumn{2}{c}{\textbf{Our Study}} \\
\cmidrule(lr){2-3} \cmidrule(lr){4-5}
& $n$ & Percent & $n$ & Percent \\
\midrule
Male/Man      & 558 & 54.39 & 410 & 52.50 \\
Female/Woman  & 468 & 45.61 & 371 & 47.50 \\
\midrule
\textbf{Total} & \textbf{1026} & \textbf{100.00} & \textbf{781} & \textbf{100.00} \\
\bottomrule
\end{tabularx}
\end{center}
\caption{Distribution of sex/gender comparing 2020 ANES Twitter population with our study population ($\chi^2(1) = 0.56$, $p = .453$).}
\end{table}

\begin{table}[H]
\begin{center}
\begin{tabularx}{\textwidth}{@{} p{8cm}cccc @{}}
\toprule
\textbf{Political Party} & \multicolumn{2}{c}{\textbf{ANES}} & \multicolumn{2}{c}{\textbf{Our Study}} \\
\cmidrule(lr){2-3} \cmidrule(lr){4-5}
& $n$ & Percent & $n$ & Percent \\
\midrule
Democrat    & 439 & 42.75 & 451 & 55.96 \\
Republican  & 222 & 21.62 & 111 & 13.77 \\
Independent & 338 & 32.91 & 205 & 25.43 \\
Other       &  28 &  2.73 &  39 &  4.84 \\
\midrule
\textbf{Total} & \textbf{1027} & \textbf{100.00} & \textbf{806} & \textbf{100.00} \\
\bottomrule
\end{tabularx}
\end{center}
\caption{Distribution of political party affiliation comparing 2020 ANES Twitter population with our study population ($\chi^2(3) = 45.56$, $p < .001$).}
\end{table}

\begin{table}[H]
\begin{center}
\begin{tabularx}{\textwidth}{@{} p{8cm}cccc @{}}
\toprule
\textbf{Education Level} & \multicolumn{2}{c}{\textbf{ANES}} & \multicolumn{2}{c}{\textbf{Our Study}} \\
\cmidrule(lr){2-3} \cmidrule(lr){4-5}
& $n$ & Percent & $n$ & Percent \\
\midrule
Some high school          &   20 &  1.94 &    5 &  0.62 \\
High school graduate      &  285 & 27.67 &  242 & 30.02 \\
Associate degree          &   97 &  9.42 &  102 & 12.66 \\
Bachelor's degree         &  341 & 33.11 &  310 & 38.46 \\
Master's degree or above  &  275 & 26.70 &  128 & 15.88 \\
Other                     &   12 &  1.17 &   17 &  2.11 \\
Prefer not to answer      &    0 &  0.00 &    2 &  0.25 \\
\midrule
\textbf{Total} & \textbf{1030} & \textbf{100.00} & \textbf{806} & \textbf{100.00} \\
\bottomrule
\end{tabularx}
\end{center}
\caption{Distribution of educational attainment in the 2020 ANES Twitter population compared to our study population ($\chi^2(6) = 43.92$, $p < .001$). For comparison purposes, we bin the ANES education categories, which are more fine-grained than ours, into our coarser categories.}
\end{table}

\begin{table}[H]
\begin{center}
\begin{tabularx}{\textwidth}{@{} p{8cm}cccc @{}}
\toprule
\textbf{Age Group} & \multicolumn{2}{c}{\textbf{ANES}} & \multicolumn{2}{c}{\textbf{Our Study}} \\
\cmidrule(lr){2-3} \cmidrule(lr){4-5}
& $n$ & Percent & $n$ & Percent \\
\midrule
18-24 years old   & 131 & 13.14 & 120 & 14.89 \\
25-34 years old   & 201 & 20.16 & 307 & 38.09 \\
35-44 years old   & 235 & 23.57 & 211 & 26.18 \\
45-54 years old   & 164 & 16.45 &  93 & 11.54 \\
55-64 years old   & 137 & 13.74 &  51 &  6.33 \\
65-74 years old   &  98 &  9.83 &  23 &  2.85 \\
75 years or older &  31 &  3.11 &   1 &  0.12 \\
\midrule
\textbf{Total} & \textbf{997} & \textbf{100.00} & \textbf{806} & \textbf{100.00} \\
\bottomrule
\end{tabularx}
\end{center}
\caption{Distribution of ages in the 2020 ANES Twitter population compared to our study population ($\chi^2(6) = 138.78$, $p < .001$). For comparison purposes, we group ages to match the age groups that we use in our study.}
\end{table}

\clearpage
\section{Pre-registered analysis} \label{appendix:full-effects}

\footnotesize

\begin{table}[!h]
\centering
\begin{tabularx}{\textwidth}{@{} lcccccc @{}}
\hline
\multirow{2}{*}{{\textbf{Outcome}}} & \textbf{Standardized} & \textbf{Unstandardized} & \textbf{Chron.} & \textbf{Eng.} & \multirow{2}{*}{\textbf{$p$-value}} & \textbf{Adjusted} \\
& \textbf{Effect} & \textbf{Effect} & \textbf{Mean} & \textbf{Mean} &\textbf{} & \textbf{$p$-value} \\ \hline

\hline
\multicolumn{7}{c}{\textbf{Emotional effects (all tweets)}} \\
\hline

Author Angry  &                0.473 &                  0.188 &       0.352 &      0.545 & 0.0002 &             0.0005 \\
  Author Sad  &                0.220 &                  0.077 &       0.293 &      0.378 & 0.0002 &             0.0005 \\
Author Anxious  &                0.232 &                  0.100 &       0.391 &      0.492 & 0.0002  &             0.0005 \\
Author Happy  &                0.016 &                  0.013 &       1.307 &      1.330 & 0.5125 &             0.1340 \\
Reader Angry  &                0.266 &                  0.107 &       0.306 &      0.412 &   0.0002 &             0.0005 \\
  Reader Sad  &                0.086 &                  0.032 &       0.316 &      0.356 &   0.0032&             0.0025 \\
Reader Anxious  &                0.119 &                  0.051 &       0.346 &      0.398 &   0.0002&             0.0005 \\
Reader Happy  &                0.119 &                  0.085 &       0.941 &      1.030 &   0.0002 &             0.0005 \\

\hline
\multicolumn{7}{c}{\textbf{Emotional effects (political tweets only)}} \\
\hline
 Author Angry  &                0.754 &                  0.299 &       1.128 &      1.438 &   0.0002 &             0.0005 \\
    Author Sad  &                0.309 &                  0.108 &       0.688 &      0.782 &   0.0056 &             0.0035 \\
Author Anxious  &                0.175 &                  0.075 &       0.824 &      0.873  &   0.0316 &             0.0129 \\
  Author Happy  &               -0.061 &                 -0.048 &       0.513 &      0.532 &   0.2138 &             0.0620 \\
  Reader Angry  &                0.377 &                  0.152 &       1.042 &      1.192 &   0.0008 &             0.0009 \\
    Reader Sad  &                0.007 &                  0.003 &       0.835 &      0.840 &   0.9363 &             0.2288 \\
Reader Anxious  &               -0.002 &                 -0.001 &       0.867 &      0.863 &   0.9679 &             0.2288 \\
  Reader Happy  &               -0.015 &                 -0.011 &       0.421 &      0.471 &   0.7761 &             0.1929 \\

\hline
\multicolumn{7}{c}{\textbf{Political effects}} \\
\hline
    Partisanship &                0.244 &                  0.051 &       0.151 &      0.202 &   0.0002 &             0.0005 \\
   Out-group Animosity &                0.236 &                  0.031 &       0.085 &      0.116 &   0.0002 &             0.0005 \\
   In-group Perc. &                0.081 &                  0.015 &       0.060 &      0.074 &   0.0014 &             0.0014 \\
  (all users) &&&&&&  \\
  Out-group Perc. &               -0.171 &                 -0.037 &      -0.108 &     -0.147 &   0.0002 &             0.0005 \\
    (all users) &&&&&&  \\
  In-group Perc.  &                0.052 &                  0.009 &       0.056 &      0.066 &   0.0954 &             0.0329 \\
(left users) &&&&&&  \\
 Out-group Perc. &               -0.157 &                 -0.031 &      -0.097 &     -0.132 &   0.0002 &             0.0005 \\
   (left users) &&&&&&  \\
 In-group Perc. &                0.151 &                  0.031 &       0.072 &      0.102  &   0.0012 &             0.0013 \\
   (right users) &&&&&&  \\
Out-group Perc. &               -0.204 &                 -0.052 &      -0.145 &     -0.200 &   0.0002 &             0.0005 \\
  (right users) &&&&&&  \\

\hline
\multicolumn{7}{c}{\textbf{Reader Preference}} \\
\hline

Reader Pref  &                0.065 &                  0.023 &       0.507 &      0.526 &   0.0226 &             0.0097 \\
(all tweets) &&&&&&  \\
Reader Pref  &               -0.180 &                 -0.065 &       0.519 &      0.465 &   0.0054 &             0.0035 \\
(political tweets) &&&&&&  \\

\hline
\end{tabularx}
\caption{Our pre-registered analysis measuring the effects of Twitter's engagement-based timeline. The table shows the average treatment effects (standardized and unstandardized), $p$-values, and FDR-adjusted $p$-values for all 26 pre-registered outcomes. The way that all statistics are calculated is described in \Cref{app:ate-estimation}.}
\label{table:full-effects}
\end{table}
\newpage
\section{Exploratory analysis}
\normalsize
Next, we describe additional, exploratory analyses that were not pre-registered.

\subsection{Descriptive statistics: tweet metadata} \label{app:metadata}

\begin{figure}[hbtp]
    \centering
    \includegraphics[width=0.95\columnwidth]{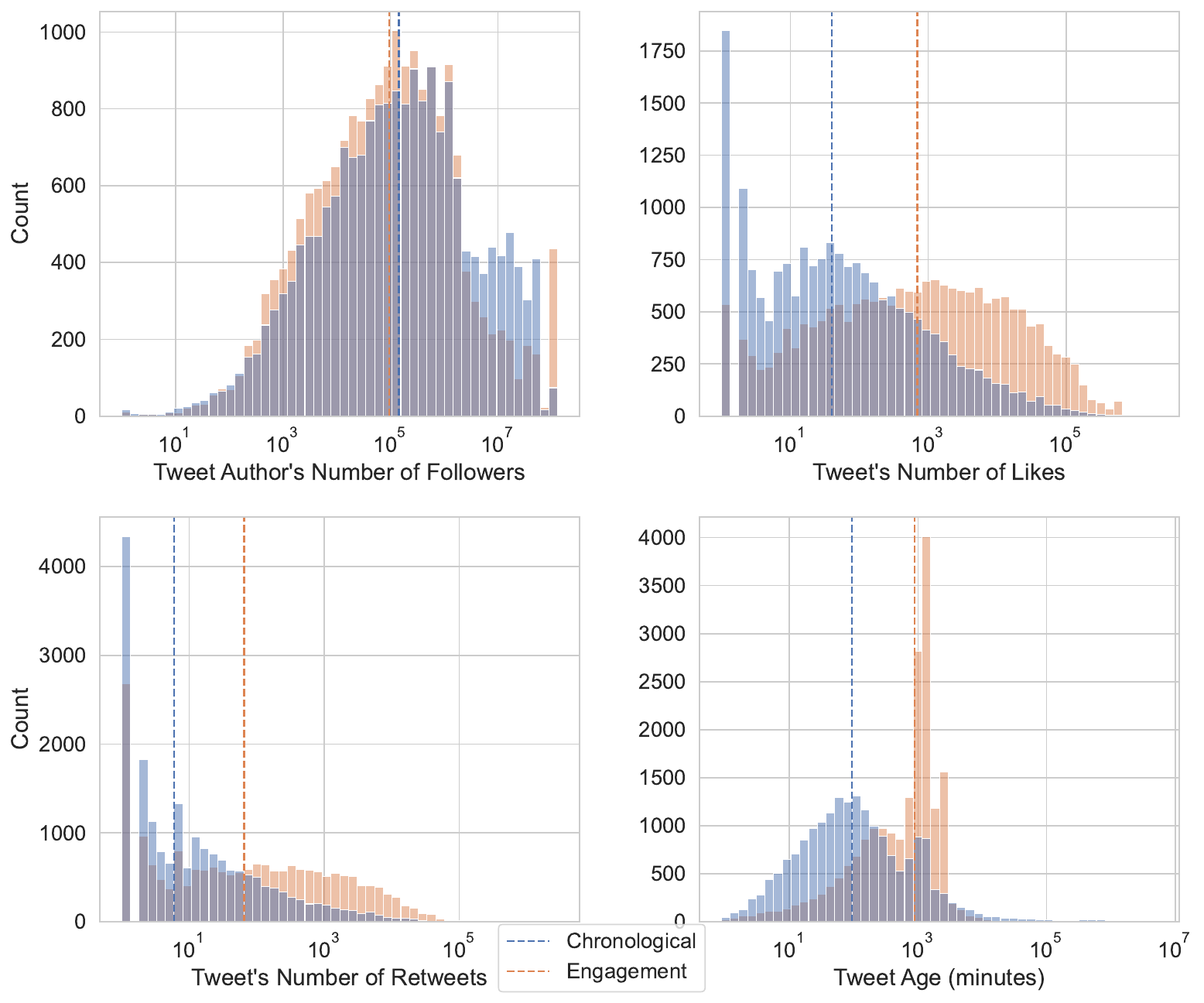}
    \caption{The graphs compare the histogram of tweets in the engagement-based and chronological timelines along four properties: the author's number of followers, the tweet's number of likes, the tweet's number of retweets, and the tweet's age. All graphs are plotted on a log-scale $x$-axis because all properties have a long-tail of extreme outliers. The dashed lines show the median value for each timeline.}
    \label{fig:hists}
\end{figure}

\Cref{table:metadata} compares the metadata of the top 10 tweets in the engagement-based and chronological timelines, pooled across all participants and waves. We include both the mean and median for each property. The median may be a more appropriate summary for attributes with extreme outliers, i.e., the tweet age (the amount of time since the tweet was created), the author's number of followers, the number of likes, and the number of retweets. Histograms for these properties are provided in \Cref{fig:hists}.

Unsurprisingly, the number of likes and retweets that each tweet has
is much higher in the engagement-based timeline. And as expected, the reader is less likely to follow the authors of tweets shown by the engagement-based algorithm compared to the chronological ranking. Interestingly, the average number of links is almost
halved, which is consistent with the results of a prior study that used eight sock puppets to audit properties of Twitter's algorithm~\cite{bandy2021more}. 

The algorithm does not
necessarily favor the most “popular” accounts: authors shown by the algorithm have a lower median number of followers and are less likely to be verified. It is important to clarify that our research was conducted before Twitter's changed its verification policy on April 1, 2023. The new policy limits verification (indicated by a blue checkmark on Twitter profiles) to Twitter Blue subscribers who pay for the service. Our data was collected under the previous policy which required users to be ``active'', ``notable'', and ``authentic'' to get verified status. Consequently, the verified users in our data set tend to be official organizational accounts, celebrities, journalists, or other well-known individuals.

\begin{figure}[tbhp]
    \centering
    \includegraphics[width=0.5\columnwidth]{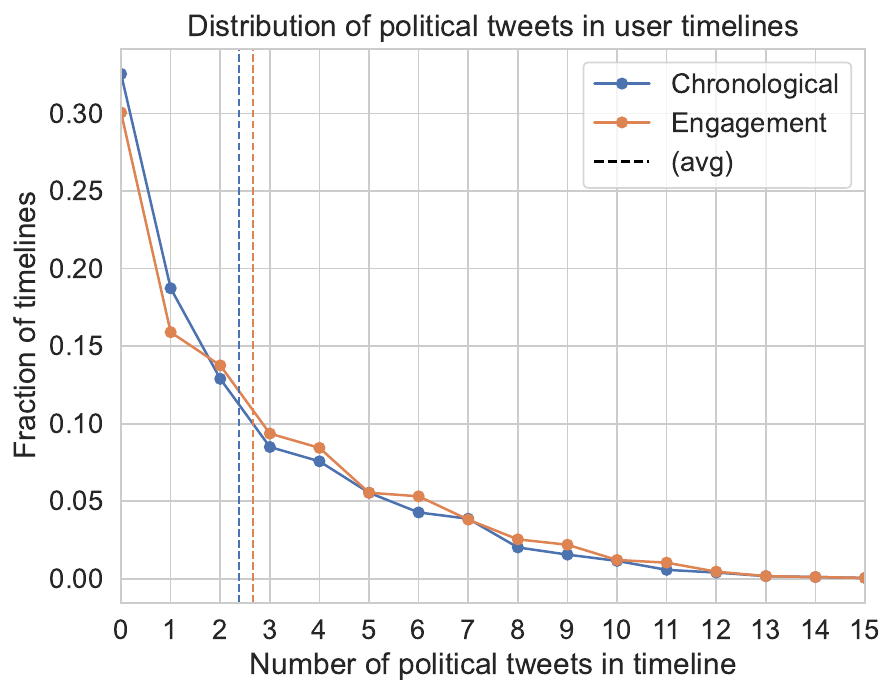}
    \caption{The distribution of political tweets in user timelines. The number of political tweets in each timeline is calculated using the participant's response to the binary question, “Is [@author-handle]’s tweet about a political or social issue?". Notably, about 30\% of participants' timelines contain no political tweets.} \label{fig:political_dist}
\end{figure}

Finally, \Cref{fig:political_dist} shows the distribution of the number of political tweets in the user timelines. There was little difference between the distribution of political tweets in the engagement-based timelines and the chronological timelines. About 70\% of participants' timelines contained at least one political tweet (out of approximately ten tweets total).

\begin{table}[hbtp]
\centering
\caption{Summary of metadata in users' engagement and chronological timelines} 
\label{table:metadata}
\resizebox{\textwidth}{!}{%
\begin{tabular}{lllllllll}
\toprule
 & Timeline & p5 & p50 & p95 & Avg. & SD & Diff \\ 
\midrule
Tweet age (minutes) & Chronological & 3.8 & \cellcolor[HTML]{C0C0C0} 92 & 2,937 & 7,812 & 126,667 &  \\ 
& Engagement & 18.4 & \cellcolor[HTML]{C0C0C0} 899 & 2,722 & 3,846 & 94,308 & p \textless{} 0.001 \\ 
Number of likes & Chronological & 0 & \cellcolor[HTML]{C0C0C0} 39 & 7,557 & 2,500 & 18,565 &  \\ 
& Engagement & 2 & \cellcolor[HTML]{C0C0C0} 744 & 83,658 & 16,912 & 62,158 & p \textless{} 0.001 \\
Number of retweets & Chronological & 0 & \cellcolor[HTML]{C0C0C0} 6 & 1,081 & 389 & 3,752 &  \\
& Engagement & 0 & \cellcolor[HTML]{C0C0C0} 68 & 8,875 & 1,787 & 13,415 & p \textless{} 0.001 \\
Number of links in tweet & Chronological & 0 & 0 & 1 & \cellcolor[HTML]{C0C0C0} 0.397 & 0.552 &  \\ 
& Engagement & 0 & 0 & 1 & \cellcolor[HTML]{C0C0C0} 0.105 & 0.326 & p \textless{} 0.001 \\ 
Number of photos in tweet & Chronological & 0 & 0 & 2 & \cellcolor[HTML]{C0C0C0} 0.457 & 0.755 &  \\
& Engagement & 0 & 0 & 2 & \cellcolor[HTML]{C0C0C0} 0.496 & 0.752 & p \textless{} 0.001 \\ 
Number of videos in tweet & Chronological & 0 & 0 & 1 & \cellcolor[HTML]{C0C0C0} 0.130 & 0.336 &  \\
& Engagement & 0 & 0 & 1 & \cellcolor[HTML]{C0C0C0} 0.184 & 0.387 & p \textless{} 0.001 \\ 
Author's \# followers & Chronological & 500 & \cellcolor[HTML]{C0C0C0} 139,874 & 25,745,780 & 3,905,313 & 12,059,120 &  \\
& Engagement & 487 & \cellcolor[HTML]{C0C0C0} 92,959 & 18,788,850 & 4,877,343 & 20,868,580 & p \textless{} 0.001 \\ 
Is the author verified? & Chronological & 0 & 1 & 1 & \cellcolor[HTML]{C0C0C0} 0.504 & 0.500 &\\ 
& Engagement & 0 & 0 & 1 & \cellcolor[HTML]{C0C0C0} 0.391 & 0.488 & p \textless{} 0.001 \\ 
Does the reader follow the author? & Chronological & 0 & 1 & 1 & \cellcolor[HTML]{C0C0C0}  0.694 & 0.461 &  \\
& Engagement & 0 & 1 & 1 & \cellcolor[HTML]{C0C0C0}  0.569 & 0.495 & p \textless{} 0.001 \\
\bottomrule
\end{tabular}
}
\caption*{\small The table compares the top ten tweets in the chronological and engagement-based timelines. Both the mean and median are shown for each attribute. The median is a more representative summary when the mean of the attribute is dominated by extreme outliers. In gray, we have highlighted which summary statistic, the mean or median, is more representative based on how long-tailed the distribution is.}
\end{table}

\subsection{Amplification of individual accounts}
In this section, we investigate which individual user accounts are the most amplified or de-amplified by Twitter's engagement-based algorithm. 

First, in \Cref{tab:ind-amp}, we show the accounts that appear most frequently in our study participants' chronological and engagement timelines. The most common accounts are quite different in the two timelines, with news outlets having much greater dominance in the chronological timeline.

Next, in \Cref{tab:compare-amplified-accounts}, we show which accounts were most and least amplified by Twitter's algorithm, where the amplification of an individual account is simply calculated as the number of tweets by that author in users' engagement-based timelines minus the number of tweets by that author in users' chronological timelines. Consistent with prior findings~\cite{bandy2021more}, the accounts that were most \emph{de-amplified} were news outlets. This is likely because news outlets post more frequently than ordinary accounts. Though the chronological timeline is susceptible to this kind of spam, the engagement-based timeline is not~\cite{ovadya_2022}. Thus, whether users are using their chronlogical or engagement-based timeline may have a large influence on their exposure to traditional news outlets~\citep{wojcieszak2024non}.

Elon Musk also stands out as a significant outlier, receiving higher amplification than any other account by a wide margin. Notably, during our study period, Platformer alleged that Twitter deployed code to artificially boost the reach of Musk's tweets~\cite{schiffer2023musk}. Though Musk denied such claims~\cite{elonmusk2023}, our data seems to suggest that he did receive an abnormally high amount of amplification, relative to other accounts. 

\begin{table}[hbtp]
\centering
\begin{tabularx}{\textwidth}{
    @{} 
    >{\arraybackslash}X 
    >{\centering\arraybackslash}X 
    >{\arraybackslash}X 
    >{\centering\arraybackslash}X 
    @{}
}
\toprule
\multicolumn{2}{c}{\textbf{Chronological}} & \multicolumn{2}{c}{\textbf{Engagement}} \\
\textbf{Top Accounts} & \textbf{Count} & \textbf{Top Accounts} & \textbf{Count} \\
\midrule
@Reuters & 532 & @elonmusk & 1387 \\
@nytimes & 529 & @POTUS & 295 \\
@FatKidDeals & 407 & @RonFilipkowski & 210 \\
@Wario64 & 365 & @JackPosobiec & 171 \\
@CNN & 365 & @fasc1nate & 152 \\
@NBA & 280 & @nytimes & 135 \\
@Forbes & 280 & @barstoolsports & 134 \\
@FoxNews & 266 & @DailyLoud & 133 \\
@AP & 251 & @hodgetwins & 126 \\
@washingtonpost & 250 & @CNN & 119 \\
@POTUS & 206 & @NBA & 115 \\
@IGN & 204 & @stillgray & 107 \\
@videogamedeals & 185 & @vidsthatgohard & 103 \\
@TMZ & 179 & @ClownWorld\_ & 102 \\
@elonmusk & 151 & @catturd2 & 97 \\
@business & 151 & @BornAKang & 96 \\
@TheOnion & 148 & @krassenstein & 95 \\
@barstoolsports & 128 & @Wario64 & 93 \\
@PlayStation & 127 & @HumansNoContext & 90 \\
@NFL & 126 & @Dexerto & 84 \\
@JackPosobiec & 123 & @CalltoActivism & 84 \\
@guardian & 122 & @mmpadellan & 82 \\
@Independent & 119 & @NFL & 82 \\
@ABC & 114 & @joncoopertweets & 81 \\
@nypost & 108 & @jordanbpeterson & 79 \\
\bottomrule
\end{tabularx}
\caption{The most common accounts seen in our study participants' chronological and engagement timelines. In both timelines, only the first ten tweets are considered.}
\label{tab:ind-amp}
\end{table}

\begin{table}[hbtp]
\centering
\begin{tabularx}{\textwidth}{
    @{} 
    >{\arraybackslash}X 
    >{\arraybackslash}X 
    >{\arraybackslash}X 
    >{\arraybackslash}X 
    @{}
}
\toprule
\textbf{Most Amplified} & \textbf{Eng-Chron Diff} & \textbf{Least Amplified} & \textbf{Eng-Chron Diff} \\
\midrule
@elonmusk & 1236.0 & @Reuters & -518.0 \\
@RonFilipkowski & 144.0 & @nytimes & -394.0 \\
@fasc1nate & 119.0 & @Wario64 & -272.0 \\
@DailyLoud & 114.0 & @Forbes & -265.0 \\
@hodgetwins & 102.0 & @CNN & -246.0 \\
@vidsthatgohard & 97.0 & @AP & -214.0 \\
@stillgray & 89.0 & @FoxNews & -198.0 \\
@POTUS & 89.0 & @washingtonpost & -188.0 \\
@BornAKang & 87.0 & @NBA & -165.0 \\
@ClownWorld\_ & 83.0 & @videogamedeals & -156.0 \\
@krassenstein & 83.0 & @IGN & -148.0 \\
@HumansNoContext & 80.0 & @TMZ & -148.0 \\
@Dexerto & 73.0 & @business & -139.0 \\
@rawsalerts & 68.0 & @TheOnion & -136.0 \\
@joncoopertweets & 67.0 & @guardian & -117.0 \\
@buitengebieden & 60.0 & @Independent & -108.0 \\
@Acyn & 59.0 & @PlayStation & -107.0 \\
@mmpadellan & 58.0 & @TheEconomist & -85.0 \\
@EndWokeness & 58.0 & @people & -82.0 \\
@CollinRugg & 58.0 & @ABC & -79.0 \\
@historyinmemes & 57.0 & @WSJ & -79.0 \\
@CalltoActivism & 53.0 & @BBCWorld & -69.0 \\
@libsoftiktok & 52.0 & @netflix & -68.0 \\
@FightHaven & 49.0 & @HuffPost & -65.0 \\
@JackPosobiec & 48.0 & @thehill & -63.0 \\
\bottomrule
\end{tabularx}
\caption{The accounts that were most and least amplified in the engagement timeline, relative to the chronological timeline. In both timelines, only the first 10 tweets are considered.}
\label{tab:compare-amplified-accounts}
\end{table}
\newpage 
\subsection{Distribution of responses to survey questions}
In \Cref{appendix:full-effects}, we reported our pre-registered results on the effects of the engagement-based timeline, relative to the chronological timeline. Here, we also additionally show how the full distribution of responses to the Likert survey questions differs between users' chronological and engagement-based timelines.

\subsubsection{Emotions}
\begin{figure}[hbtp]
    \centering
    \includegraphics[width=\columnwidth]{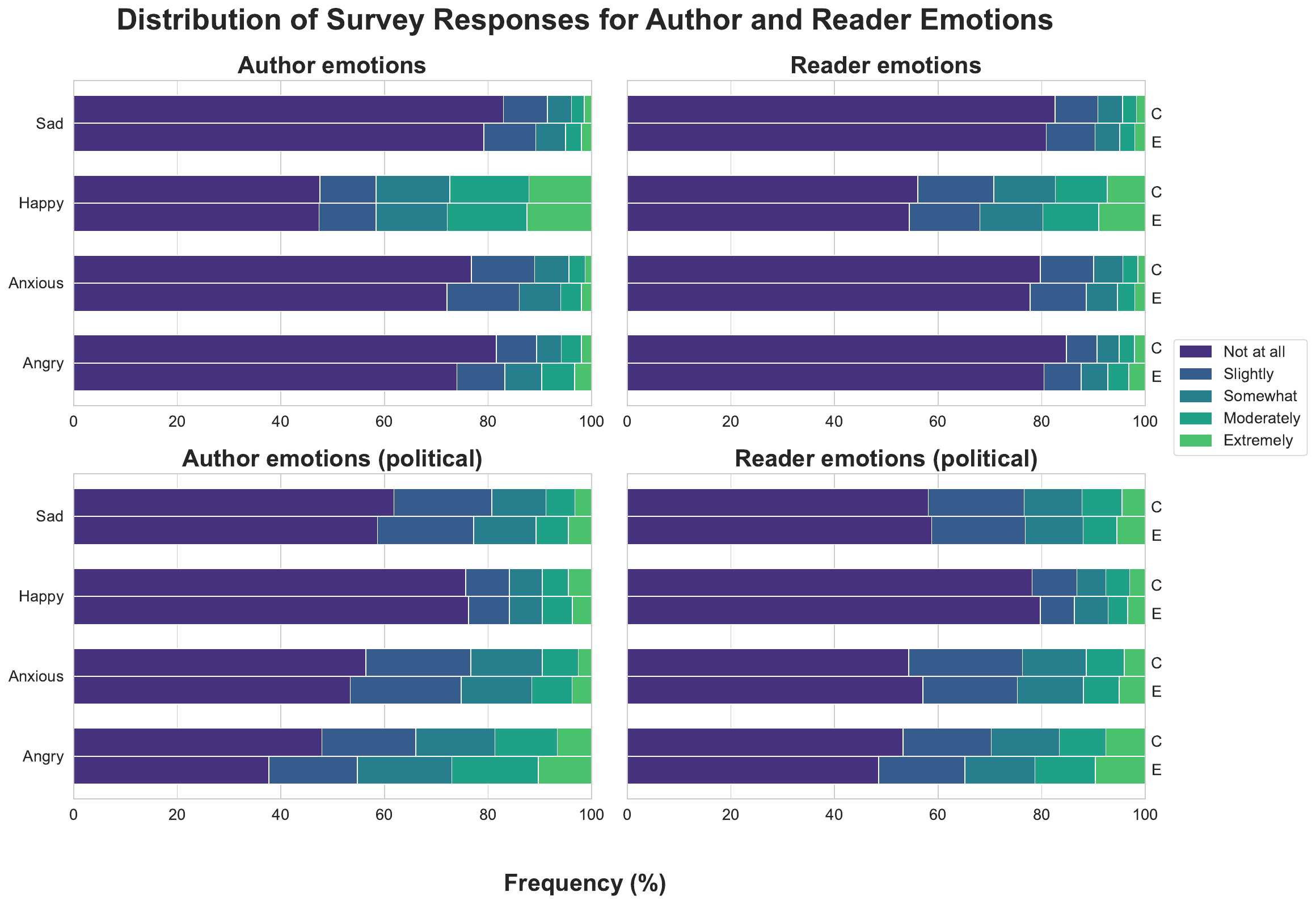}
    \caption{Distribution of author and reader emotions in the chronological (C) and engagement-based (E) timelines}
    \label{fig:likert_author_reader_ems}
\end{figure}

\begin{figure}[hbtp]
    \centering
    \includegraphics[width=\columnwidth]{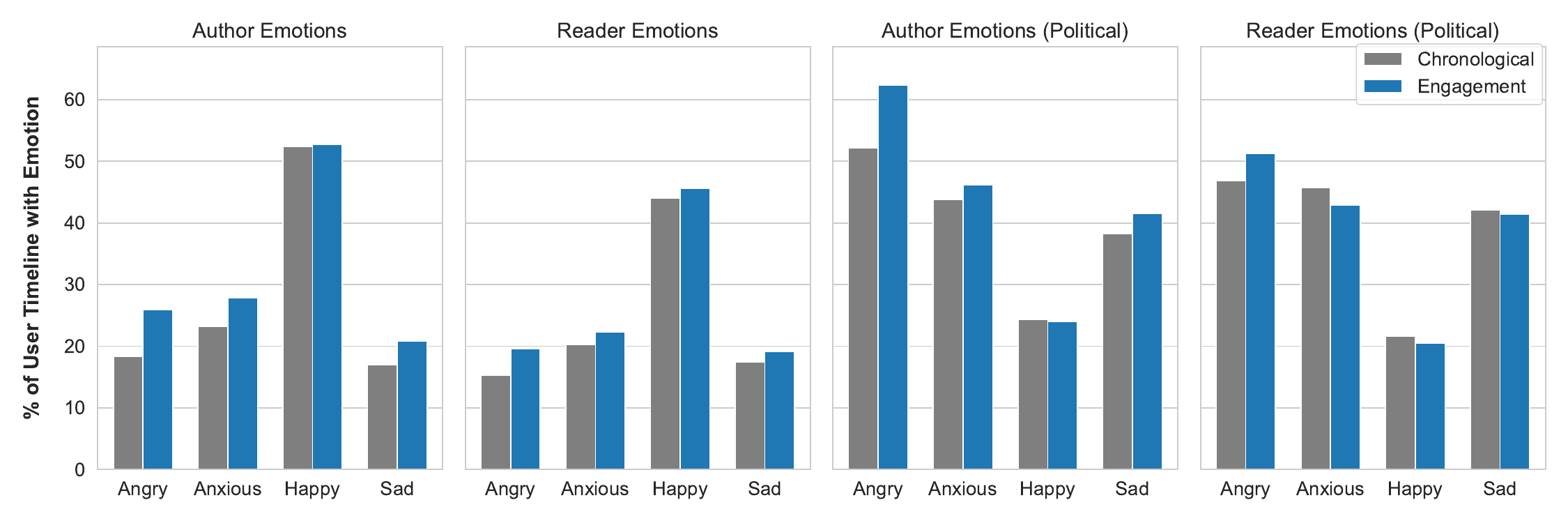}
    \caption{The percentage of each (binarized) emotion in the engagement and chronological timelines.}
    \label{fig:binarized_ems}
\end{figure}

First, we show the distribution of survey responses for questions gauging the author and reader's emotions (sad, happy, anxious, and angry), considering both the distribution for tweets overall and for only political tweets, in \Cref{fig:likert_author_reader_ems} and Tables \ref{tab:author_emotions_overall}-\ref{tab:reader_emotions_political}.  For further ease of interpretation, in \Cref{fig:binarized_ems}, we also show the mean level of author and reader emotions in the chronological and engagement-based timelines after binarizing responses. In particular, if a reader responds ``Not at all'' to an emotion, that is coded as zero, while any response between ``Slightly'' and ``Extremely'' is coded as one. Notably, there was a large increase in tweets expressing anger between the engagement and chronological timeline. In particular, 62 percent of political tweets in the engagement timeline expressed anger, compared to 52 percent in the chronological timeline.

\begin{table}[hbtp]
    \centering
    \small
    \begin{tabularx}{\textwidth}{@{} lllllllll @{}}
    \toprule
    \textbf{Answer} & \multicolumn{2}{c}{\textbf{Angry}} & \multicolumn{2}{c}{\textbf{Anxious}} & \multicolumn{2}{c}{\textbf{Happy}} & \multicolumn{2}{c}{\textbf{Sad}} \\
    \textbf{} & \textbf{\% Chron} & \textbf{\% Eng} & \textbf{\% Chron} & \textbf{\% Eng} & \textbf{\% Chron} & \textbf{\% Eng} & \textbf{\% Chron} & \textbf{\% Eng} \\
    \midrule
    Not at all (0) & 81.57 & 73.99 & 76.81 & 72.10 & 47.55 & 47.38 & 82.96 & 79.21 \\
    Slightly (1)  & 7.80  & 9.28  & 12.20 & 13.94 & 10.85 & 11.04 & 8.50  & 10.01 \\
    Somewhat (2)  & 4.79  & 7.09  & 6.61  & 7.99  & 14.18 & 13.72 & 4.67  & 5.78  \\
    Moderately (3)& 3.89  & 6.36  & 3.12  & 4.01  & 15.33 & 15.40 & 2.43  & 3.06  \\
    Extremely (4)& 1.95  & 3.28  & 1.26  & 1.96  & 12.08 & 12.47 & 1.44  & 1.95  \\
    \bottomrule
    \end{tabularx}
    \caption{Author Emotions (Overall)}
    \label{tab:author_emotions_overall}
\end{table}

\begin{table}[hbtp]
    \centering
    \small
    \begin{tabularx}{\textwidth}{@{} lllllllll @{}}
    \toprule
    \textbf{Answer} & \multicolumn{2}{c}{\textbf{Angry}} & \multicolumn{2}{c}{\textbf{Anxious}} & \multicolumn{2}{c}{\textbf{Happy}} & \multicolumn{2}{c}{\textbf{Sad}} \\
    \textbf{} & \textbf{\% Chron} & \textbf{\% Eng} & \textbf{\% Chron} & \textbf{\% Eng} & \textbf{\% Chron} & \textbf{\% Eng} & \textbf{\% Chron} & \textbf{\% Eng} \\
    \midrule
    Not at all (0) & 84.75 & 80.44 & 79.76 & 77.73 & 56.06 & 54.42 & 82.58 & 80.88 \\
    Slightly (1)  & 5.94  & 7.17  & 10.30 & 10.87 & 14.71 & 13.66 & 8.28  & 9.42 \\
    Somewhat (2)  & 4.25  & 5.13  & 5.63  & 6.04  & 11.88 & 12.16 & 4.77  & 4.77 \\
    Moderately (3) & 3.01  & 4.09  & 2.86  & 3.34  & 10.04 & 10.81 & 2.67  & 2.89 \\
    Extremely (4) & 2.05  & 3.17  & 1.45  & 2.02  & 7.31  & 8.95  & 1.70  & 2.04 \\
    \bottomrule
    \end{tabularx}
    \caption{Reader Emotions (Overall)}
    \label{tab:reader_emotions_overall}
\end{table}

\begin{table}[hbtp]
    \centering
    \small
    \begin{tabularx}{\textwidth}{@{} lllllllll @{}}
    \toprule
    \textbf{Answer} & \multicolumn{2}{c}{\textbf{Angry}} & \multicolumn{2}{c}{\textbf{Anxious}} & \multicolumn{2}{c}{\textbf{Happy}} & \multicolumn{2}{c}{\textbf{Sad}} \\
    \textbf{} & \textbf{\% Chron} & \textbf{\% Eng} & \textbf{\% Chron} & \textbf{\% Eng} & \textbf{\% Chron} & \textbf{\% Eng} & \textbf{\% Chron} & \textbf{\% Eng} \\
    \midrule
    Not at all (0) & 47.93 & 37.65 & 56.43 & 53.38 & 75.66 & 76.24 & 61.82 & 58.64 \\
    Slightly (1)  & 18.15 & 17.13 & 20.26 & 21.45 & 8.44  & 7.90  & 18.94 & 18.55 \\
    Somewhat (2)  & 15.23 & 18.21 & 13.76 & 13.61 & 6.40  & 6.33  & 10.43 & 12.12 \\
    Moderately (3)& 12.09 & 16.70 & 7.00  & 7.80  & 4.99  & 5.79  & 5.59  & 6.17  \\
    Extremely (4) & 6.60  & 10.31 & 2.55  & 3.76  & 4.52  & 3.74  & 3.23  & 4.52  \\
    \bottomrule
    \end{tabularx}
    \caption{Author Emotions (Political)}
    \label{tab:author_emotions_political}
\end{table}

\begin{table}[hbtp]
    \centering
    \small
    \begin{tabularx}{\textwidth}{@{} lllllllll @{}}
    \toprule
    \textbf{Answer} & \multicolumn{2}{c}{\textbf{Angry}} & \multicolumn{2}{c}{\textbf{Anxious}} & \multicolumn{2}{c}{\textbf{Happy}} & \multicolumn{2}{c}{\textbf{Sad}} \\
    \textbf{} & \textbf{\% Chron} & \textbf{\% Eng} & \textbf{\% Chron} & \textbf{\% Eng} & \textbf{\% Chron} & \textbf{\% Eng} & \textbf{\% Chron} & \textbf{\% Eng} \\
    \midrule
    Not at all (0) & 53.24 & 48.51 & 54.30 & 57.09 & 78.17 & 79.70 & 58.10 & 58.75 \\
    Slightly (1)  & 17.00 & 16.68 & 21.99 & 18.22 & 8.65  & 6.61  & 18.52 & 18.09 \\
    Somewhat (2)  & 13.16 & 13.51 & 12.35 & 12.76 & 5.55  & 6.51  & 11.15 & 11.17 \\
    Moderately (3)& 8.96  & 11.68 & 7.32  & 6.91  & 4.61  & 3.77  & 7.77  & 6.54  \\
    Extremely (4) & 7.63  & 9.62  & 4.04  & 5.03  & 3.02  & 3.40  & 4.45  & 5.45  \\
    \bottomrule
    \end{tabularx}
    \caption{Reader Emotions (Political)}
    \label{tab:reader_emotions_political}
\end{table}

\begin{table}[hbtp]
    \centering
    \begin{tabularx}{\textwidth}{@{} lllllllll @{}}
    \toprule
    \textbf{Timeline} & \multicolumn{4}{c}{\textbf{Author}} & \multicolumn{4}{c}{\textbf{Reader}} \\
    \textbf{} & \textbf{Angry} & \textbf{Anxious} & \textbf{Happy} & \textbf{Sad} & \textbf{Angry} & \textbf{Anxious} & \textbf{Happy} & \textbf{Sad} \\
    \midrule
        Chronological & 18.396 & 23.186 & 52.448 & 17.020 & 15.243 & 20.225 & 43.973 & 17.414 \\
        Engagement & 25.963 & 27.897 & 52.687 & 20.786 & 19.540 & 22.272 & 45.642 & 19.119 \\
    \bottomrule
\end{tabularx}
\caption{Binarized Emotions (Overall). The mean percent of each binarized emotion for readers and authors across each timeline.}
\label{tab:binarized_ems_all}
\end{table}

\begin{table}[hbtp]
    \centering
    \begin{tabularx}{\textwidth}{@{} lllllllll @{}}
    \toprule
    \textbf{Timeline} & \multicolumn{4}{c}{\textbf{Author}} & \multicolumn{4}{c}{\textbf{Reader}} \\
    \textbf{} & \textbf{Angry} & \textbf{Anxious} & \textbf{Happy} & \textbf{Sad} & \textbf{Angry} & \textbf{Anxious} & \textbf{Happy} & \textbf{Sad} \\
    \midrule
        Chronological & 52.151 & 43.752 & 24.341 & 38.281 & 46.893 & 45.729 & 21.663 & 42.065 \\
        Engagement & 62.312 & 46.123 & 23.944 & 41.538 & 51.279 & 42.929 & 20.536 & 41.383 \\
    \bottomrule
\end{tabularx}
\caption{Binarized Emotions (Political). The mean percent of each binarized emotion for readers and authors across each timeline, limited to just political tweets.}
\label{tab:binarized_ems_pol}
\end{table}

\newpage 
\subsubsection{Political outcomes}

\begin{figure}[hbtp]
    \centering
    \includegraphics[width=\columnwidth]{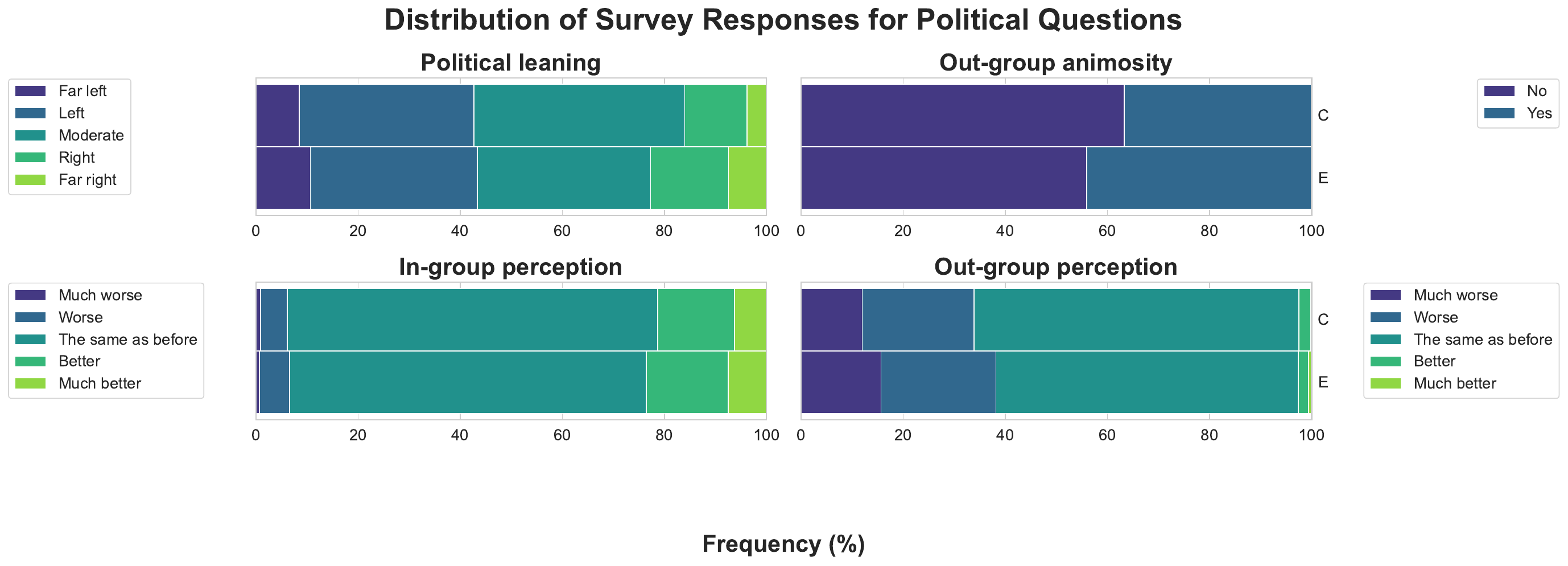}
    \caption{Distribution of political outcomes outcomes in the chronological (C) and engagement-based (E) timelines for all users}
    \label{fig:likert_pol_all}
\end{figure}

\begin{figure}[hbtp]
    \centering
    \includegraphics[width=\columnwidth]{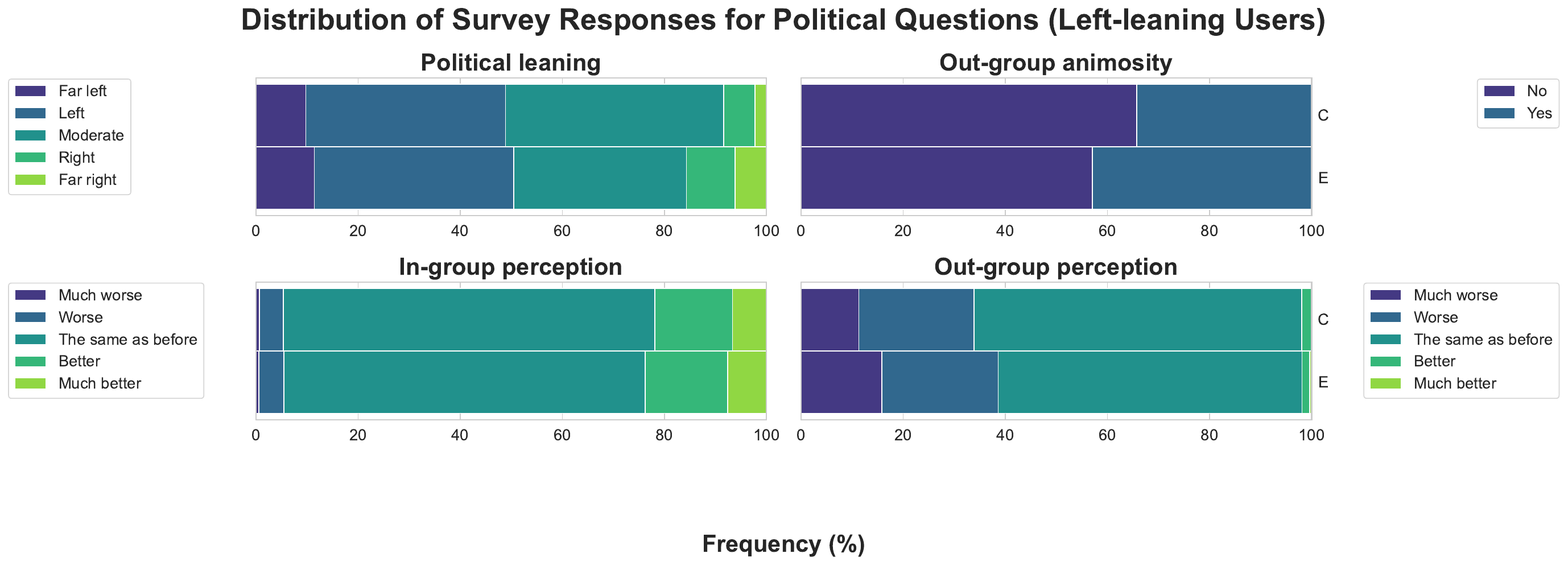}
    \caption{Distribution of political outcomes in the chronological (C) and engagement-based (E) timelines for left-leaning users only}
    \label{fig:likert_pol_left}
\end{figure}

\begin{figure}[hbtp]
    \centering
    \includegraphics[width=\columnwidth]{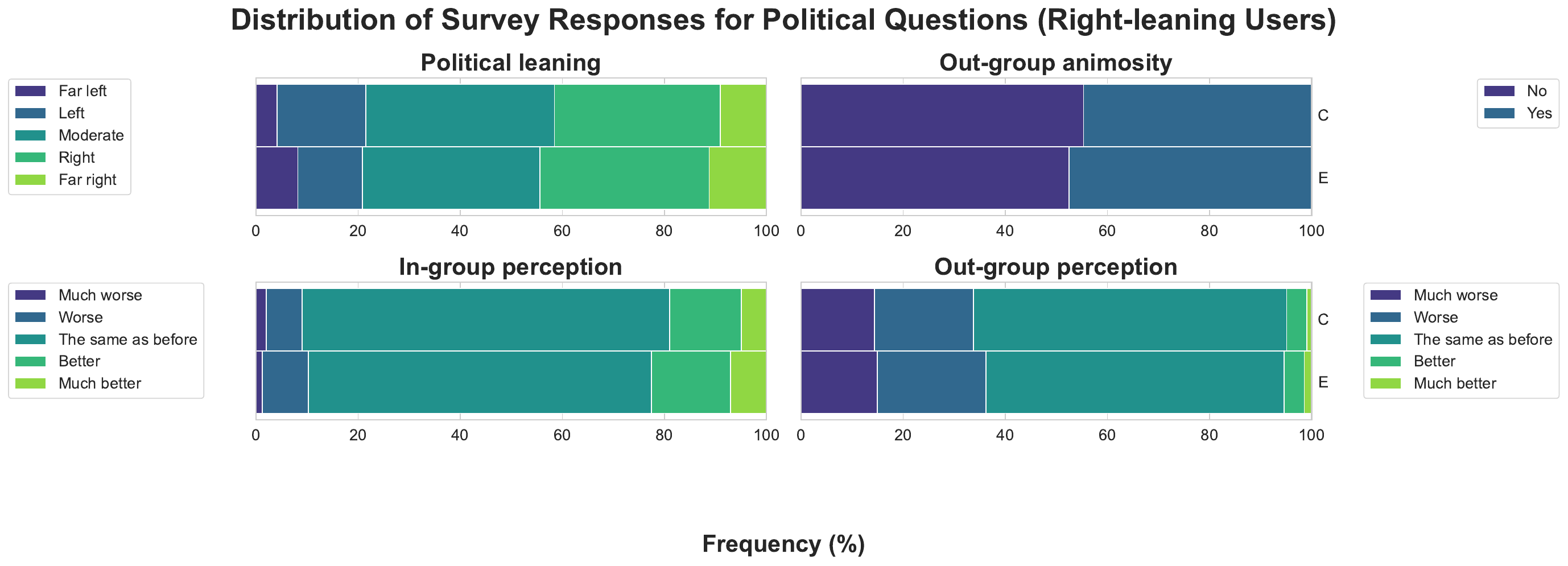}
    \caption{Distribution of political outcomes in the chronological (C) and engagement-based (E) timelines for right-leaning users only}
    \label{fig:likert_pol_right}
\end{figure}

Next, we show the distribution of survey responses for political questions (political leaning, out-group animosity, in-group perception, and out-group perception). We show the distributions for users overall (\Cref{fig:likert_pol_all}, Tables \ref{tab:political_leaning_all_users}-\ref{tab:out_group_animosity_all_users}) as well as separately for left-leaning  (\Cref{fig:likert_pol_left}, Tables \ref{tab:political_leaning_left_users}-\ref{tab:out_group_animosity_left_users}) and right-leaning users (\Cref{fig:likert_pol_right}, Tables \ref{tab:political_leaning_right_users}-\ref{tab:out_group_animosity_right_users}).

Interestingly, in the chronological timeline, right-leaning users are exposed to much more cross-cutting content than left-leaning users;  21 percent of the political tweets shown to right-leaning users are left-leaning while only 8 percent of the political tweets shown to left-leaning users are right-leaning. This may reflect the fact the population of Twitter users tends to be predominantly left-leaning (see \Cref{appendix:user-demographics}). The engagement-based algorithm doubles the amount of cross-cutting content that left-leaning users see (16 percent compared to 8 percent in the chronological timeline). For right-leaning users, the proportion of cross-cutting content remains the same, i.e., 21 percent in both timelines.

\begin{table}[hbtp]
    \centering
    \begin{tabularx}{\textwidth}{ @{} XXX @{} }
    \toprule
    \textbf{Answer} & \textbf{\% Chronological} & \textbf{\% Engagement} \\
    \midrule
    Far left (-2) & 8.48 & 10.64 \\
    Left (-1) & 34.20 & 32.70 \\
    Moderate (0) & 41.27 & 33.94 \\
    Right (1) & 12.20 & 15.28 \\
    Far right (2) & 3.85 & 7.44 \\
    \bottomrule
    \end{tabularx}
    \caption{Political leaning (all users)}
    \label{tab:political_leaning_all_users}
\end{table}

\begin{table}[hbtp]
    \centering
    \begin{tabularx}{\textwidth}{
    @{} 
    >{\hsize=1.2\hsize}X 
    >{\hsize=.9\hsize}X 
    >{\hsize=.9\hsize}X 
    >{\hsize=.9\hsize}X 
    >{\hsize=.9\hsize}X 
    @{}
    }
    \toprule
    \textbf{Answer} & \multicolumn{2}{l}{\textbf{In-group perc.}} & \multicolumn{2}{l}{\textbf{Out-group perc.}} \\
    & \textbf{\% Chron} & \textbf{\% Eng} & \textbf{\% Chron} & \textbf{\% Eng}\\
    \midrule
    Much worse (-2) & 0.87 & 0.64 & 11.97 & 15.63 \\
    Worse (-1) & 5.27 & 5.91 & 21.90 & 22.51 \\
    The same as before (0)& 72.56 & 69.91 & 63.58 & 59.21 \\
    Better (1) & 15.00 & 16.07 & 2.33 & 2.07 \\
    Much better (2) & 6.30 & 7.47 & 0.22 & 0.58 \\
    \bottomrule
    \end{tabularx}
    \caption{In-group and out-group perception (all users)}
    \label{tab:in_out_group_affect_all_users}
\end{table}

\begin{table}[hbtp]
    \centering
    \begin{tabularx}{\textwidth}{ @{} XXX @{} }
    \toprule
    \textbf{Answer} & \textbf{\% Chronological} & \textbf{\% Engagement} \\
    \midrule
    No (0) & 63.26 & 55.92 \\
    Yes (1) & 36.74 & 44.08 \\
    \bottomrule
    \end{tabularx}
    \caption{Out-group animosity (all users)}
    \label{tab:out_group_animosity_all_users}
\end{table}

\begin{table}[hbtp]
    \centering
    \begin{tabularx}{\textwidth}{ @{} XXX @{} }
    \toprule
    \textbf{Answer} & \textbf{\% Chronological} & \textbf{\% Engagement} \\
    \midrule
    Far left (-2) & 9.77 & 11.42 \\
    Left (-1) & 39.09 & 39.05 \\
    Moderate (0) & 42.76 & 33.85 \\
    Right (1) & 6.12 & 9.54 \\
    Far right (2) & 2.26 & 6.15 \\
    \bottomrule
    \end{tabularx}
    \caption{Political leaning for left-leaning users only}
    \label{tab:political_leaning_left_users}
\end{table}

\begin{table}[hbtp]
    \centering
    \begin{tabularx}{\textwidth}{
    @{} 
    >{\hsize=1.2\hsize}X 
    >{\hsize=.9\hsize}X 
    >{\hsize=.9\hsize}X 
    >{\hsize=.9\hsize}X 
    >{\hsize=.9\hsize}X 
    @{}
    }
    \toprule
    \textbf{Answer} & \multicolumn{2}{l}{\textbf{In-group perc. (left users)}} & \multicolumn{2}{l}{\textbf{Out-group perc. (left users)}} \\
    & \textbf{\% Chron} & \textbf{\% Eng} & \textbf{\% Chron} & \textbf{\% Eng}\\
    \midrule
    Much worse (-2) & 0.64 & 0.58 & 11.30 & 15.84 \\
    Worse (-1) & 4.74 & 4.91 & 22.59 & 22.78 \\
    The same as before (0) & 72.76 & 70.76 & 64.19 & 59.49 \\
    Better (1) & 15.19 & 16.16 & 1.82 & 1.48 \\
    Much better (2) & 6.68 & 7.59 & 0.10 & 0.41 \\
    \bottomrule
    \end{tabularx}
    \caption{In-group and out-group perception for left-leaning users only}
    \label{tab:in_out_affect_left_users}
\end{table}

\begin{table}[hbtp]
    \centering
    \begin{tabularx}{\textwidth}{ @{} XXX @{} }
    \toprule
    \textbf{Answer} & \textbf{\% Chronological} & \textbf{\% Engagement} \\
    \midrule
    No (0) & 65.72 & 57.04 \\
    Yes (1) & 34.28 & 42.96 \\
    \bottomrule
    \end{tabularx}
    \caption{Out-group animosity for left-leaning users only}
    \label{tab:out_group_animosity_left_users}
\end{table}

\begin{table}[hbtp]
    \centering
    \begin{tabularx}{\textwidth}{ @{} XXX @{} }
    \toprule
    \textbf{Answer} & \textbf{\% Chronological} & \textbf{\% Engagement} \\
    \midrule
    Far left (-2) & 4.11 & 8.19 \\
    Left (-1) & 17.37 & 12.65 \\
    Moderate (0) & 36.96 & 34.74 \\
    Right (1) & 32.46 & 33.17 \\
    Far right (2) & 9.10 & 11.24 \\
    \bottomrule
    \end{tabularx}
    \caption{Political leaning for right-leaning users only}
    \label{tab:political_leaning_right_users}
\end{table}

\begin{table}[hbtp]
    \centering
    \begin{tabularx}{\textwidth}{
    @{} 
    >{\hsize=1.2\hsize}X 
    >{\hsize=.9\hsize}X 
    >{\hsize=.9\hsize}X 
    >{\hsize=.9\hsize}X 
    >{\hsize=.9\hsize}X 
    @{}
    }
    \toprule
    \textbf{Answer} & \multicolumn{2}{c}{\textbf{In-group}} & \multicolumn{2}{c}{\textbf{Out-group}} \\
    & \textbf{\% Chron} & \textbf{\% Eng} & \textbf{\% Chron} & \textbf{\% Eng}\\
    \midrule
    Much worse (-2) & 2.01 & 1.21 & 14.35 & 14.94 \\
    Worse (-1) & 7.02 & 9.06 & 19.44 & 21.29 \\
    The same as before (0) & 71.97 & 67.17 & 61.32 & 58.40 \\
    Better (1) & 14.07 & 15.47 & 3.97 & 3.91 \\
    Much better (2) & 4.92 & 7.09 & 0.93 & 1.46 \\
    \bottomrule
    \end{tabularx}
    \caption{In-group and out-group perception for right-leaning users only}
    \label{tab:in_out_group_affect_right_users}
\end{table}

\begin{table}[hbtp]
    \centering
    \begin{tabularx}{\textwidth}{ @{} XXX @{} }
    \toprule
    \textbf{Answer} & \textbf{\% Chronological} & \textbf{\% Engagement} \\
    \midrule
    No (0) & 55.33 & 52.44 \\
    Yes (1) & 44.67 & 47.56 \\
    \bottomrule
    \end{tabularx}
    \caption{Out-group animosity for right-leaning users only}
    \label{tab:out_group_animosity_right_users}
\end{table}

\newpage
\subsubsection{Reader's stated preference}
Next, we show how the distribution of users' stated preference for tweets in their chronological and engagement-based timelines differs. Here we examine stated preference across all tweets and just political tweets. Notably, when considering all tweets, readers see similar percentages of unwanted tweets in both timelines (13 percent for both the chronological and engagement timelines). However, when we restrict to only political tweets, 22 percent of tweets in the engagement timeline are unwanted while only 16 percent of tweets in the chronological timeline are unwanted.

\begin{figure}[hbtp]
    \centering
    \includegraphics[width=\columnwidth]{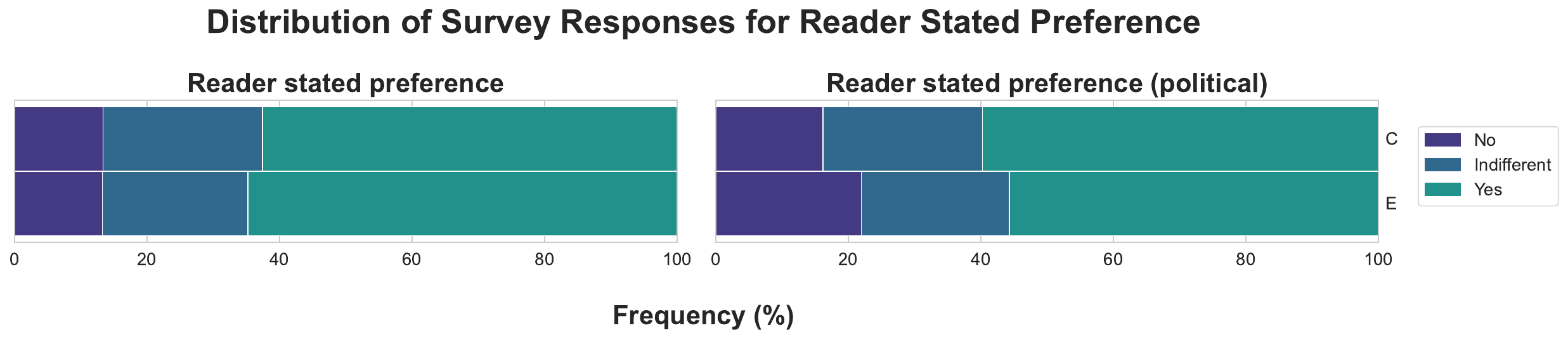}
    \caption{Distribution of readers' stated preference in the chronological (C) and engagement-based (E) timelines across all tweets (left figure) and political tweets only (right figure)}
    \label{fig:likert_reader_stated_pref}
\end{figure}

\begin{table}[hbtp]
    \centering
    \begin{tabularx}{\textwidth}{ @{} XXX @{} }
    \toprule
    \textbf{Answer} & \textbf{\% Chronological} & \textbf{\% Engagement} \\
    \midrule
    No (-1) & 13.38 & 13.29 \\
    Indifferent (0) & 24.03 & 21.94 \\
    Yes (1) & 62.59 & 64.78 \\
    \bottomrule
    \end{tabularx}
    \caption{Reader stated preference}
    \label{tab:reader_stated_preference}
\end{table}

\begin{table}[hbtp]
    \centering
    \begin{tabularx}{\textwidth}{ @{} XXX @{} }
    \toprule
    \textbf{Answer} & \textbf{\% Chronological} & \textbf{\% Engagement} \\
    \midrule
    No (-1) & 16.21 & 21.96 \\
    Indifferent (0) & 24.02 & 22.37 \\
    Yes (1) & 59.77 & 55.68 \\
    \bottomrule
    \end{tabularx}
    \caption{Reader stated preference after restricting to political tweets only}
    \label{tab:reader_stated_preference_political}
\end{table}

\pagebreak

\subsection{Outcomes by tweet rank} \label{app:outcomes-by-rank}
\begin{figure}[hbtp]
    \centering
    \includegraphics[width=1.04\columnwidth]{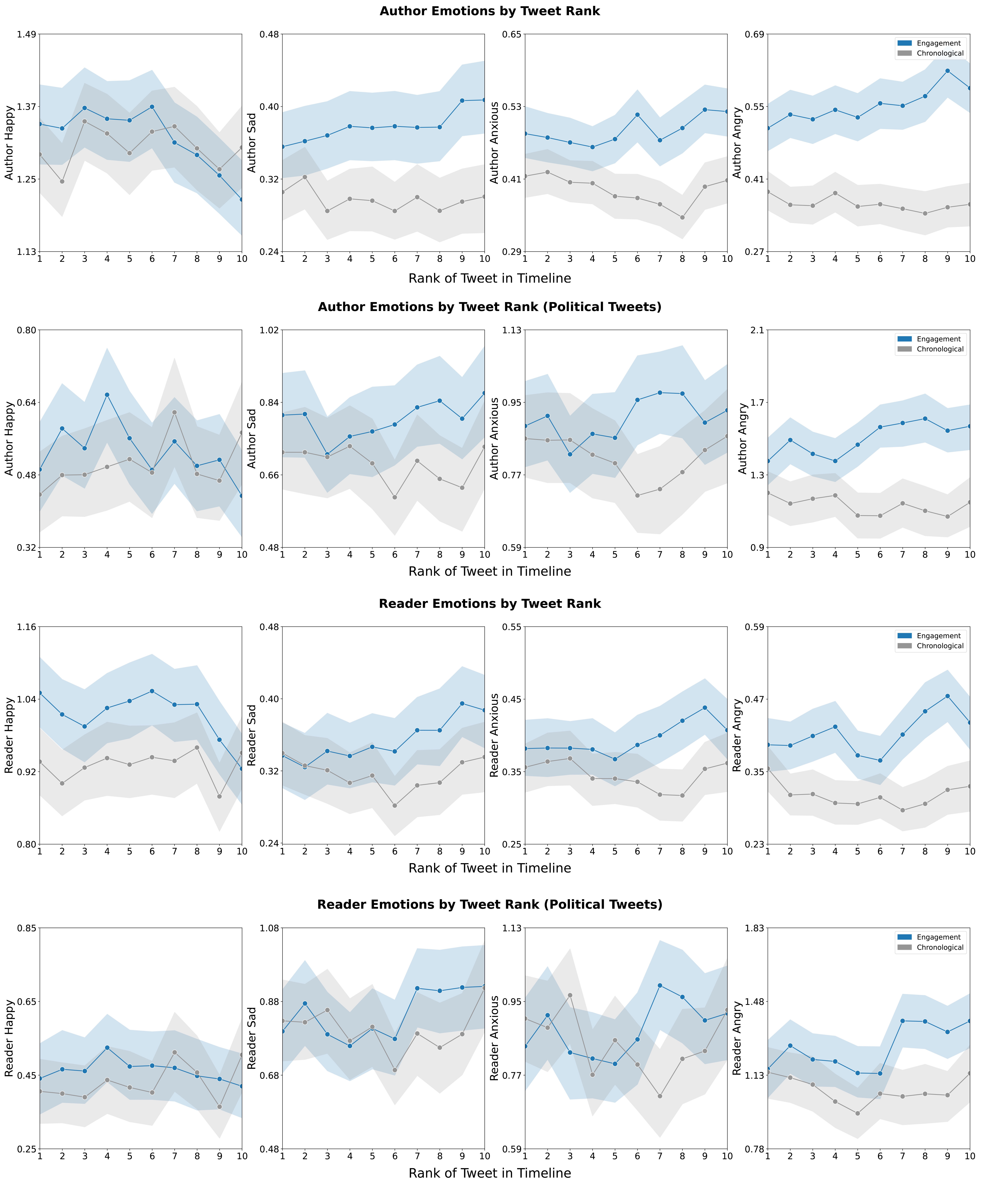}
    \caption{\footnotesize The mean of the author and reader's emotions broken down by the rank of the tweet in the timeline. The figures show outcomes both for tweets overall and specifically for political tweets. There is no statistically significant difference between high-ranked tweets and low-ranked tweets, although there appears to be a trend in which higher-ranked tweets are less likely to express negative emotions. The 95 percent confidence intervals are generated through bootstrapping.}
    \label{fig:emotions-by-rank}
\end{figure}

\begin{figure}[hbtp]
    \centering
    \includegraphics[width=0.98\columnwidth]{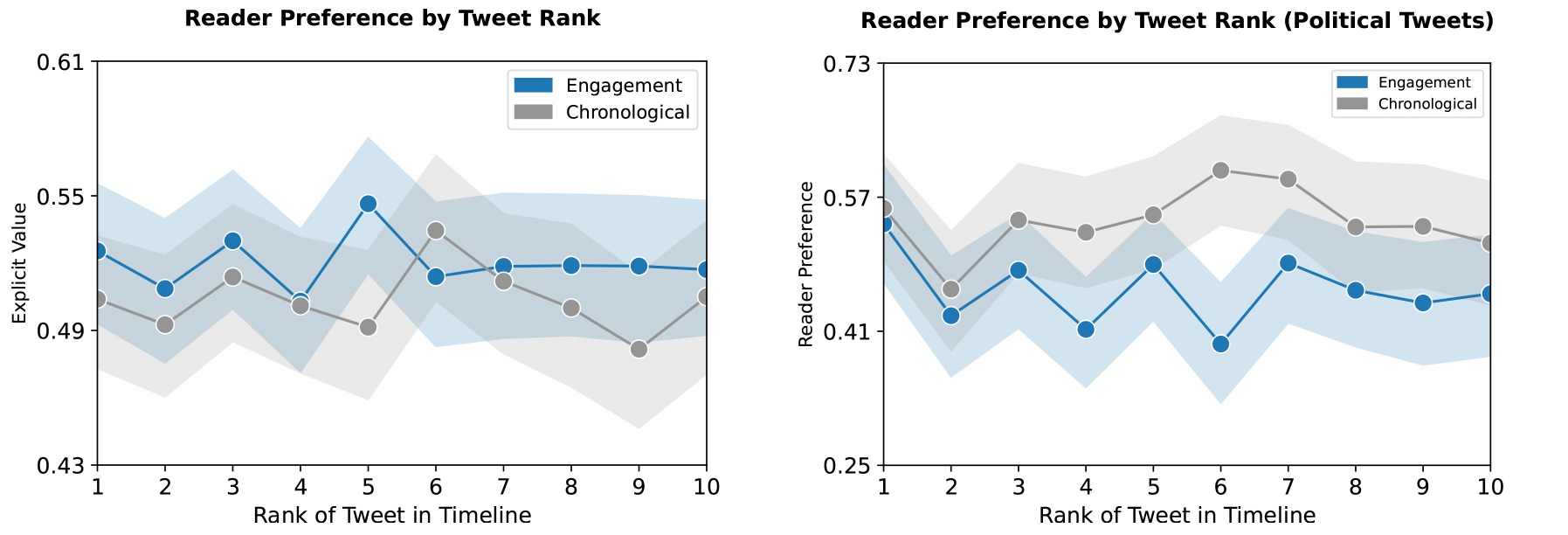}
    \caption{\small The mean of the readers' explicit preference as a function of the rank of the tweet in the timeline. The figures show outcomes both for tweets overall and specifically for political tweets. There is no statistically significant difference between high-ranked tweets and low-ranked tweets. The 95 percent confidence intervals are generated through bootstrapping.}
    \label{fig:reader_preference_by_rank}
\end{figure}

\begin{figure}[hbtp]
    \centering
    \includegraphics[width=0.98\columnwidth]{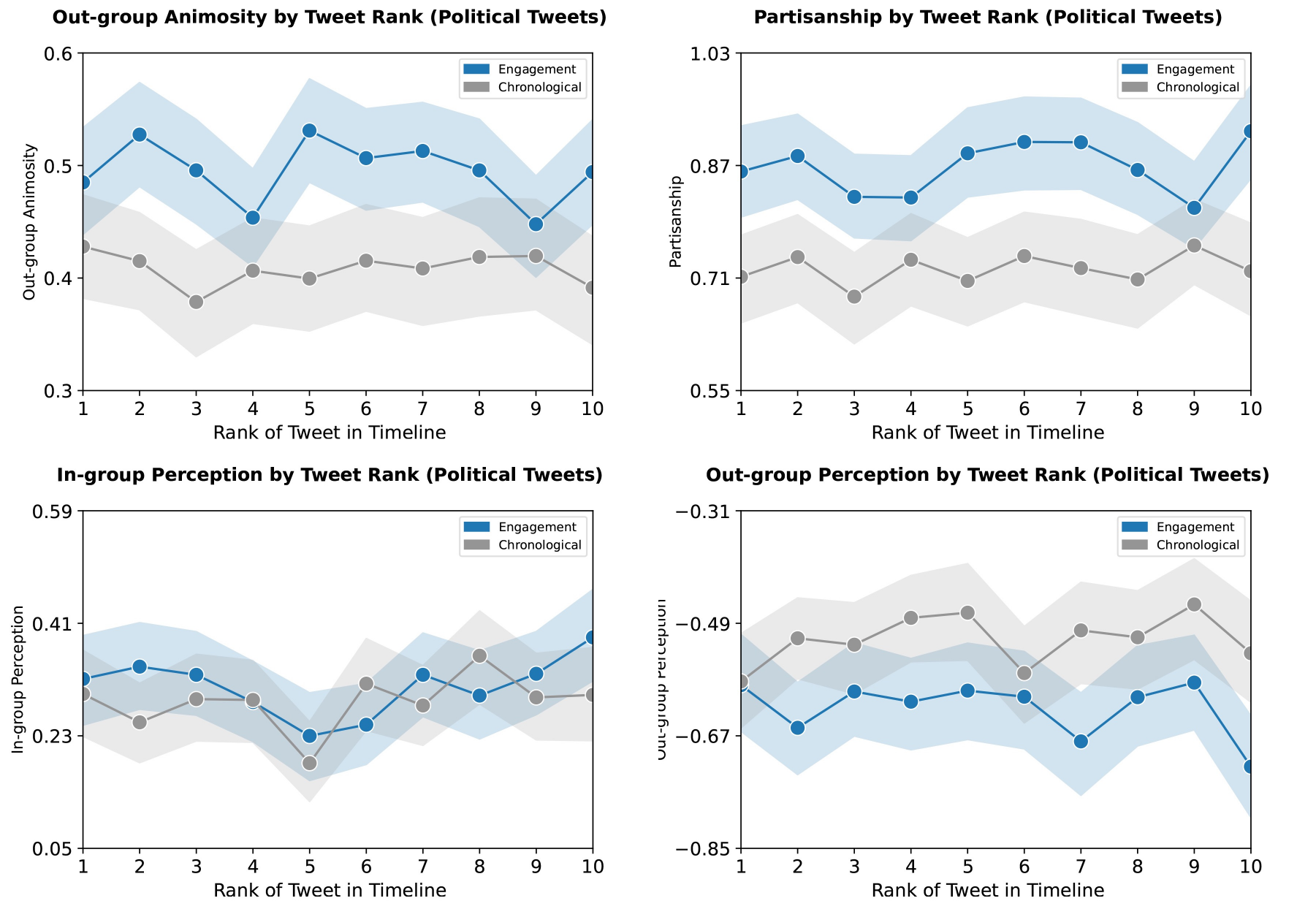}
    \caption{\small Out-group animosity, in-group perception, out-group perception, and partisanship broken down by the rank of the tweet in timeline. There is no statistically significant difference between high-ranked tweets and low-ranked tweets. The 95 percent confidence intervals are generated through bootstrapping.}
    \label{fig:political_effects_by_rank}
\end{figure}

Here, we explore how the outcomes measured change based on the position of the tweet in each timeline (Figures \ref{fig:emotions-by-rank}-\ref{fig:political_effects_by_rank}). A priori, we expect that the outcomes in the chronological timeline have no correlation with tweet position since the chronological timeline is simply ordered by time. However, it would be reasonable to hypothesize that higher-ranked tweets in the engagement-based timeline would exhibit differences from lower-ranked tweets from the timeline. That being said, we did not find any statistically significant differences between the high-ranked tweets and low-ranked tweets, even in the engagement-based timeline. However, it is important to note that we only consider the top 10 tweets in the timeline. It could be the case that if we considered more tweets, we would find larger differences between high-ranked tweets and low-ranked tweets.

\label{appendix:rank}

\newpage
\subsection{Effects of Engagement Timeline with Varied Tweet Threshold} \label{appendix:robustness_to_tweet_threshold}
We only survey users about the first 10 tweets in their engagement-based timeline and in their chronological timeline. Thus, when computing the average treatment effect (ATE) of Twitter's engagement-based timeline, we only consider the first 10 tweets in both. To evaluate the robustness of our results to this choice of threshold, here we also calculate the ATE across a range of thresholds, spanning from 5 to 10 tweets. Across these thresholds, our findings were consistently quite robust. This consistency aligns with our observations in \Cref{app:outcomes-by-rank}, where no significant disparities were identified between high and low-ranked tweets concerning the outcomes we evaluated. Nonetheless, potential differences could emerge further down the timeline ranking. As future work, it would be interesting to execute a broader robustness analysis that encompasses thresholds beyond the initial 10 tweets.

\begin{figure}[htbp]
    \centering
    \includegraphics[width=0.98\columnwidth]{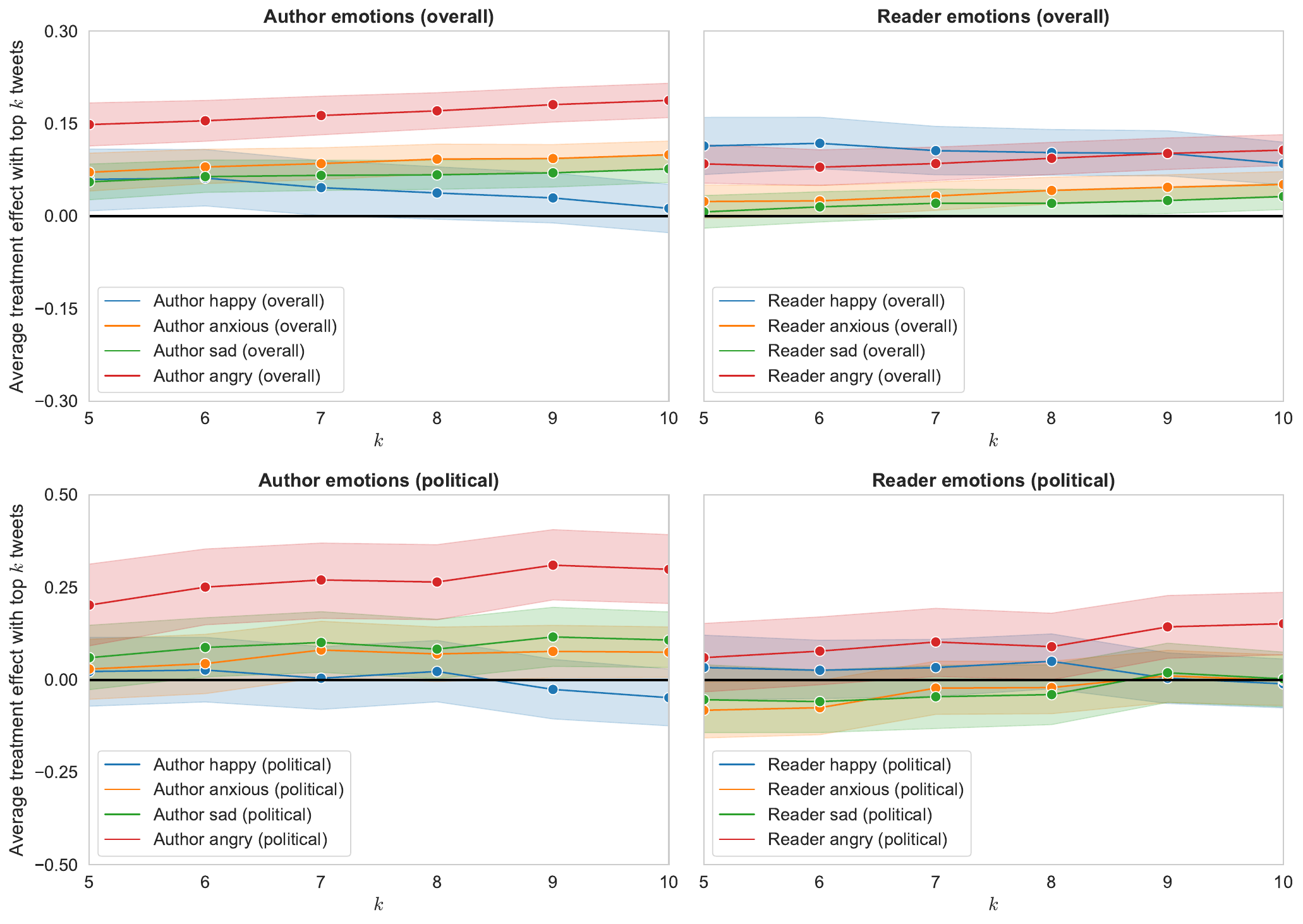}
    \caption{\small The ATE, based on different tweet thresholds, for the author and readers' emotions. The top two plots show these outcomes for all tweets, and the bottom plots restrict to only political tweets. The effect sizes are in the original units and are not standardized.}
    \label{}
\end{figure}

\begin{figure}[htbp]
    \centering
    \includegraphics[width=0.98\columnwidth]{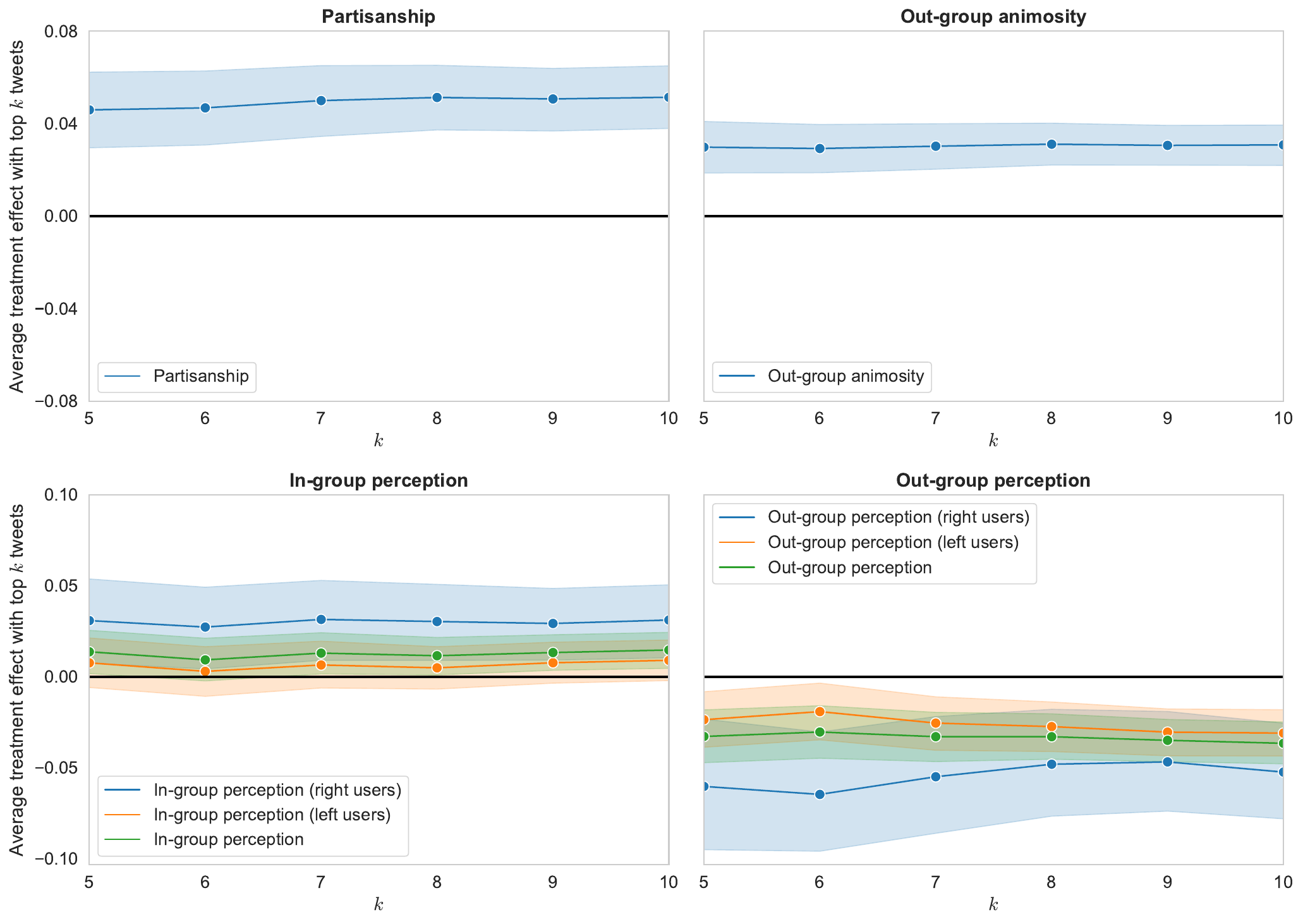}
    \caption{\small The ATE, based on different tweet thresholds, for partisanship, out-group animosity, in-group perception, and out-group perception. The effect sizes are in the original units and are not standardized.}
    \label{fig:political_effects}
\end{figure}

\begin{figure}[htbp]
    \centering
    \includegraphics[width=0.55\columnwidth]{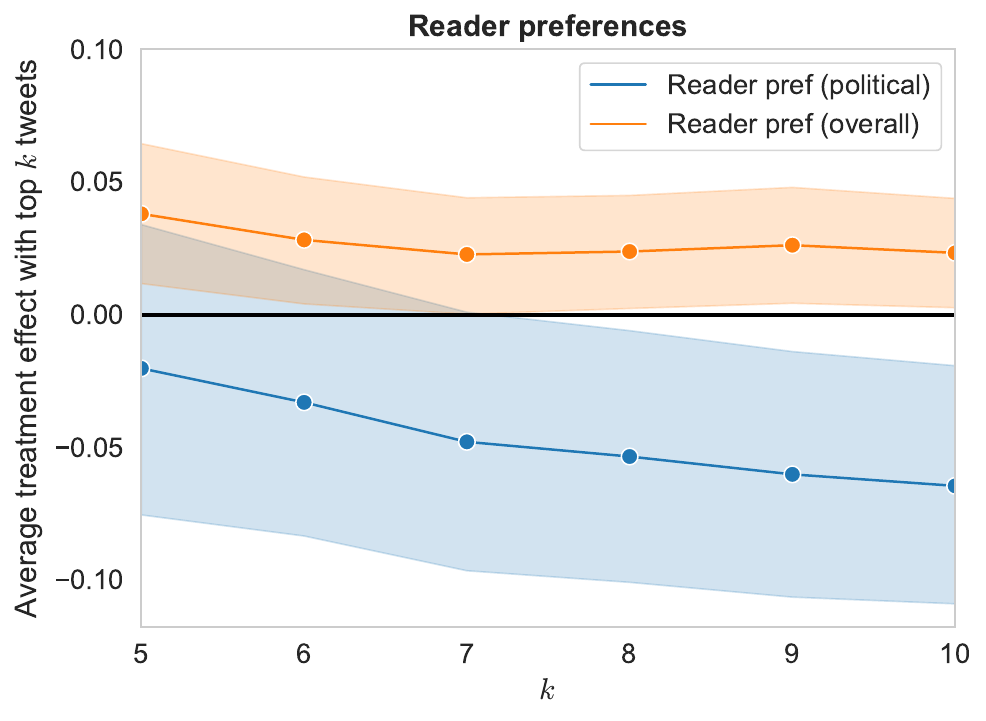}
    \caption{\small The ATE, based on different tweet thresholds, for the readers' explicit preference for the tweet. The effect sizes are in the original units and are not standardized.}
    \label{fig:reader_prefs}
\end{figure}
\newpage
\subsection{Effects of engagement timeline with GPT-4 labels} \label{appendix:gpt-effects}

\begin{figure}
    \centering
    \includegraphics[width=\textwidth]{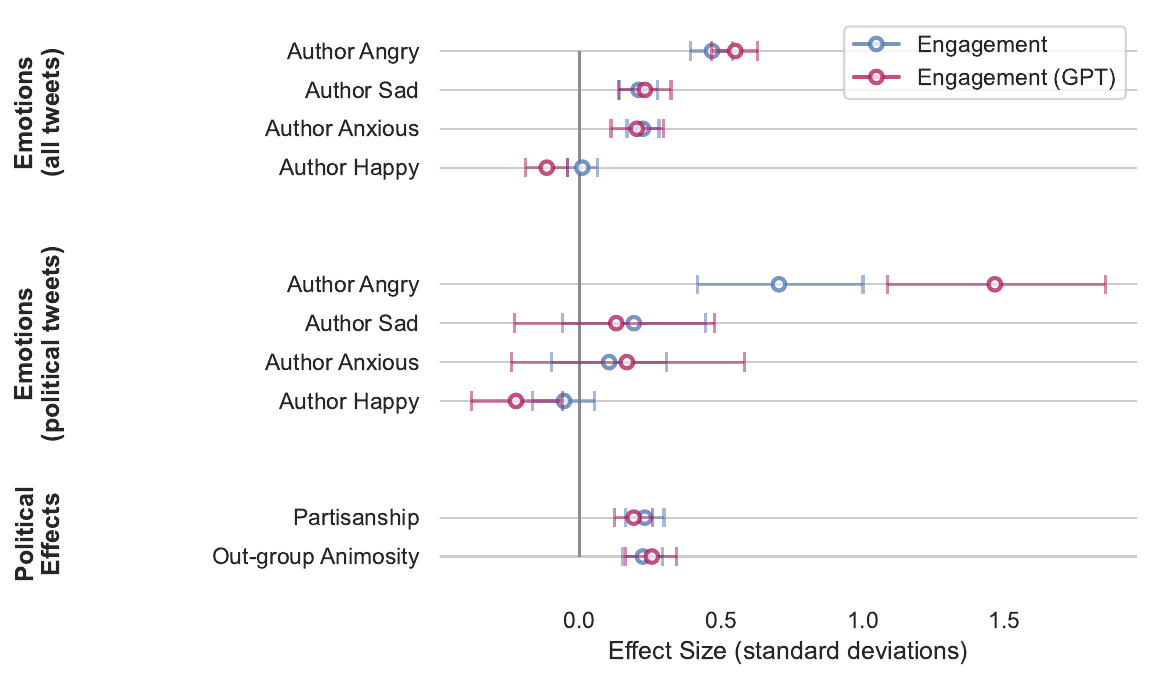}
    \caption{A comparison of the average treatment effects for tweet-based outcomes, using both the labels given by our participants and labels given by GPT-4. For this comparison, only tweets that GPT-4 provided a validly formatted response to were used to estimate the average treatment effects (for both the ATEs calculated with the human labels and the ATEs calculated with the GPT-4 labels).}
    \label{fig:gpt}
\end{figure}

\begin{center}
\begin{table}[t]
\setlength{\tabcolsep}{4pt}
\begin{tabularx}{\textwidth}{lccccc}
\toprule
\multirow{2}{*}{{\textbf{Outcome}}} & \textbf{Standardized} & \textbf{Unstandardized} & \textbf{Chronological} & \textbf{Engagement} & \multirow{2}{*}{\textbf{$p$-value}} \\
& \textbf{Effect} & \textbf{Effect} & \textbf{Mean} & \textbf{Mean} &\textbf{}  \\
\hline
\multicolumn{6}{c}{\textbf{Emotional effects (all tweets)}} \\
\hline
Author Angry &               0.5490 &                 0.1710 &      0.2495 &     0.4200 &   0.0002  \\
Author Sad &                 0.2302 &                 0.0508 &      0.2285 &     0.2845 &   0.0002  \\
Author Anxious &             0.2021 &                 0.0283 &      0.1067 &     0.1372 &   0.0004  \\
Author Happy &              -0.1158 &                -0.0670 &      1.3921 &     1.3303 &   0.0012  \\
\hline
\multicolumn{6}{c}{\textbf{Emotional effects (political tweets only)}} \\
\hline
Author Angry &              1.4653 &                 0.4565 &      1.2030 &     1.6045 &   0.0002  \\
Author Sad &                0.1294 &                 0.0285 &      0.5129 &     0.5338 &   0.4938  \\
Author Anxious &            0.1668 &                 0.0233 &      0.2060 &     0.2169 &   0.4408  \\
Author Happy&              -0.2244 &                -0.1299 &      0.6099 &     0.4841 &   0.0066  \\
\hline
\multicolumn{6}{c}{\textbf{Political effects}} \\
\hline
Partisanship &              0.1909 &                 0.0283 &      0.0950 &     0.1264 &   0.0002  \\
Out-group Animosity &       0.2548 &                 0.0311 &      0.0627 &     0.0922 &   0.0002  \\
\bottomrule
\end{tabularx}
\caption{Average treatment effects on tweet-based outcomes using GPT-4 judgments.}
\label{tab:gpt}
\end{table}
\end{center}

In our main analysis, all outcomes are measured through survey questions given to readers; this includes both outcomes that explicitly ask about the reader's perspective (i.e., the reader's emotions, in-group perceptions, out-group perceptions, and stated preferences) as well as outcomes that pertain to the tweet itself (the author's emotions, the ideological leaning of the tweet, and whether the tweet contains out-group animosity). Here, we elicit labels for the tweet-based outcomes from GPT-4~\cite{openai2023gpt} and report effects based on these machine learning labels (\Cref{tab:gpt} and \Cref{fig:gpt}).

All the results based on GPT-4 judgments are notably similar to the results based on reader judgments. Even under the GPT-4 labels, the algorithm is found to amplify content that is more emotional (especially angry), partisan, and likely to contain out-group animosity. The main difference in results is that GPT-4 judges the political tweets chosen by the engagement-based algorithm to be even angrier than how humans judged the tweets to be.

\subsubsection{Prompts and method}
We had GPT-4 (\texttt{gpt-4-1106-preview} in the OpenAI API) label the 28,301 unique tweets shown to users in our study. We asked GPT-4 to label the author's emotions in the tweet (along the four dimensions of angry, sad, happiness, and anxiety), whether the tweet was political, and if so, to label the ideological leaning of the tweet and whether the tweet had out-group animosity. All questions were phrased in the same way as the questions asked to our study participants. 

Out of the 28,301 unique tweets that we submitted to GPT-4 to label, it returned validly formatted responses for 24,998 tweets. To estimate the average treatment effects displayed in \Cref{fig:gpt}, we used only the tweets with valid GPT-4 responses (for both the ATEs based on human labels and those based on GPT-4 labels). This was done to ensure that any observed difference in ATEs between the human and GPT-4 labels was not influenced by using different sets of tweets for each. After filtering to tweets with valid GPT-4 responses, the ATEs were computed through the procedure described in \Cref{app:ate-estimation}, simply replacing human labels with GPT-4 labels when relevant.

Below is the prompt given to GPT-4 when a tweet is not a quote tweet or a reply to another tweet.

\begin{lstlisting}
Consider the given tweet by @{author_main}: `{main_tweet_text}`

1. How is @{author_main} feeling in their tweet?
    - Angry
    - Anxious
    - Happy
    - Sad
2. Is @{author_main}'s tweet about a political or social issue?

Answer NULL to the subsequent questions if your response to the previous 
question is "No".
3. How does @{author_main}'s tweet lean politically?
4. Is @{author_main}'s tweet expressing anger, frustration, or hostility
    towards a person or group on the Left?
5. Is @{author_main}'s tweet expressing anger, frustration, or hostility
    towards a person or group on the Right?

Respond with a JSON, of the format:
{{
    "author_emotions": {{
        "angry": [VALUE], 
        "anxious": [VALUE], 
        "happy": [VALUE], 
        "sad": [VALUE]
    }},
    "is_political": [VALUE],
    "political_leaning": [VALUE],
    "animosity_left": [VALUE],
    "animosity_right": [VALUE]
}}
where [VALUE] is an integer key from:
1. {{
        "Not at all": 0, 
        "Slightly": 1, 
        "Somewhat": 2, 
        "Moderately": 3, 
        "Extremely": 4
    }}
2. {{"No": 0, "Yes": 1}}
3. {{
        "Far Left": -2, 
        "Left": -1, 
        "Moderate": 0, 
        "Right": 1, 
        "Far Right": 2, 
        NULL
    }}
4. {{"No": 0, "Yes": 1, NULL}}
5. {{"No": 0, "Yes": 1, NULL}}
\end{lstlisting}

Below is the prompt used when a tweet is a quote tweet or a reply to another tweet. Like our participants, GPT-4 is given both tweets for context and asked to label both of them. In the following prompt, the variable \texttt{other\_tweet\_type} is either equal to ``quote tweet of'' or ``reply to.''

\begin{lstlisting}
Consider the following tweets, where @{author_main}'s tweet is a
{other_tweet_type} @{author_other}'s tweet: 
@{author_main}: `{main_tweet_text}`
@{author_other}: `{other_tweet_text}`

Note that there are two tweets. 
We will first be asking you questions about @{author_main}'s tweet. 
You can use @{author_other}'s tweet for context, but answer the following 
questions while focusing on @{author_main}'s tweet.
1. How is @{author_main} feeling in their tweet?
    - Angry
    - Anxious
    - Happy
    - Sad
2. Is @{author_main}'s tweet about a political or social issue?

Answer NULL to the subsequent questions if your response to the previous 
question is "No".
3. How does @{author_main}'s tweet lean politically?
4. Is @{author_main}'s tweet expressing anger, frustration, or hostility
towards a person or group on the Left?
5. Is @{author_main}'s tweet expressing anger, frustration, or hostility
towards a person or group on the Right?

Next, we will be asking you questions about @{author_other}'s tweet.
6. How is @{author_other} feeling in their tweet?
    - Angry
    - Anxious
    - Happy
    - Sad
7. Is @{author_other}'s tweet about a political or social issue?

Answer NULL to the subsequent questions if your response to the previous 
question is "No".
8. How does @{author_other}'s tweet lean politically?
9. Is @{author_other}'s tweet expressing anger, frustration, or hostility
towards a person or group on the Left?
10. Is @{author_other}'s tweet expressing anger, frustration, or hostility
towards a person or group on the Right?
    
Respond with a JSON, of the format:
{{
    "@{author_main}'s tweet": {{
        "author_emotions": {{
            "angry": [VALUE], 
            "anxious": [VALUE], 
            "happy": [VALUE], 
            "sad": [VALUE]
        }},
        "is_political": [VALUE],
        "political_leaning": [VALUE],
        "animosity_left": [VALUE],
        "animosity_right": [VALUE]
    }},
    "@{author_other}'s tweet": {{
        "author_emotions": {{
            "angry": [VALUE], 
            "anxious": [VALUE], 
            "happy": [VALUE], 
            "sad": [VALUE]
        }},
        "is_political": [VALUE],
        "political_leaning": [VALUE],
        "animosity_left": [VALUE],
        "animosity_right": [VALUE]
    }}
}}
where [VALUE] is an integer key from: 
1. {{
        "Not at all": 0, 
        "Slightly": 1, 
        "Somewhat": 2, 
        "Moderately": 3, 
        "Extremely": 4
    }}
2. {{"No": 0, "Yes": 1}}
3. {{
        "Far Left": -2, 
        "Left": -1, 
        "Moderate": 0, 
        "Right": 1, 
        "Far Right": 2, 
        NULL
    }}
4. {{"No": 0, "Yes": 1, NULL}}
5. {{"No": 0, "Yes": 1, NULL}}
6. {{
        "Not at all": 0, 
        "Slightly": 1, 
        "Somewhat": 2, 
        "Moderately": 3, 
        "Extremely": 4
    }}
7. {{"No": 0, "Yes": 1}}
8. {{
        "Far Left": -2, 
        "Left": -1, 
        "Moderate": 0, 
        "Right": 1, 
        "Far Right": 2, 
        NULL
    }}
9. {{"No": 0, "Yes": 1, NULL}}
10. {{"No": 0, "Yes": 1, NULL}}
\end{lstlisting}

\newpage

\subsection{Heterogeneous effects} \label{appendix:heterogenous-effects}

Here, we explore how the effects shown in \Cref{fig:effects} change when restricting our analysis to subpopulations of users, among demographic lines (or using other of our survey questions). We subset users the demographic subgroups described in \Cref{appendix:user-demographics}, and additionally by two survey questions (the users' self-reported main reason for using Twitter and main category of content seen during the user study):
\begin{itemize}
    \item User political leaning (``Political Leaning'')
    \item User political party (``Political Party'')
    \item User age group (``Age'')
    \item User gender (``Gender'')
    \item User race (``Race'')
    \item User ethicity (``Ethnicity'')
    \item User education level (``Education Level'')
    \item User annual household income (``Household Income'')
    \item User primary reason for using Twitter (``Why Twitter'')
    \item User category of content seen (``Primary Category of Content'')
\end{itemize}

For all questions we consider here that allow multiple answers (race, ethnicity, category of content), we double-count participants as being present in each subpopulation that they selected.

The phrasings of all questions, and answer formats can be found in \Cref{app:survey}. For easier reference we include below that of the non-demographic survey questions. 

We ask participants about the primary reason they use Twitter as follows: ``What would you say is the main reason you use Twitter?'' The options to select from are: ``A way to stay informed,'' ``Entertainment,'' ``Keeping me connected to other people,'' ``It's useful for my job or school,'' ``Let's me see different points of view,'' ``A way to express my opinions.''

We ask participants about the content shown to them as follows: ``What were the tweets we showed you today predominantly about? Select a maximum of two.'' The options to select from are ``News,'' ``Politics,'' ``Work,'' ``Entertainment,'' ``Hobbies.''

We report these effects by first displaying them for each outcome (\Cref{appendix:heterogenous-effects-by-outcome}) and then for each demographic attribute or survey question response (\Cref{appendix:heterogenous-effects-by-subgroup}).



\newpage

\subsubsection{Outcome effects for each subgroup, grouped by outcome}\label{appendix:heterogenous-effects-by-outcome}

\begin{figure}[H]
    \centering
    \begin{adjustbox}{center}
        \includegraphics[width=1.2\textwidth]{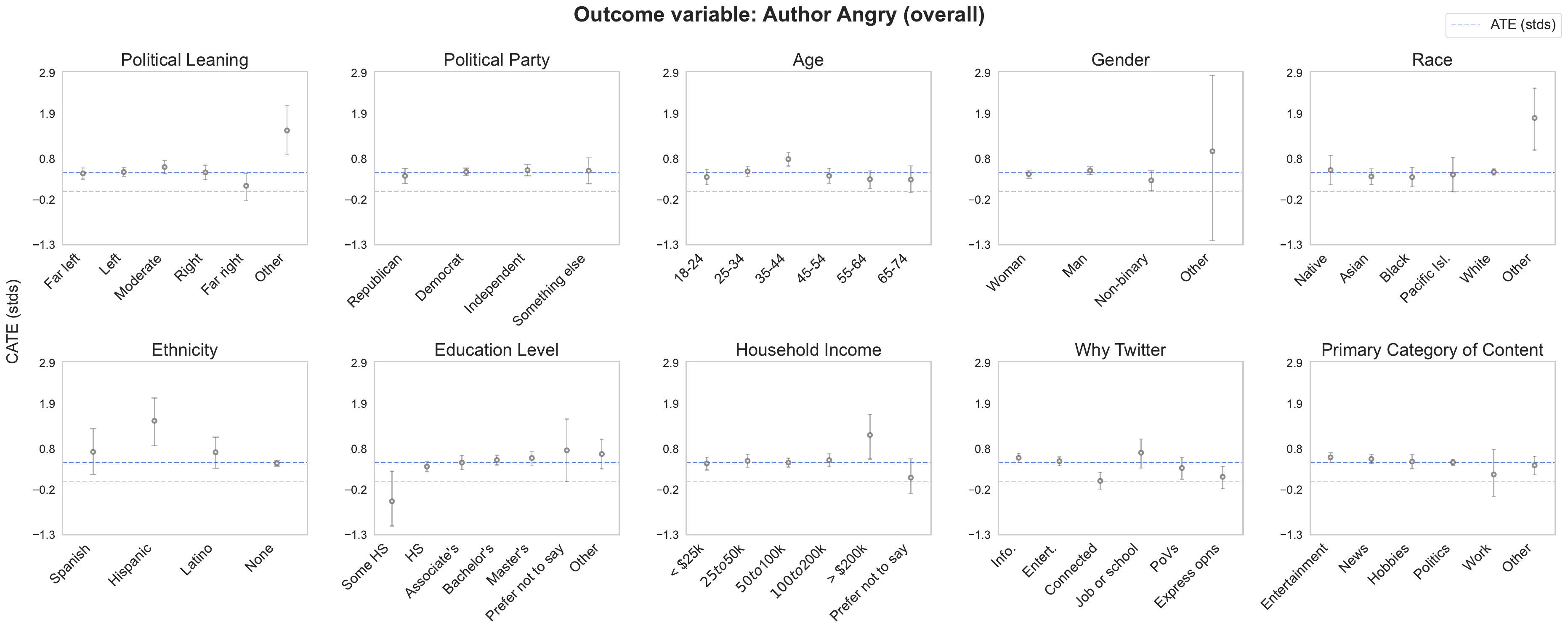}
    \end{adjustbox}
    \caption{\small Conditional average treatment effect (CATE) across subgroups for the outcome variable ‘author angry.’ The blue line shows the average treatment effect (ATE).}
\end{figure}

\begin{figure}[H]
    \centering
    \begin{adjustbox}{center}
        \includegraphics[width=1.2\textwidth]{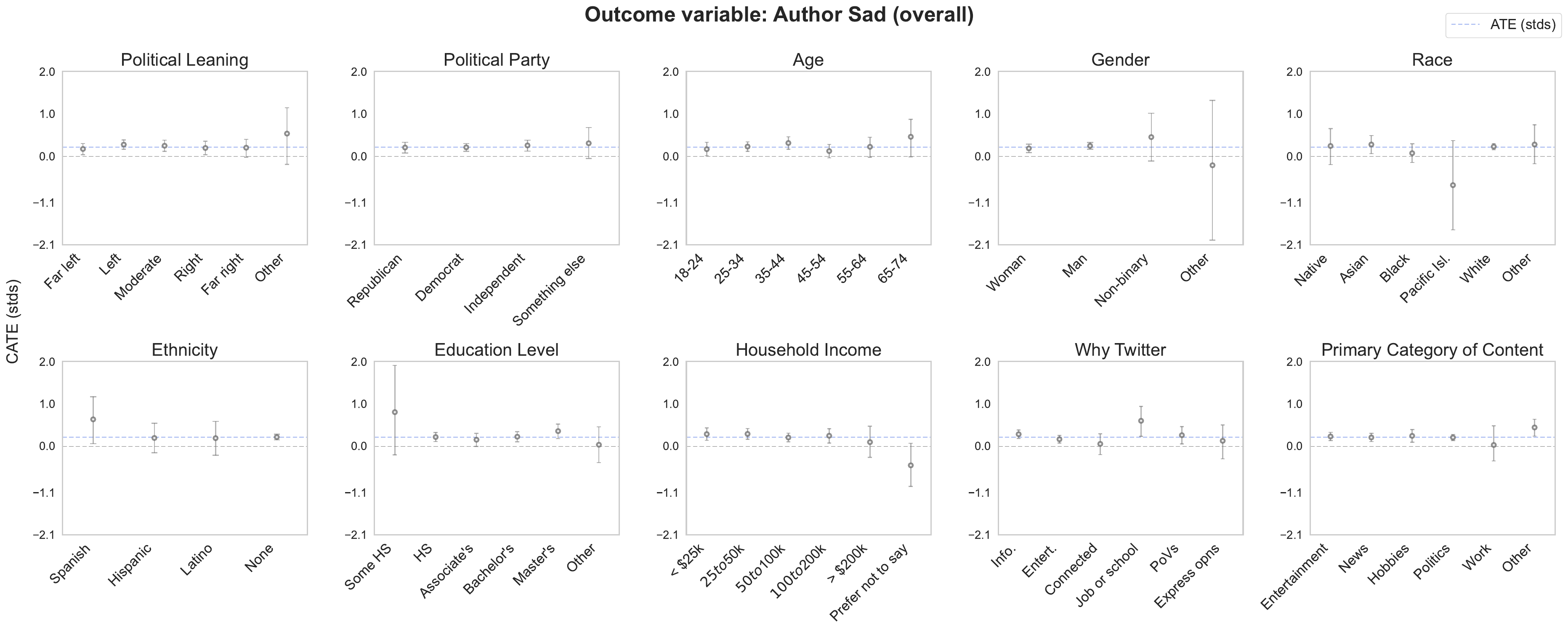}
    \end{adjustbox}
    \caption{\small Conditional average treatment effect (CATE) across subgroups for the outcome variable ‘author sad.’ The blue line shows the average treatment effect (ATE).}
\end{figure}

\begin{figure}[H]
    \centering
    \begin{adjustbox}{center}
        \includegraphics[width=1.2\textwidth]{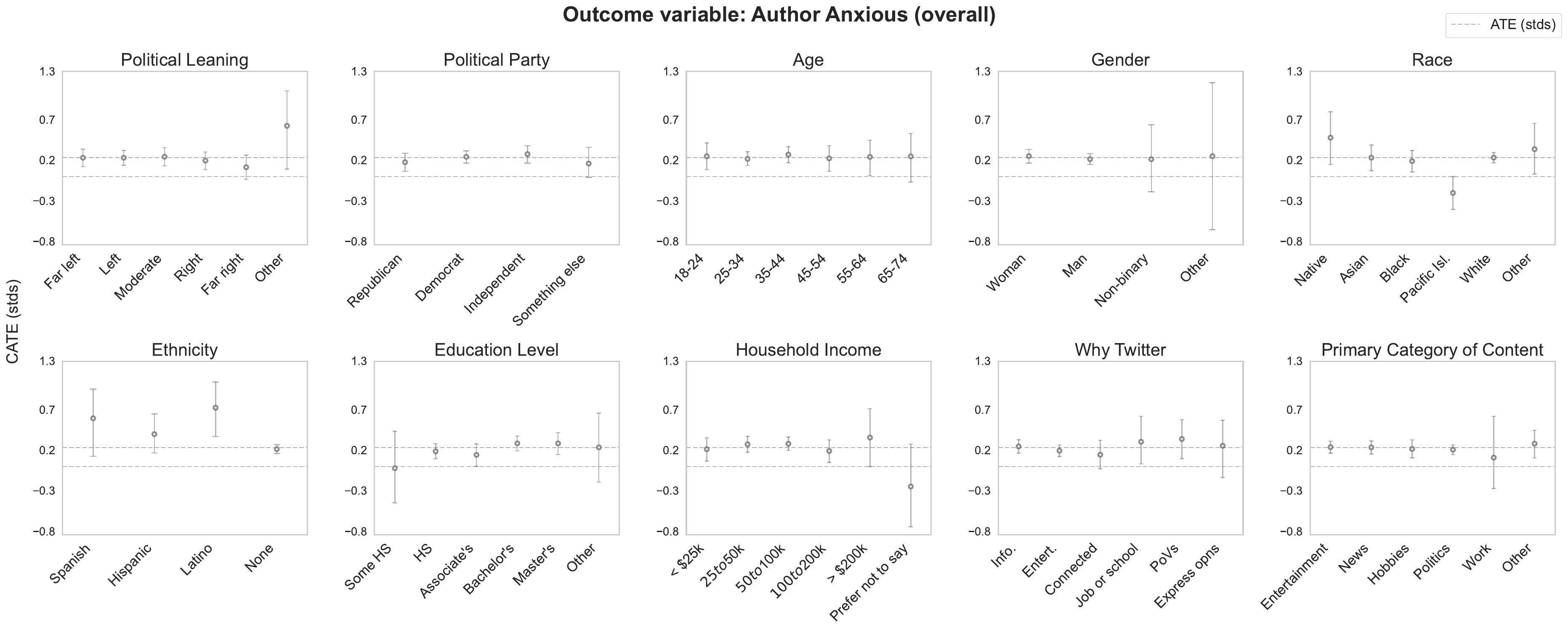}
    \end{adjustbox}
    \caption{\small Conditional average treatment effect (CATE) across subgroups for the outcome variable ‘author anxious.’ The blue line shows the average treatment effect (ATE).}
\end{figure}

\begin{figure}[H]
    \centering
    \begin{adjustbox}{center}
        \includegraphics[width=1.2\textwidth]{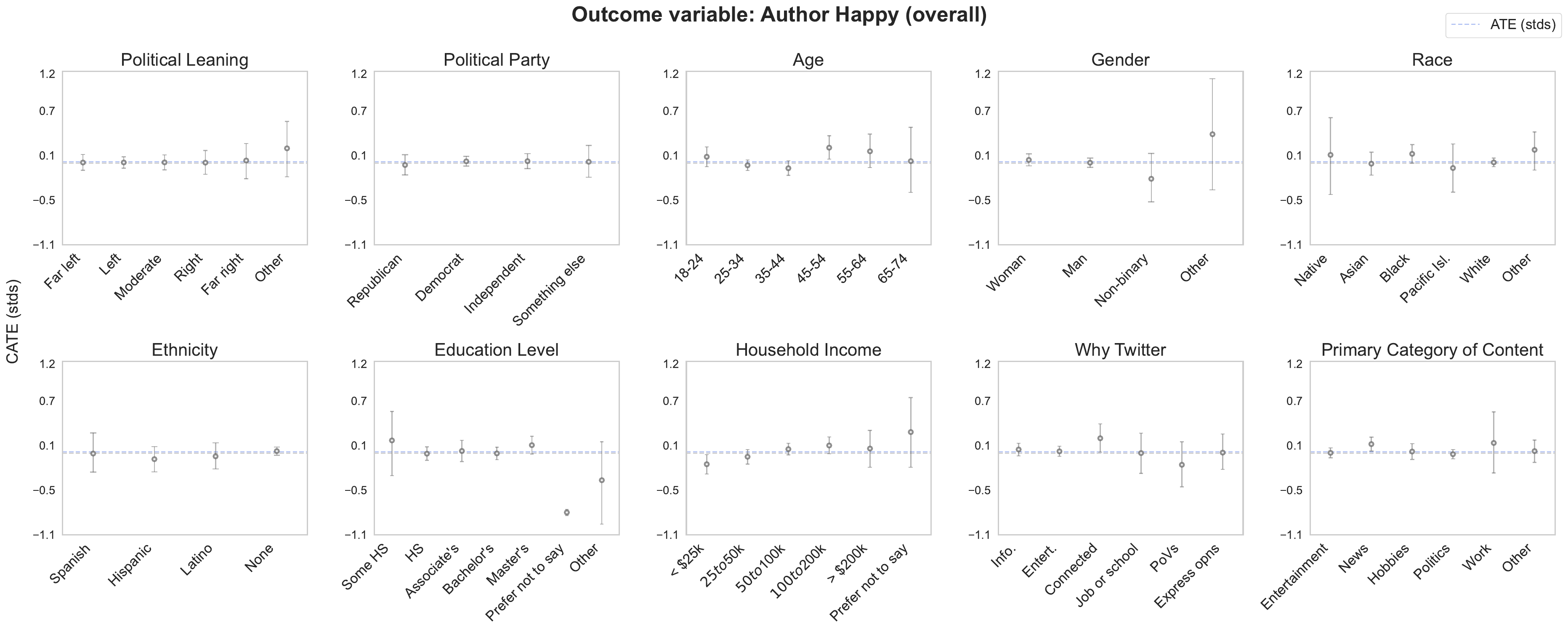}
    \end{adjustbox}
    \caption{\small Conditional average treatment effect (CATE) across subgroups for the outcome variable ‘author happy.’ The blue line shows the average treatment effect (ATE).}
\end{figure}

\begin{figure}[H]
    \centering
    \begin{adjustbox}{center}
        \includegraphics[width=1.2\textwidth]{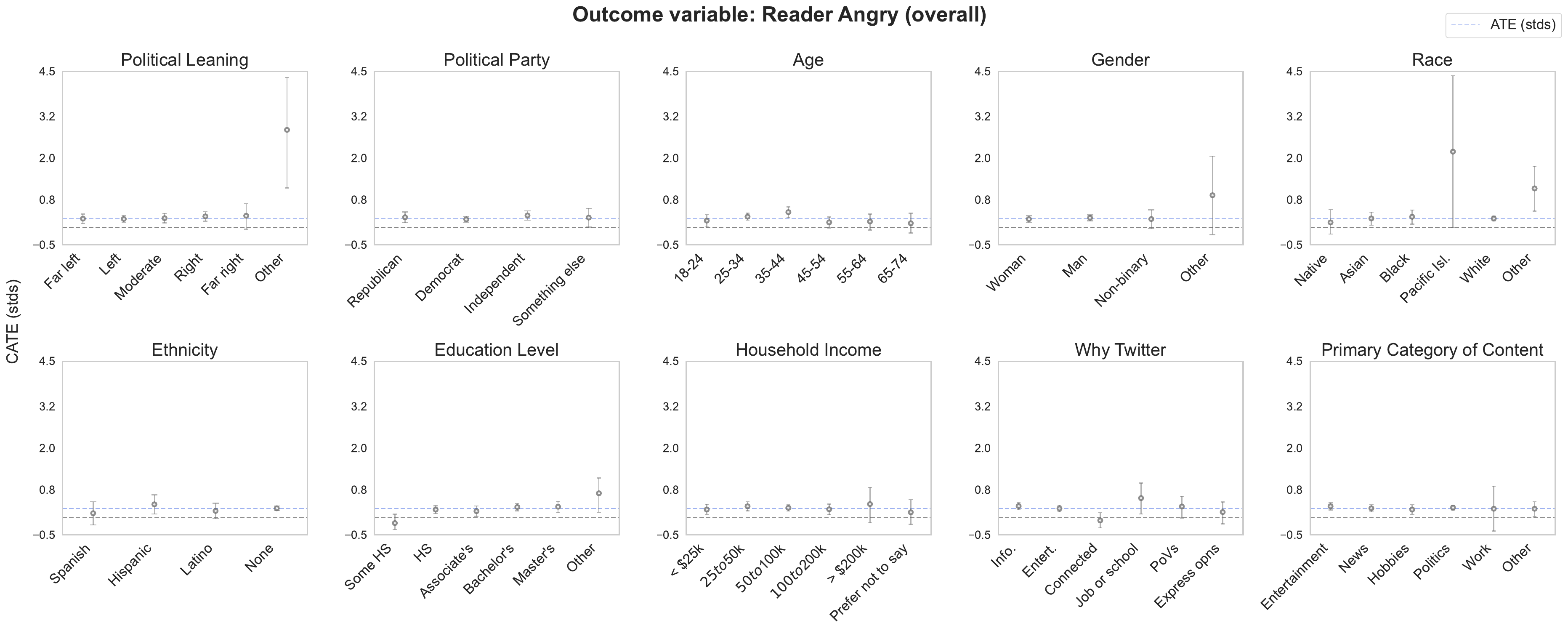}
    \end{adjustbox}
    \caption{\small Conditional average treatment effect (CATE) across subgroups for the outcome variable ‘reader angry (overall)’. The blue line shows the average treatment effect (ATE).}
\end{figure}

\begin{figure}[H]
    \centering
    \begin{adjustbox}{center}
        \includegraphics[width=1.2\textwidth]{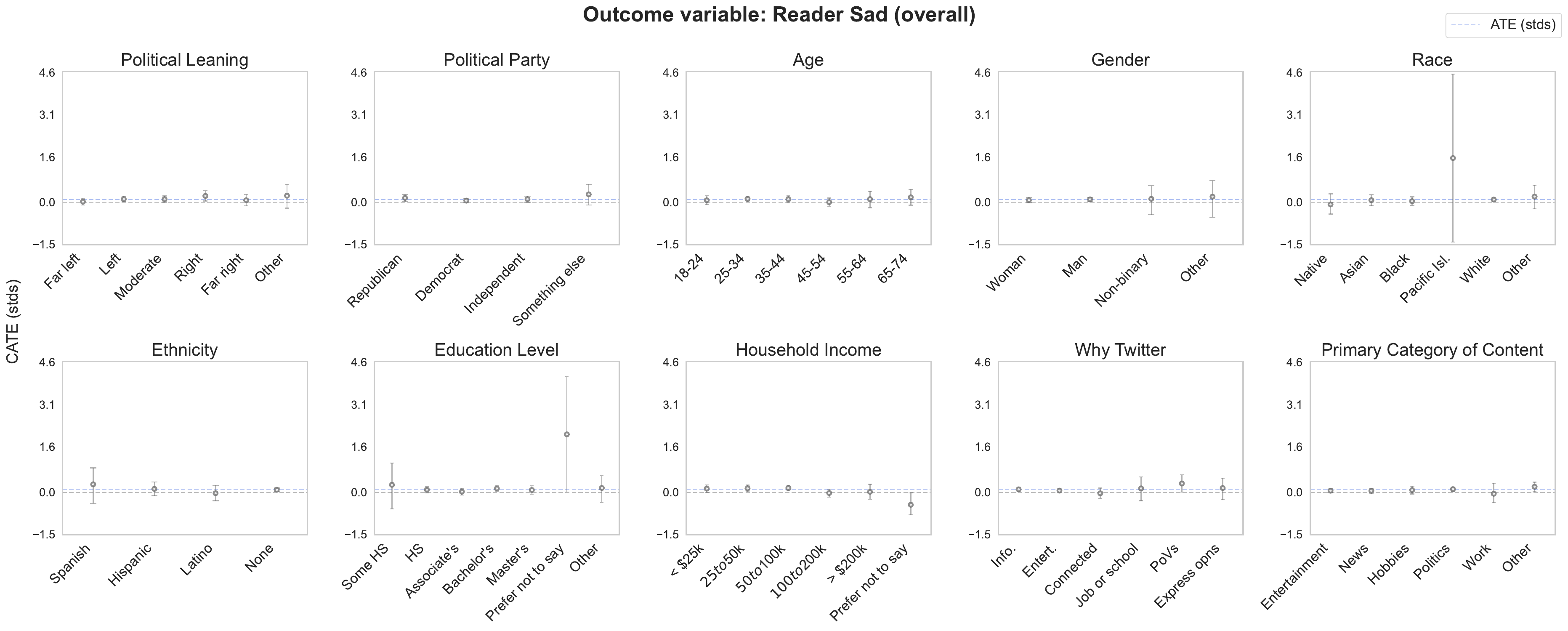}
    \end{adjustbox}
    \caption{\small Conditional average treatment effect (CATE) across subgroups for the outcome variable ‘reader sad (overall).’ The blue line shows the average treatment effect (ATE).}
\end{figure}

\begin{figure}[H]
    \centering
    \begin{adjustbox}{center}
        \includegraphics[width=1.2\textwidth]{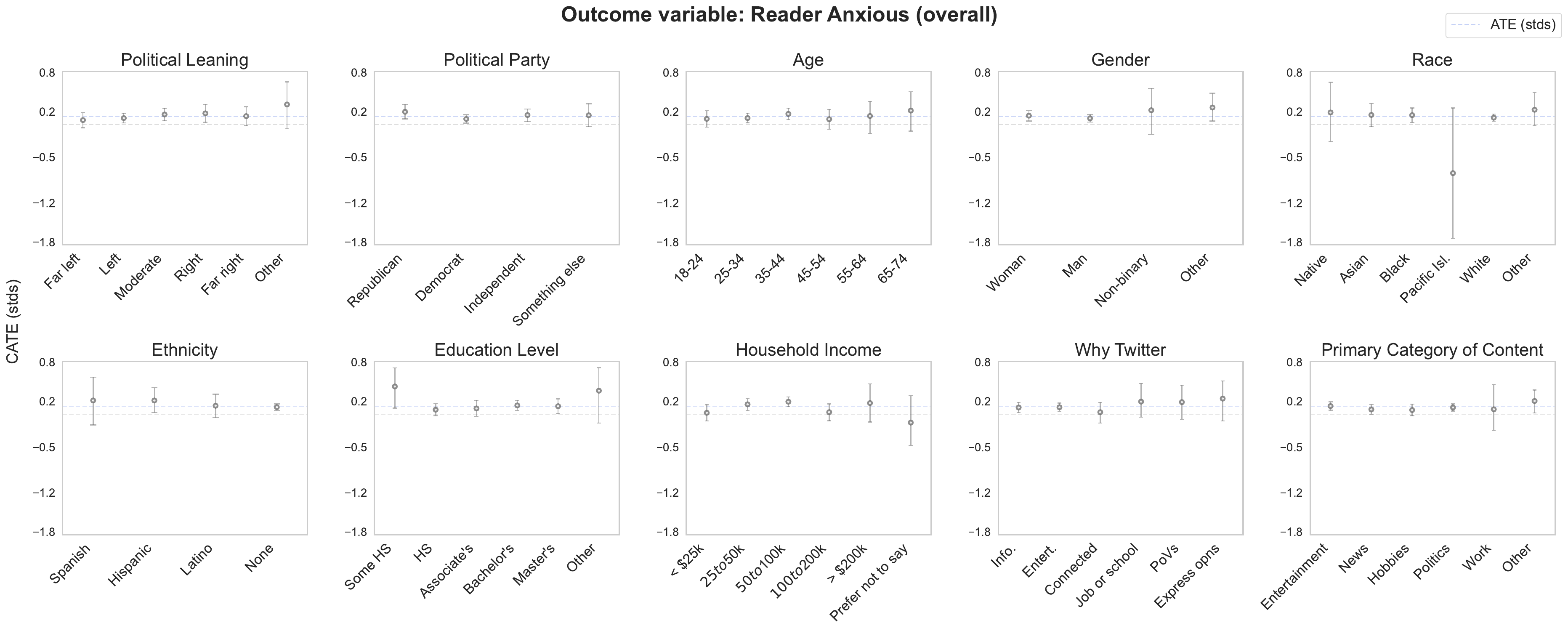}
    \end{adjustbox}
    \caption{\small Conditional average treatment effect (CATE) across subgroups for the outcome variable ‘reader anxious (overall).’ The blue line shows the average treatment effect (ATE).}
\end{figure}

\begin{figure}[H]
    \centering
    \begin{adjustbox}{center}
        \includegraphics[width=1.2\textwidth]{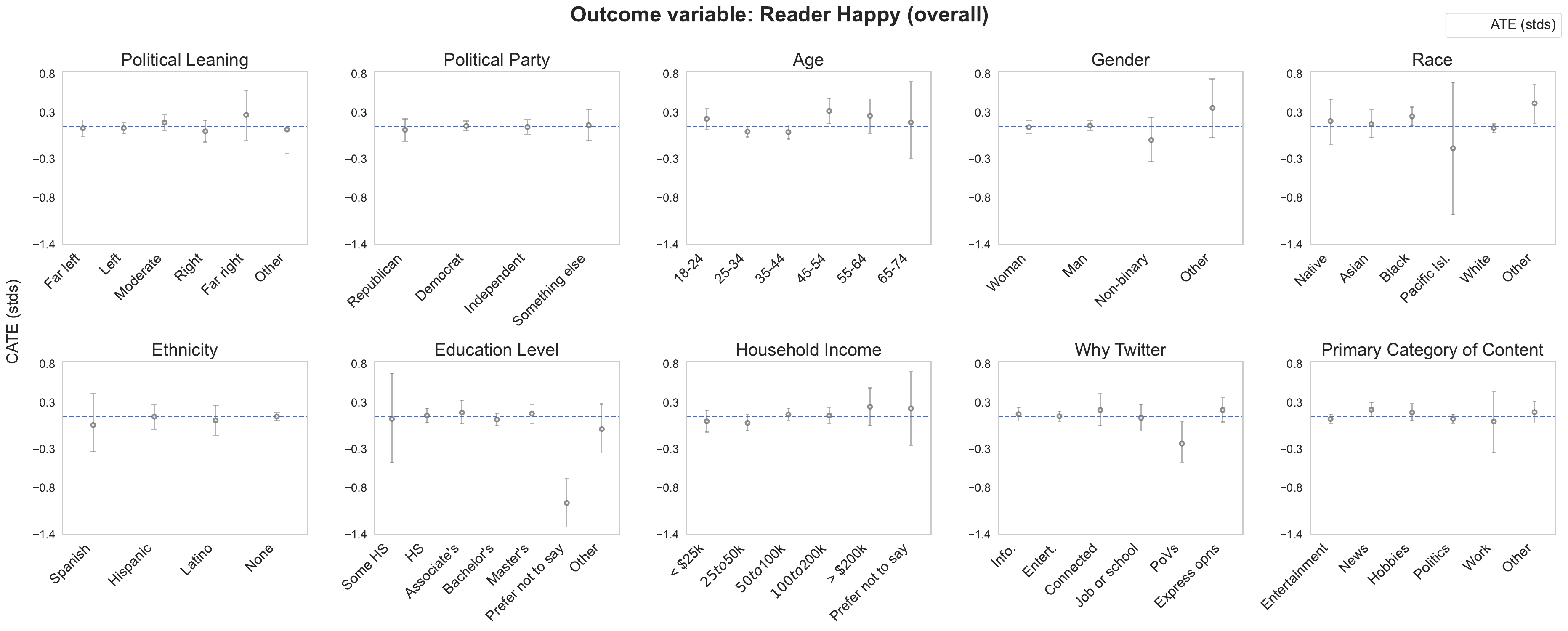}
    \end{adjustbox}
    \caption{\small Conditional average treatment effect (CATE) across subgroups for the outcome variable ‘reader happy (overall).’ The blue line shows the average treatment effect (ATE).}
\end{figure}

\begin{figure}[H]
    \centering
    \begin{adjustbox}{center}
        \includegraphics[width=1.2\textwidth]{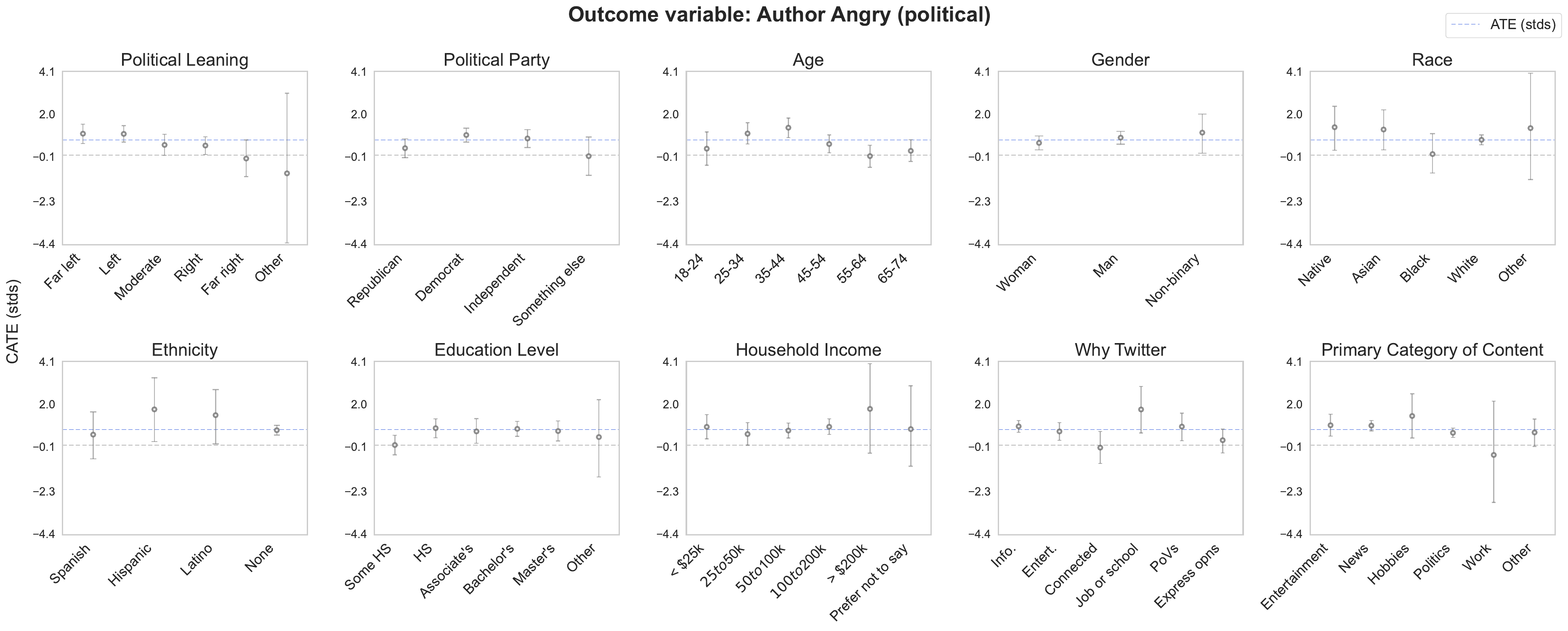}
    \end{adjustbox}
    \caption{\small Conditional average treatment effect (CATE) across subgroups for the outcome variable ‘author angry (political).’ The blue line shows the average treatment effect (ATE).}
\end{figure}

\begin{figure}[H]
    \centering
    \begin{adjustbox}{center}
        \includegraphics[width=1.2\textwidth]{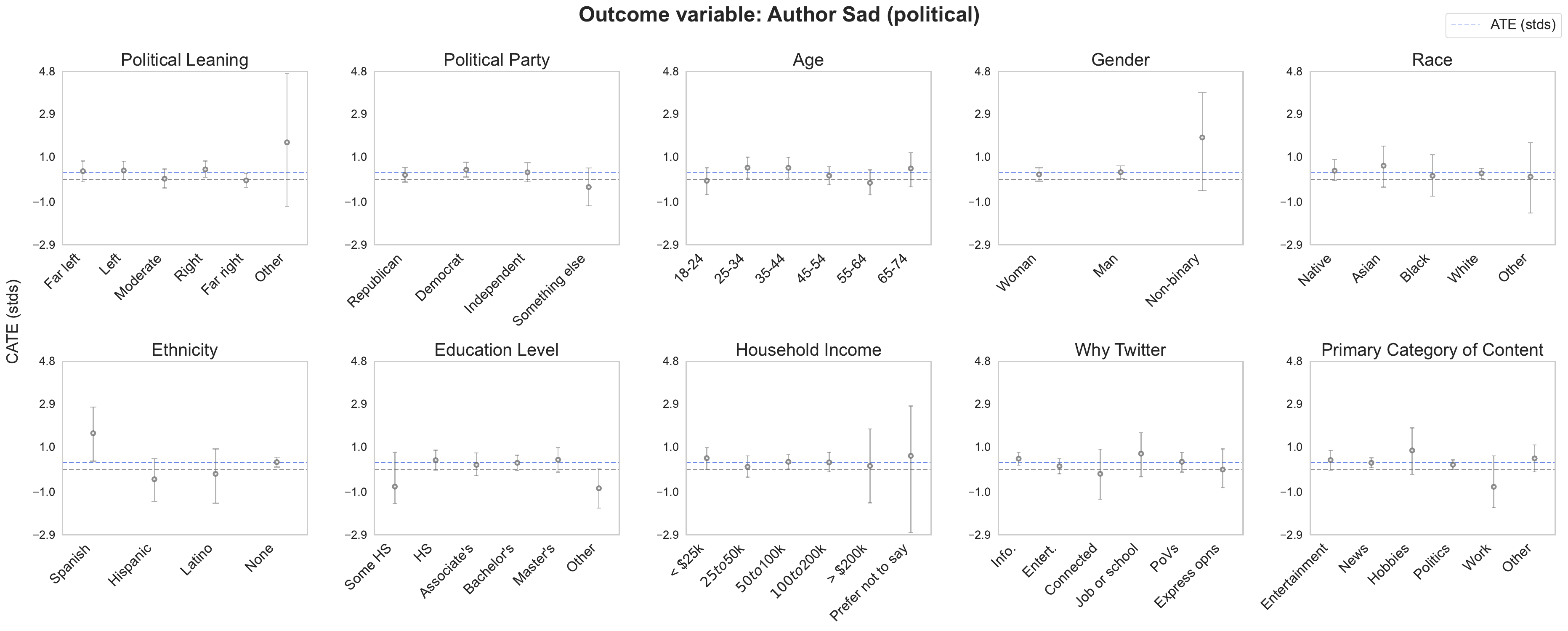}
    \end{adjustbox}
    \caption{\small Conditional average treatment effect (CATE) across subgroups for the outcome variable ‘author sad (political).’ The blue line shows the average treatment effect (ATE).}
\end{figure}

\begin{figure}[H]
    \centering
    \begin{adjustbox}{center}
        \includegraphics[width=1.2\textwidth]{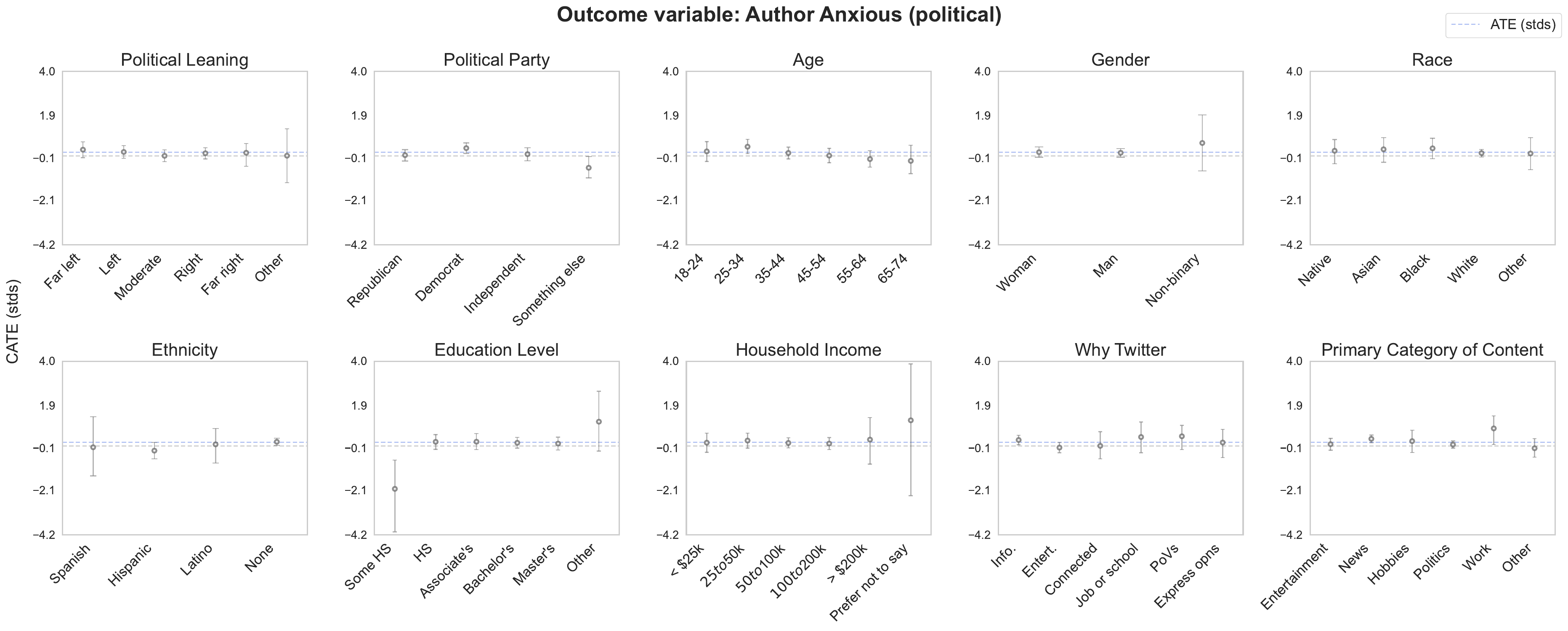}
    \end{adjustbox}
    \caption{\small Conditional average treatment effect (CATE) across subgroups for the outcome variable ‘author anxious (political).’ The blue line shows the average treatment effect (ATE).}
\end{figure}

\begin{figure}[H]
    \centering
    \begin{adjustbox}{center}
        \includegraphics[width=1.2\textwidth]{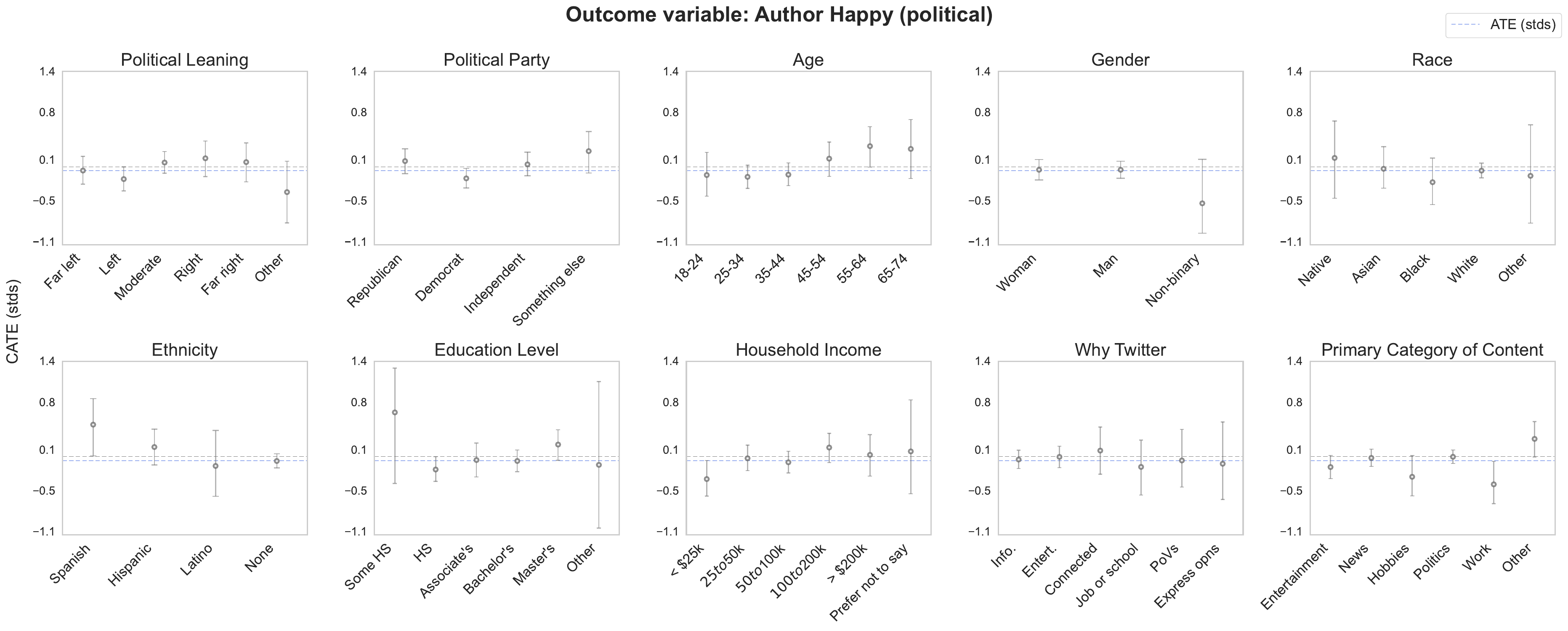}
    \end{adjustbox}
    \caption{\small Conditional average treatment effect (CATE) across subgroups for the outcome variable ‘author happy (political).’ The blue line shows the average treatment effect (ATE).}
\end{figure}

\begin{figure}[H]
    \centering
    \begin{adjustbox}{center}
        \includegraphics[width=1.2\textwidth]{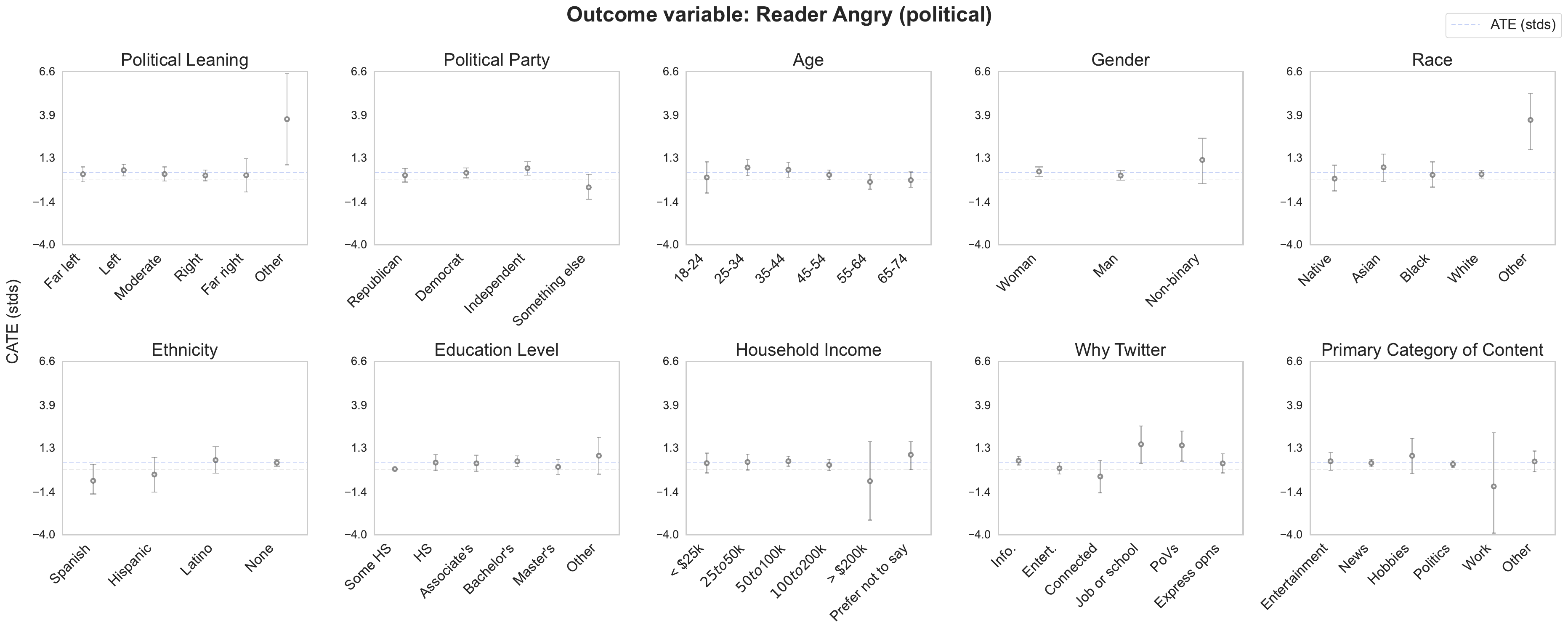}
    \end{adjustbox}
    \caption{\small Conditional average treatment effect (CATE) across subgroups for the outcome variable ‘reader angry (political).’ The blue line shows the average treatment effect (ATE).}
\end{figure}

\begin{figure}[H]
    \centering
    \begin{adjustbox}{center}
        \includegraphics[width=1.2\textwidth]{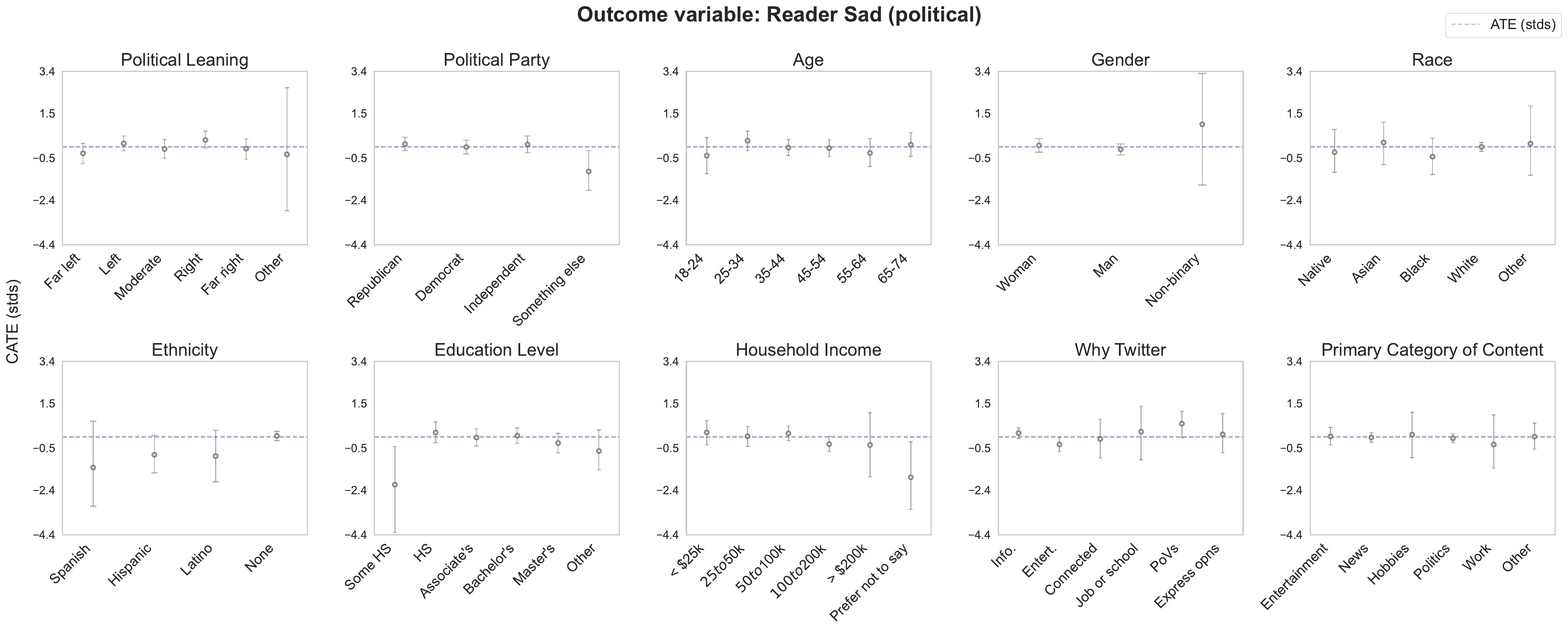}
    \end{adjustbox}
    \caption{\small Conditional average treatment effect (CATE) across subgroups for the outcome variable ‘reader sad (political).’ The blue line shows the average treatment effect (ATE).}
\end{figure}

\begin{figure}[H]
    \centering
    \begin{adjustbox}{center}
        \includegraphics[width=1.2\textwidth]{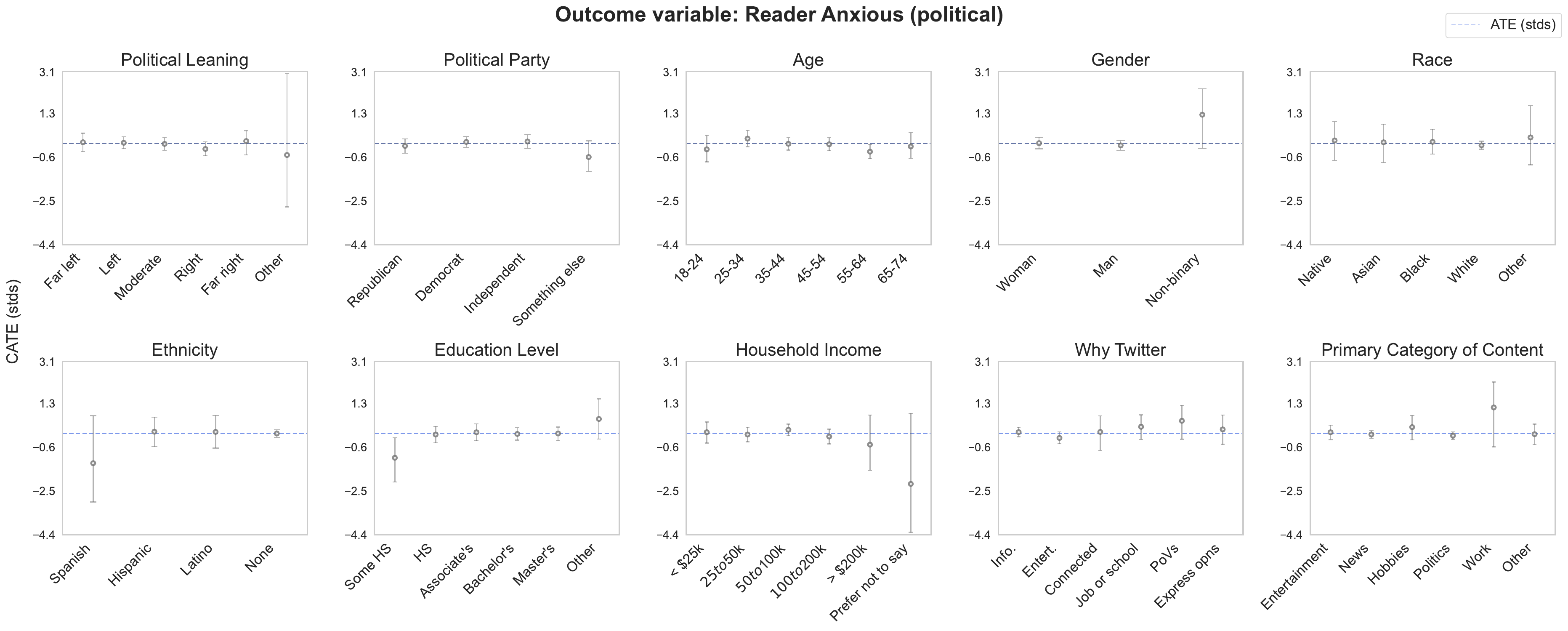}
    \end{adjustbox}
    \caption{\small Conditional average treatment effect (CATE) across subgroups for the outcome variable ‘reader anxious (political).’ The blue line shows the average treatment effect (ATE).}
\end{figure}

\begin{figure}[H]
    \centering
    \begin{adjustbox}{center}
        \includegraphics[width=1.2\textwidth]{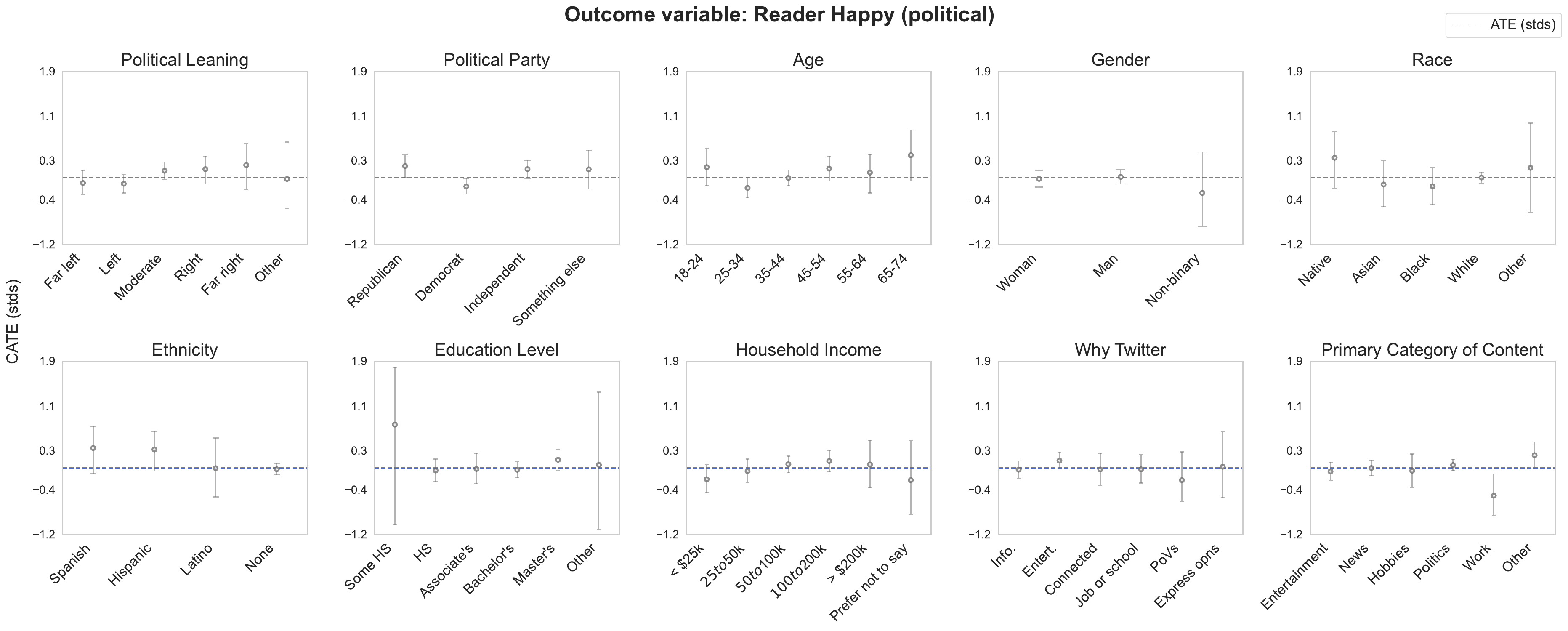}
    \end{adjustbox}
    \caption{\small Conditional average treatment effect (CATE) across subgroups for the outcome variable ‘reader happy (political).’ The blue line shows the average treatment effect (ATE).}
\end{figure}

\begin{figure}[H]
    \centering
    \begin{adjustbox}{center}
        \includegraphics[width=1.2\textwidth]{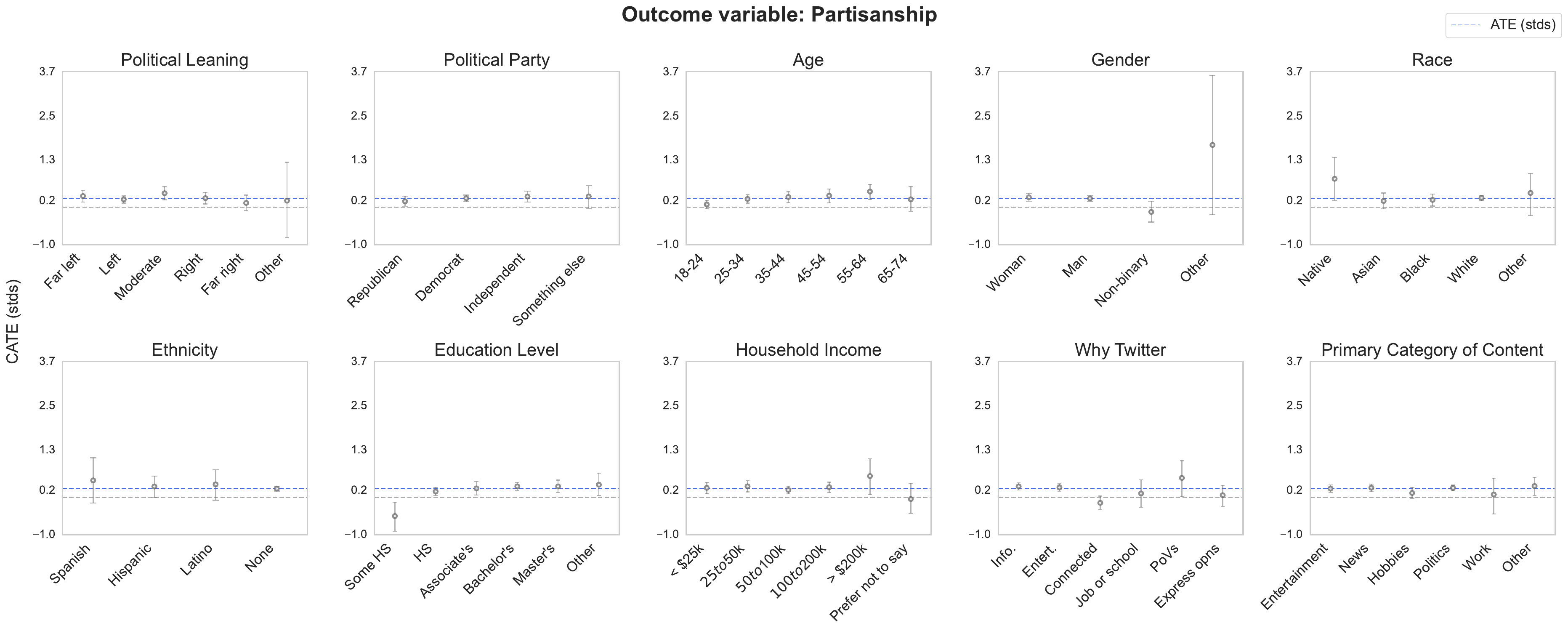}
    \end{adjustbox}
    \caption{\small Conditional average treatment effect (CATE) for the outcome variable ‘partisanship.’ The blue line shows the average treatment effect (ATE).}
\end{figure}

\begin{figure}[H]
    \centering
    \begin{adjustbox}{center}
        \includegraphics[width=1.2\textwidth]{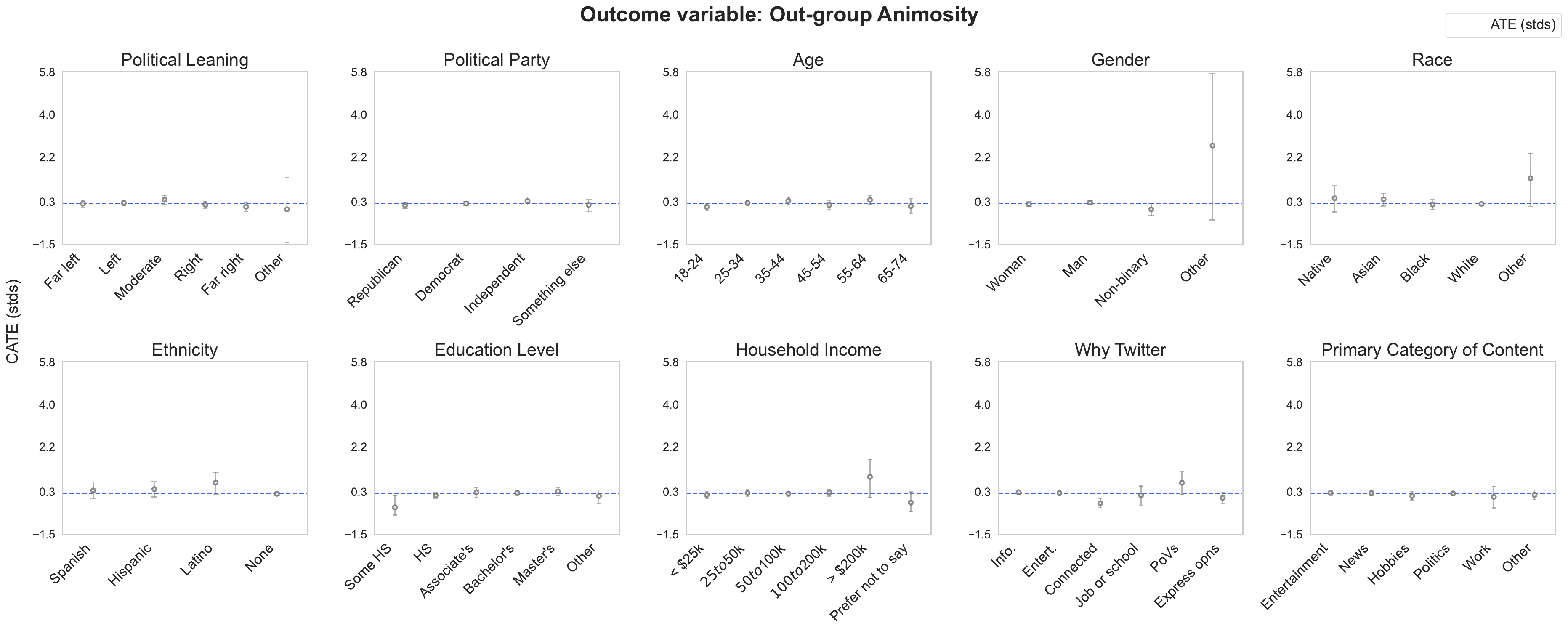}
    \end{adjustbox}
    \caption{\small Conditional average treatment effect (CATE) for the outcome variable ‘out-group animosity.’ The blue line shows the average treatment effect (ATE).}
\end{figure}

\begin{figure}[H]
    \centering
    \begin{adjustbox}{center}
        \includegraphics[width=1.2\textwidth]{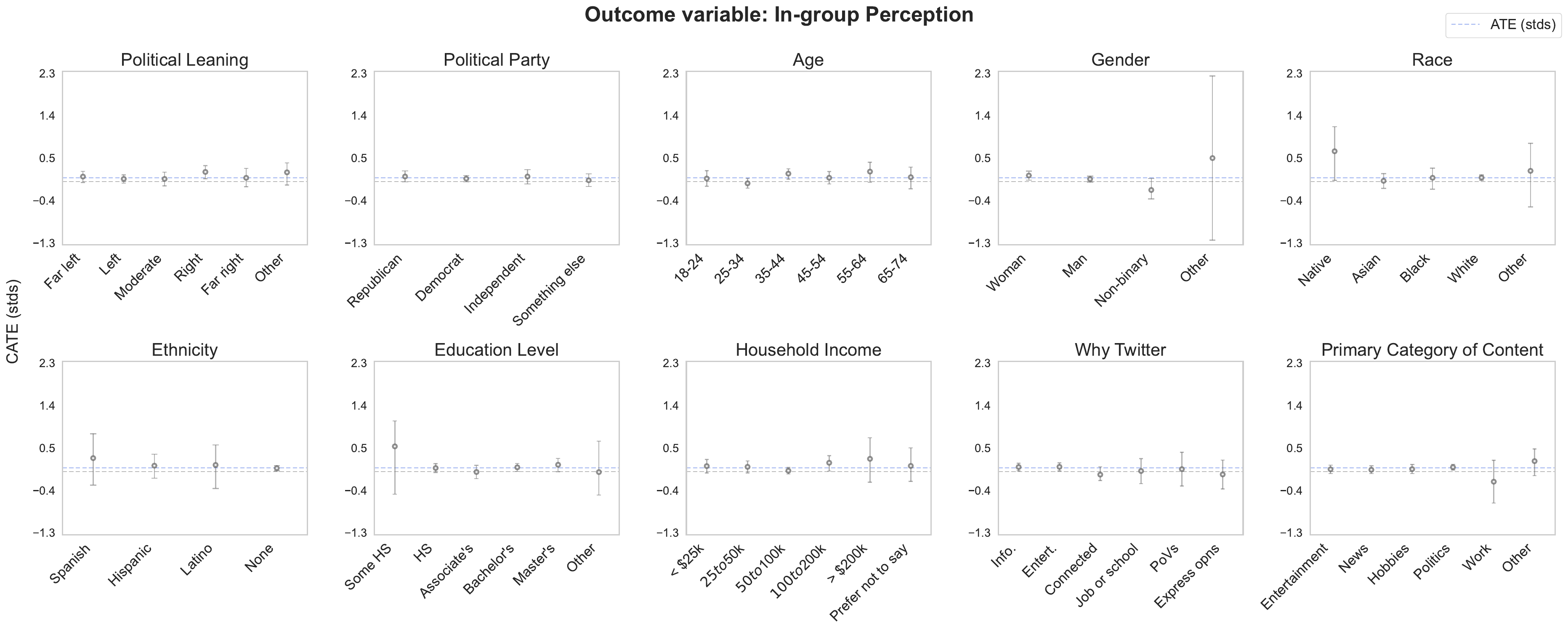}
    \end{adjustbox}
    \caption{\small Conditional average treatment effect (CATE) for the outcome variable ‘in-group perception.’ The blue line shows the average treatment effect (ATE).}
\end{figure}

\begin{figure}[H]
    \centering
    \begin{adjustbox}{center}
        \includegraphics[width=1.2\textwidth]{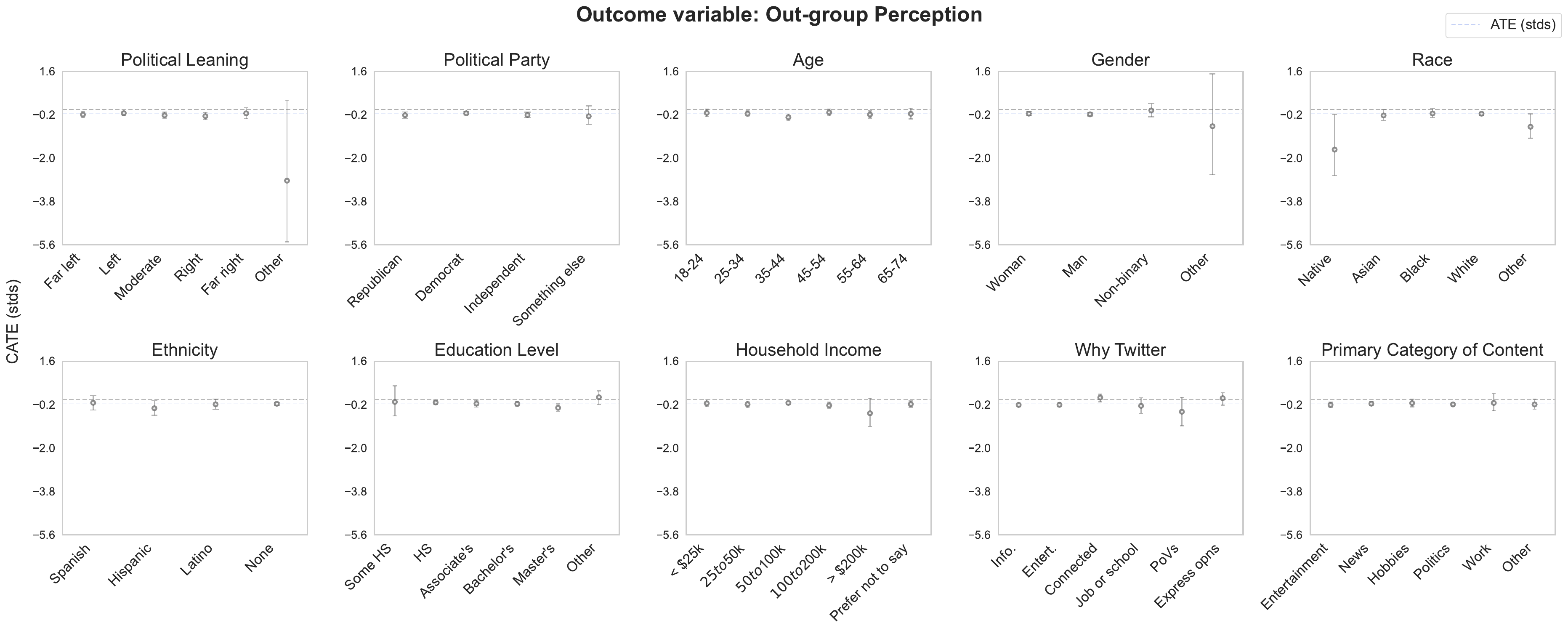}
    \end{adjustbox}
    \caption{\small Conditional average treatment effect (CATE) for the outcome variable ‘out-group perception.’ The blue line shows the average treatment effect (ATE).}
\end{figure}

\begin{figure}[H]
    \centering
    \begin{adjustbox}{center}
        \includegraphics[width=1.2\textwidth]{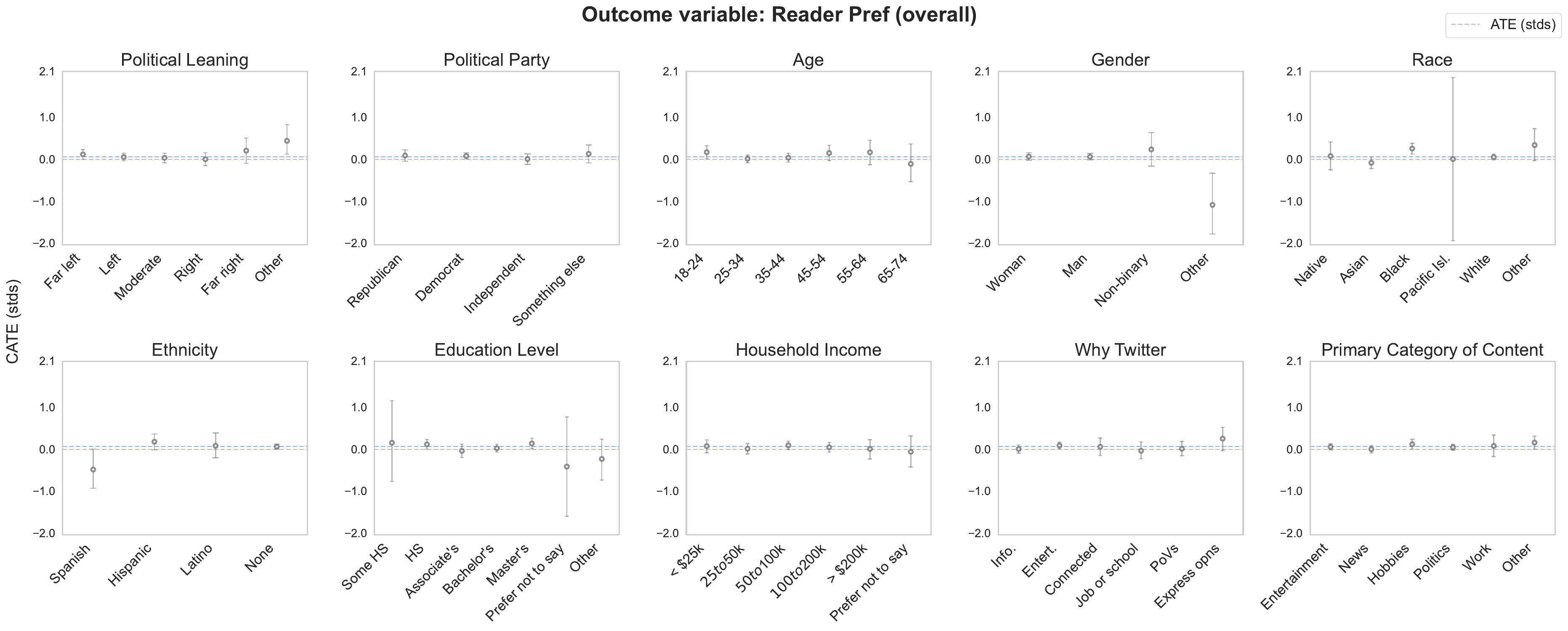}
    \end{adjustbox}
    \caption{\small Conditional average treatment effect (CATE) across subgroups for the outcome variable ‘reader preference (overall).’ The blue line shows the average treatment effect (ATE).}
\end{figure}

\begin{figure}[H]
    \centering
    \begin{adjustbox}{center}
        \includegraphics[width=1.2\textwidth]{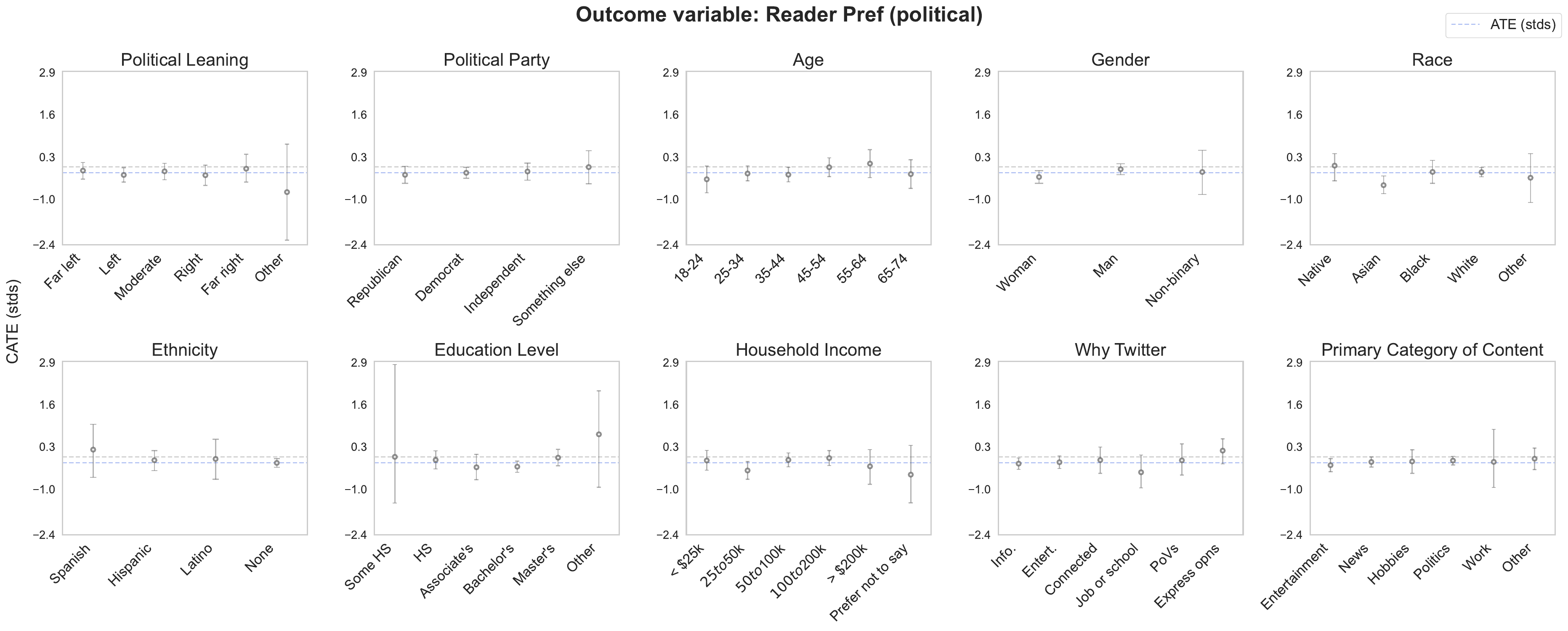}
    \end{adjustbox}
    \caption{\small Conditional average treatment effect (CATE) across subgroups for the outcome variable ‘reader preference (political).’ The blue line shows the average treatment effect (ATE).}
\end{figure}

\subsubsection{Outcome effects for each subgroup, grouped by subgroup} \label{appendix:heterogenous-effects-by-subgroup}

\begin{figure}[H]
    \centering
    \includegraphics[width=0.78\columnwidth]{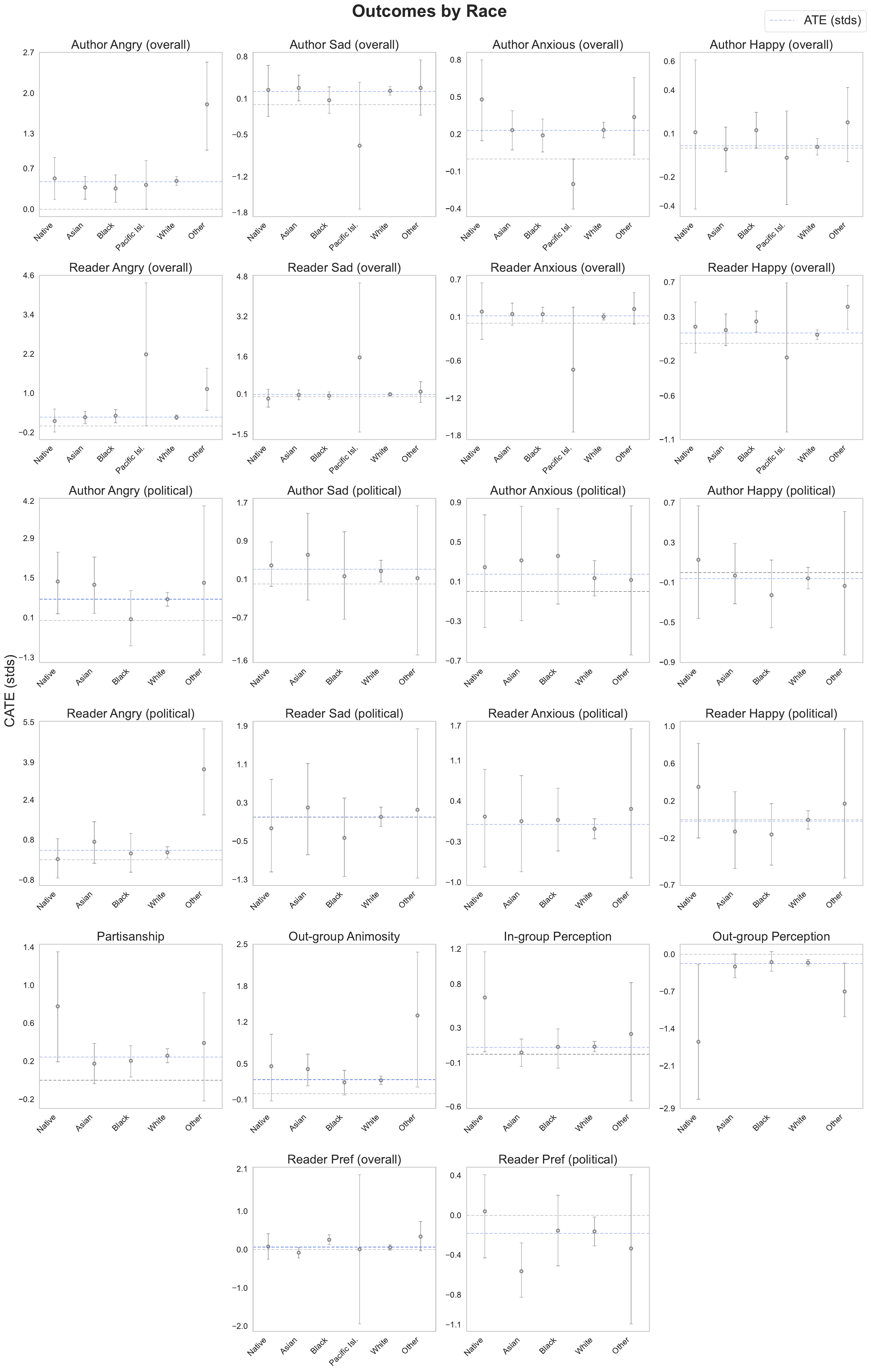}
    \caption{\small Conditional average treatment effect (CATE) for all outcomes when conditioned on different races. The blue line shows the average treatment effect (ATE).}
\end{figure}

\begin{figure}[H]
    \centering
    \includegraphics[width=0.78\columnwidth]{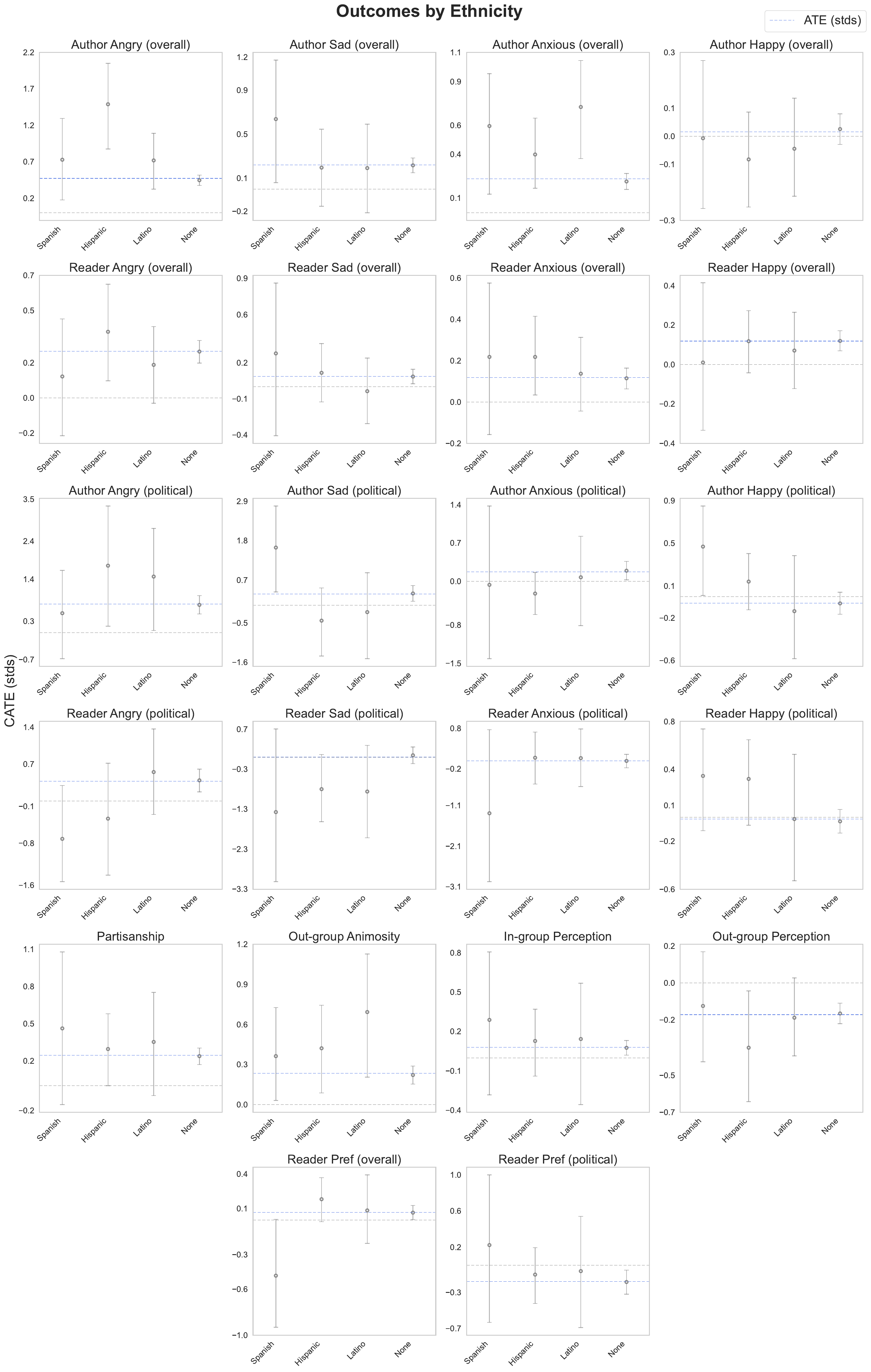}
    \caption{\small Conditional average treatment effect (CATE) for all outcomes when conditioned on different ethnicities. The blue line shows the average treatment effect (ATE).}
\end{figure}

\begin{figure}[H]
    \centering
    \includegraphics[width=0.78\columnwidth]{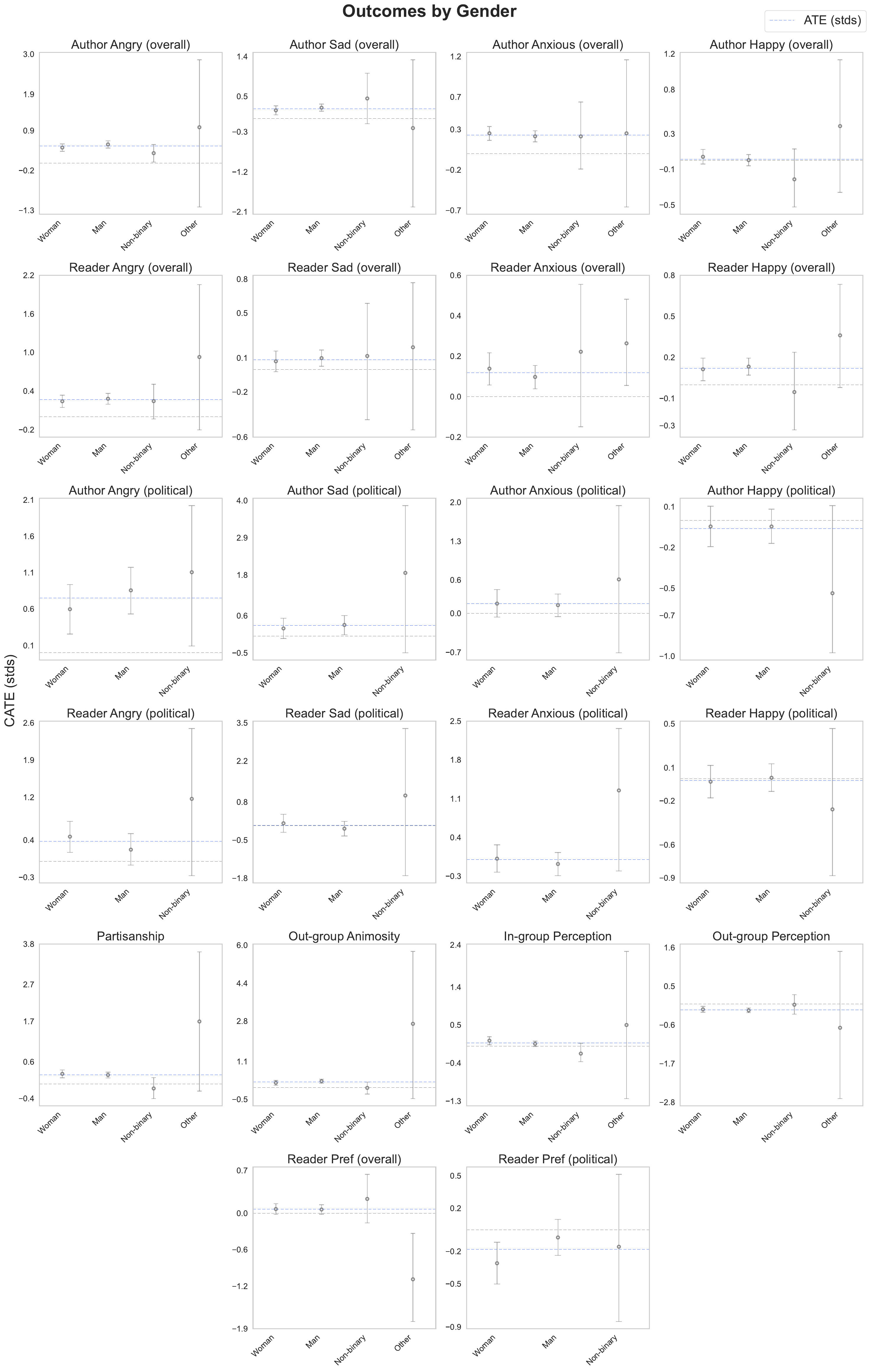}
    \caption{\small Conditional average treatment effect (CATE) for all outcomes when conditioned on different genders. The blue line shows the average treatment effect (ATE).}
\end{figure}

\begin{figure}[H]
    \centering
    \includegraphics[width=0.78\columnwidth]{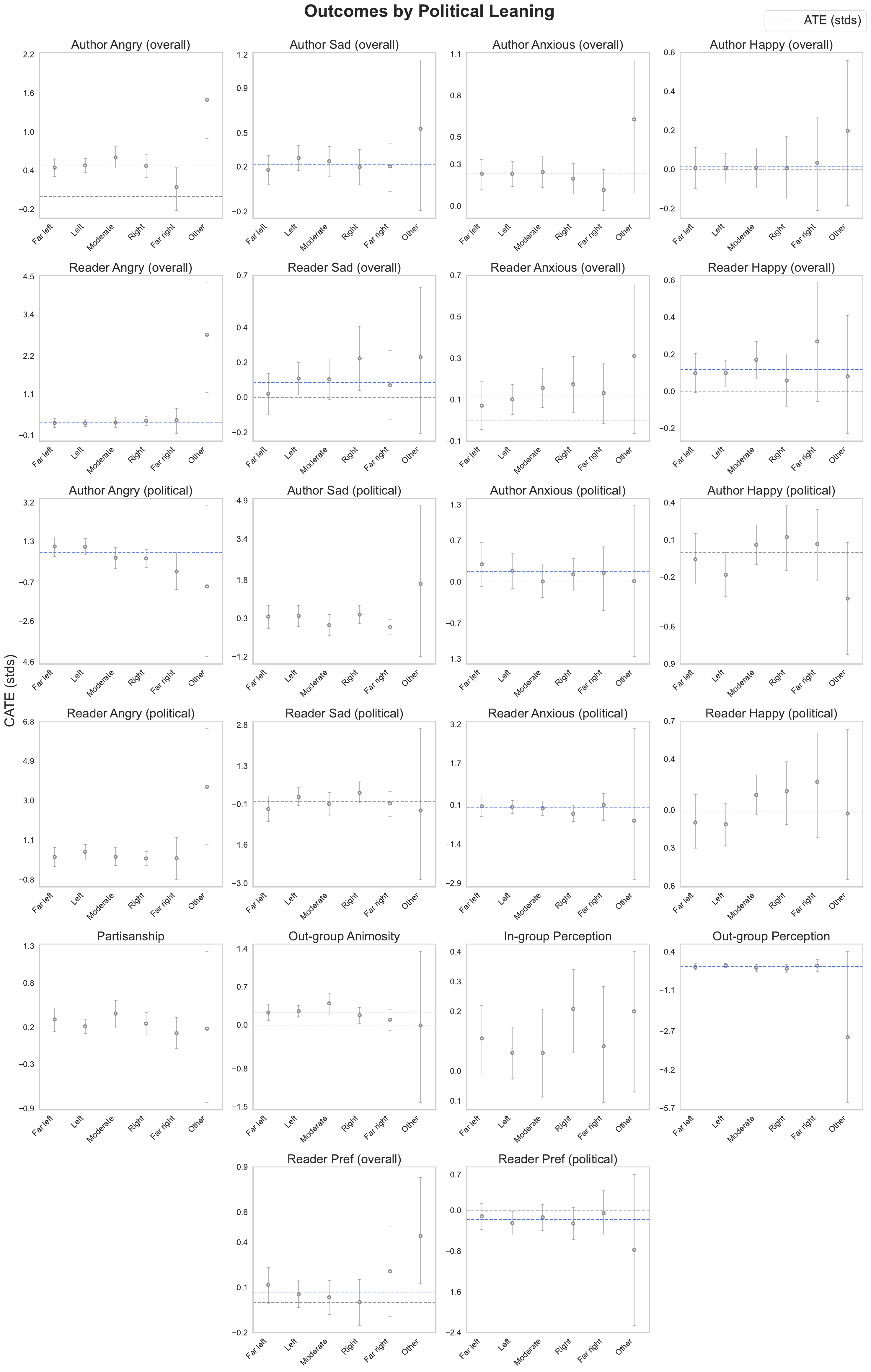}
    \caption{\small Conditional average treatment effect (CATE) for all outcomes when conditioned on different ideological political leanings. The blue line shows the average treatment effect (ATE).}
\end{figure}

\begin{figure}[H]
    \centering
    \includegraphics[width=0.78\columnwidth]{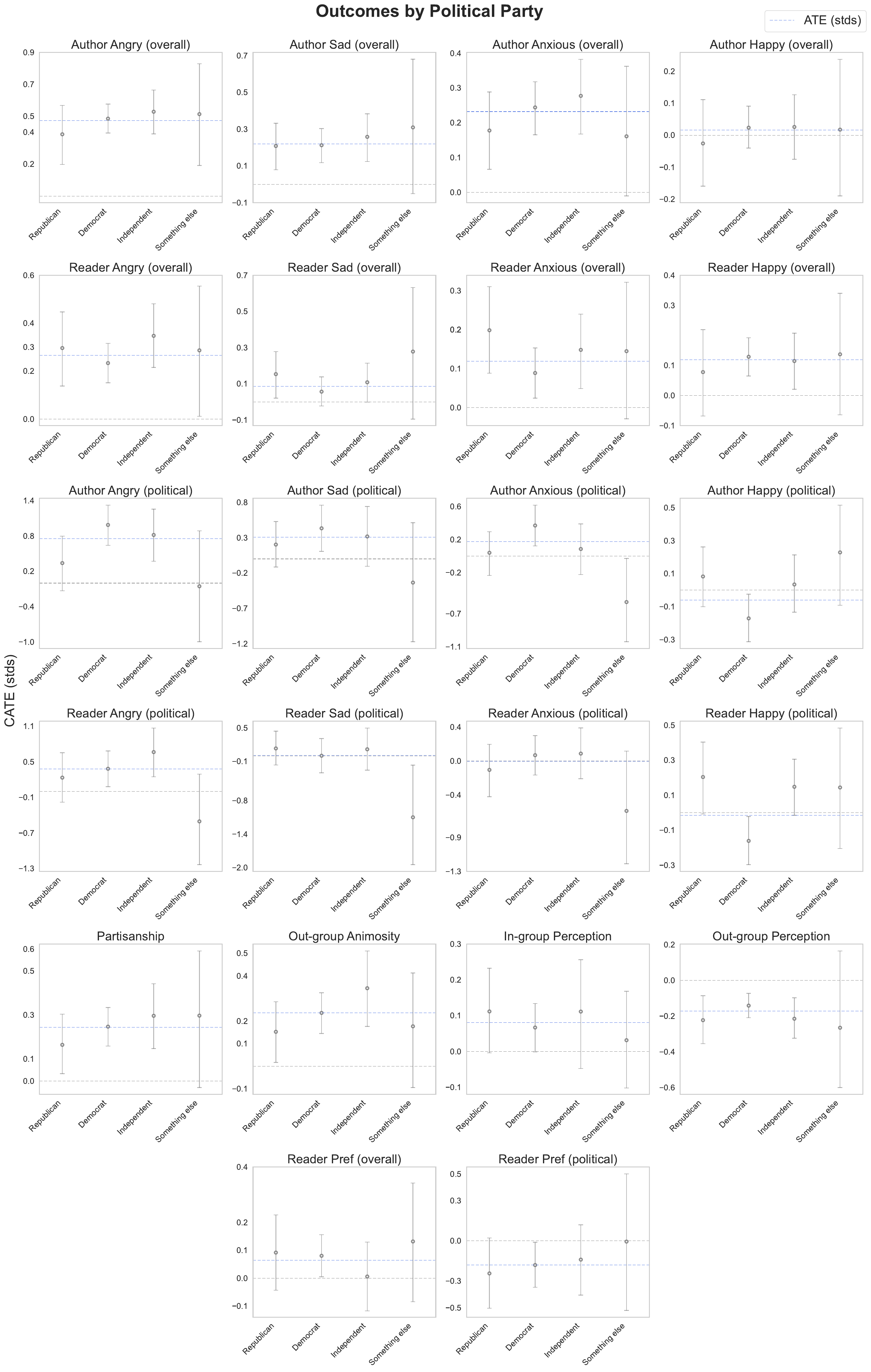}
    \caption{\small Conditional average treatment effect (CATE) for all outcomes when conditioned on different ideological political parties. The blue line shows the average treatment effect (ATE).}
\end{figure}

\begin{figure}[H]
    \centering
    \includegraphics[width=0.78\columnwidth]{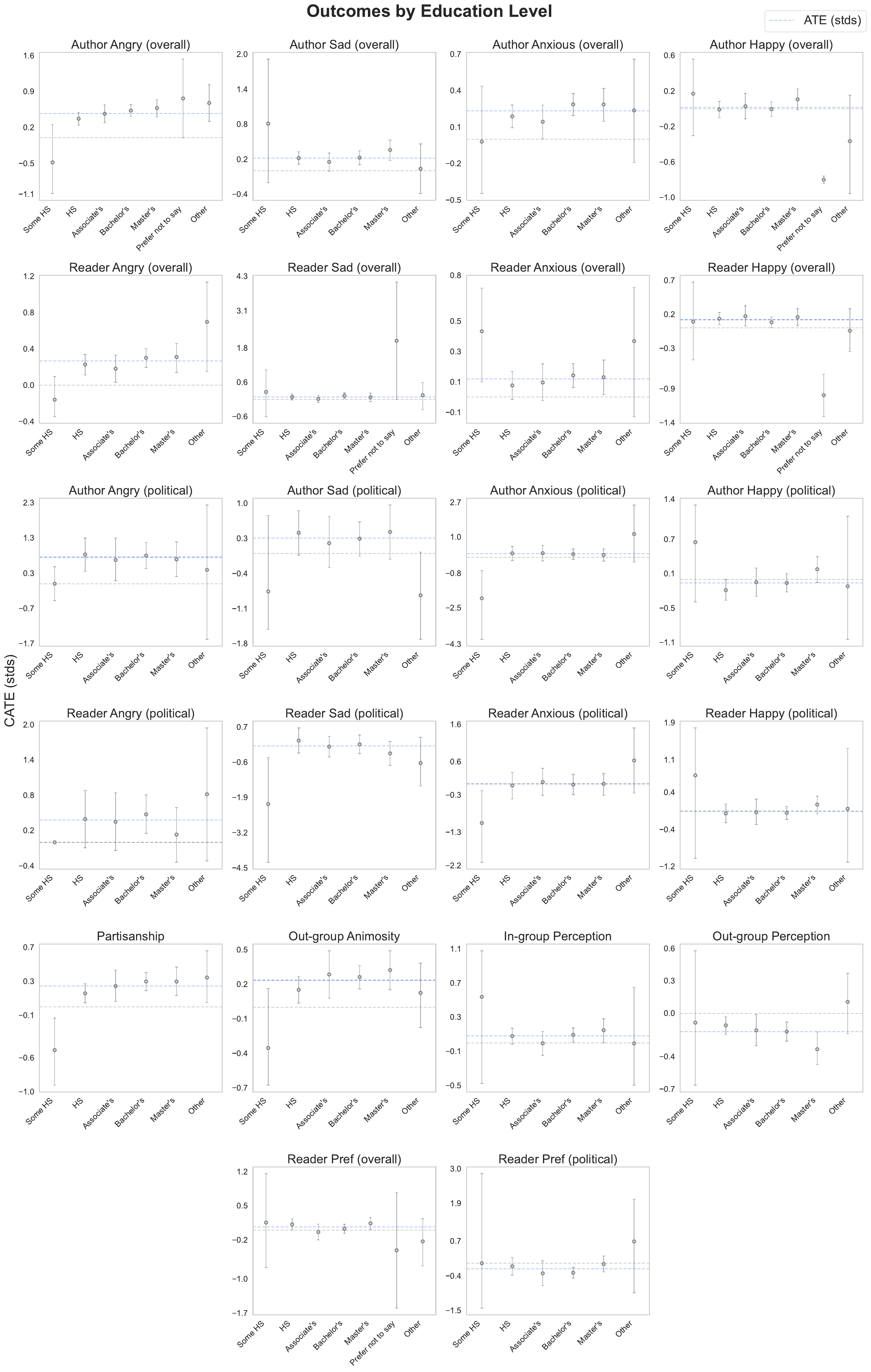}
    \caption{\small Conditional average treatment effect (CATE) for all outcomes when conditioned on different education levels. The blue line shows the average treatment effect (ATE).}
\end{figure}

\begin{figure}[H]
    \vspace{-1em}
    \centering
    \includegraphics[width=0.78\columnwidth]{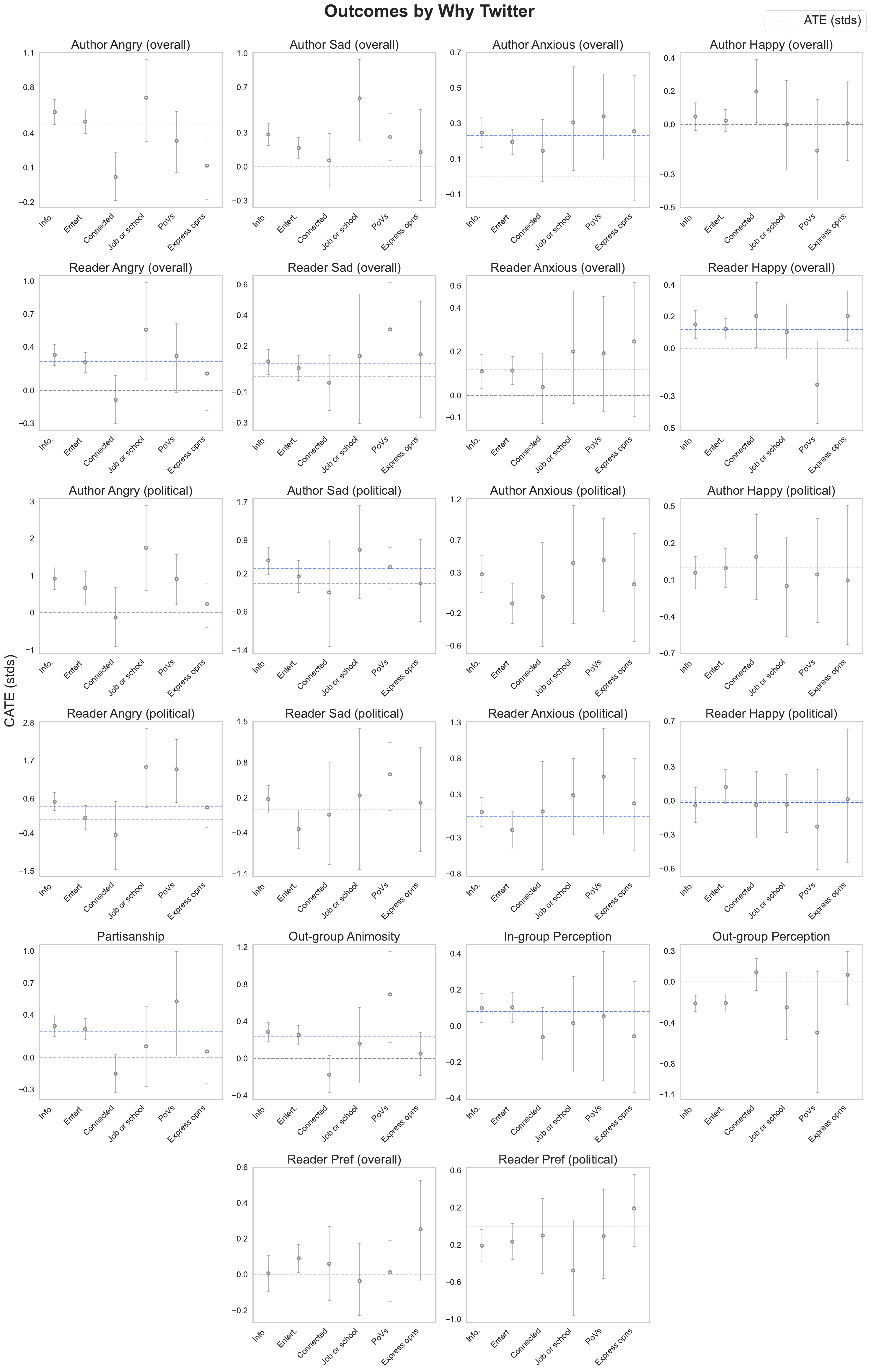}
    \vspace{-1em}
    \caption{\small Conditional average treatment effect (CATE) for all outcomes when conditioned on different main reasons for using Twitter. The blue line shows the average treatment effect (ATE). We ask participants about the main reason they use Twitter as follows: ``What would you say is the main reason you use Twitter?'' (\Cref{app:survey}). The options to select from are: ``A way to stay informed,'' ``Entertainment'' ``Keeping me connected to other people,'' ``It's useful for my job or school,'' ``Let's me see different points of view,'' ``A way to express my opinions''.}

\end{figure}

\begin{figure}[H]
    \centering
    \includegraphics[width=0.78\columnwidth]{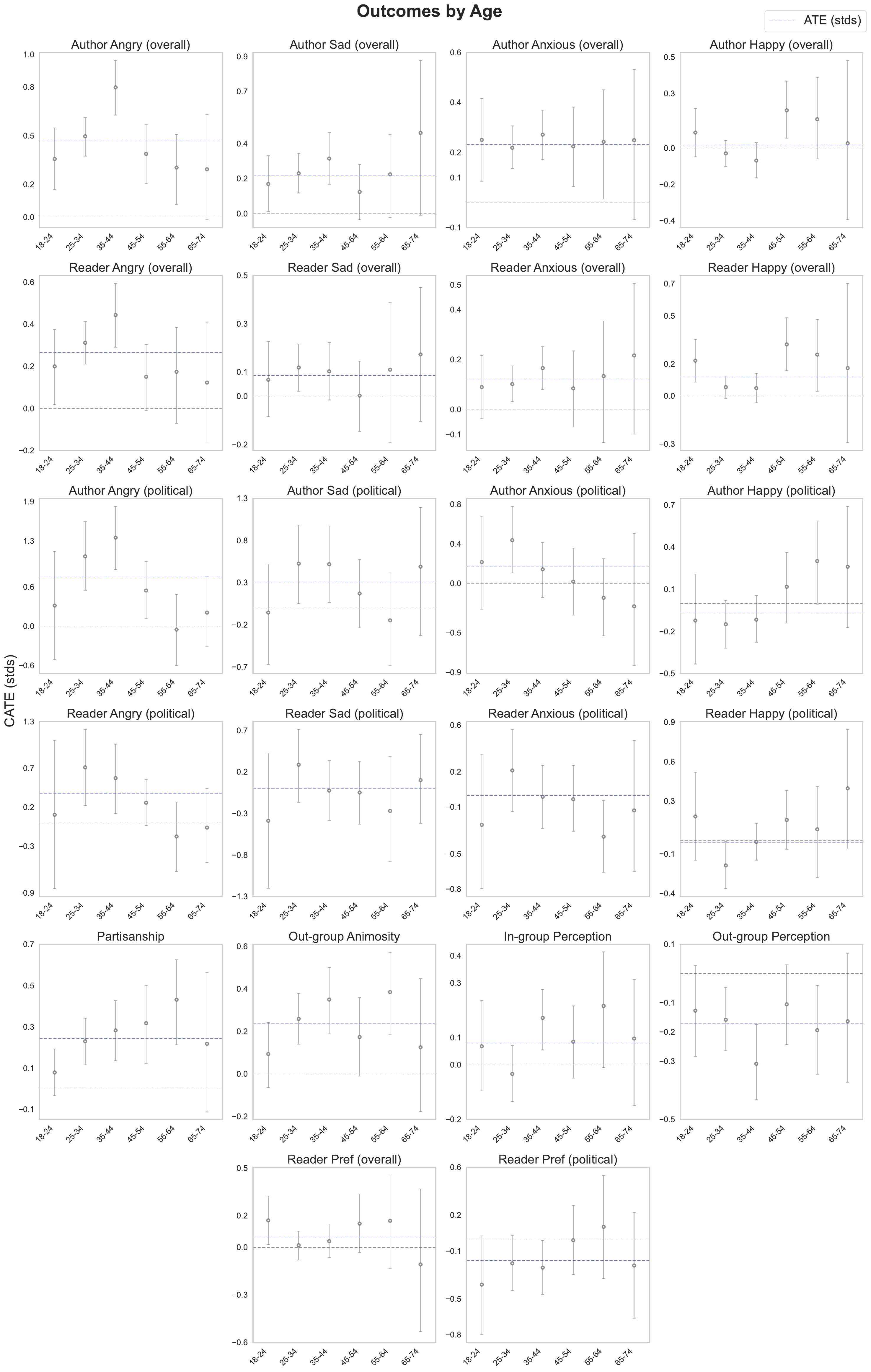}
    \caption{\small Conditional average treatment effect (CATE) for all outcomes when conditioned on different age groups. The blue line shows the average treatment effect (ATE).}
\end{figure}

\begin{figure}[H]
    \centering
    \includegraphics[width=0.78\columnwidth]{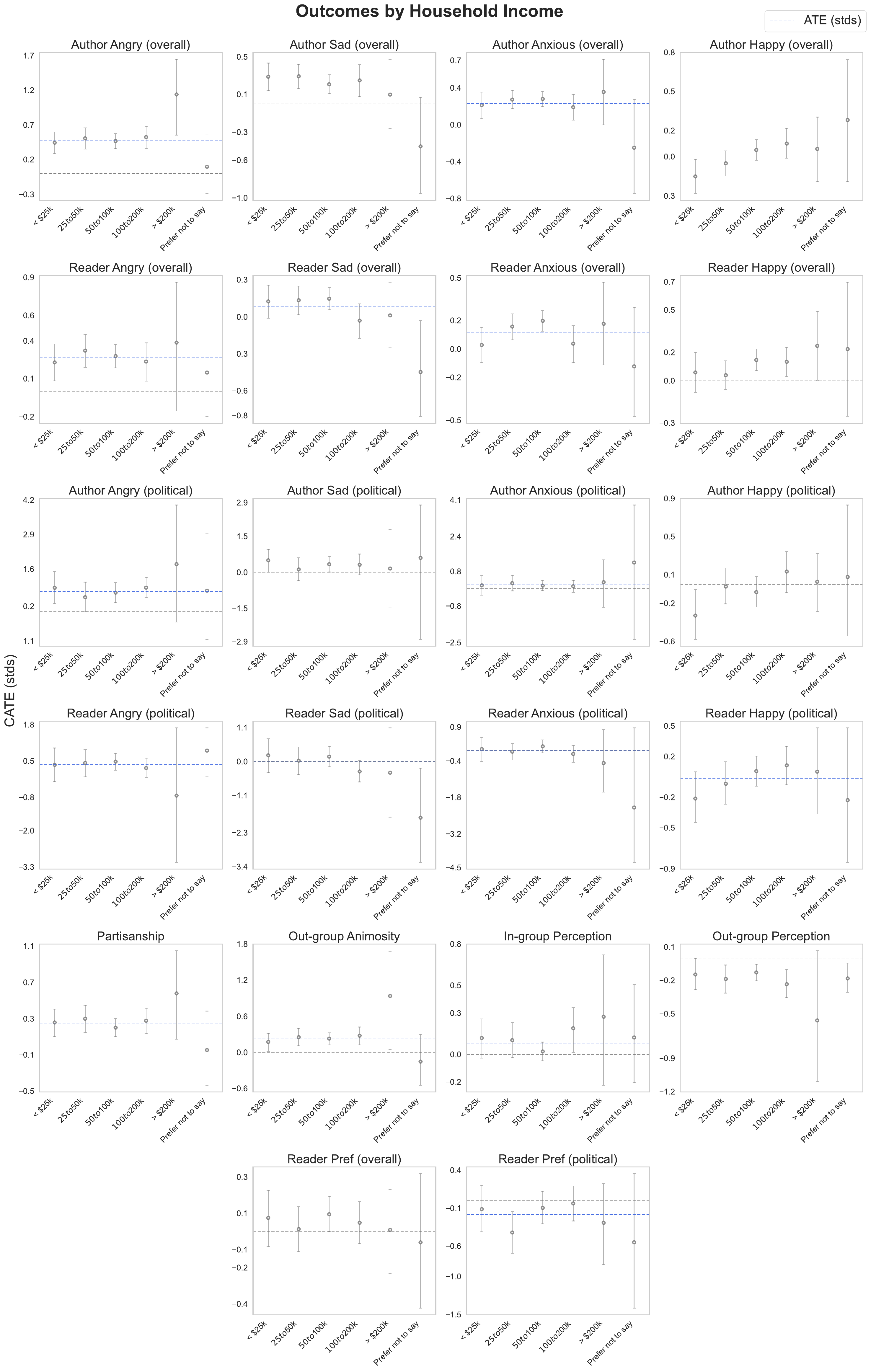}
    \caption{\small Conditional average treatment effect (CATE) for all outcomes when conditioned on different annual household income levels. The blue line shows the average treatment effect (ATE).}
\end{figure}

\begin{figure}[H]
    \centering
    \includegraphics[width=0.78\columnwidth]{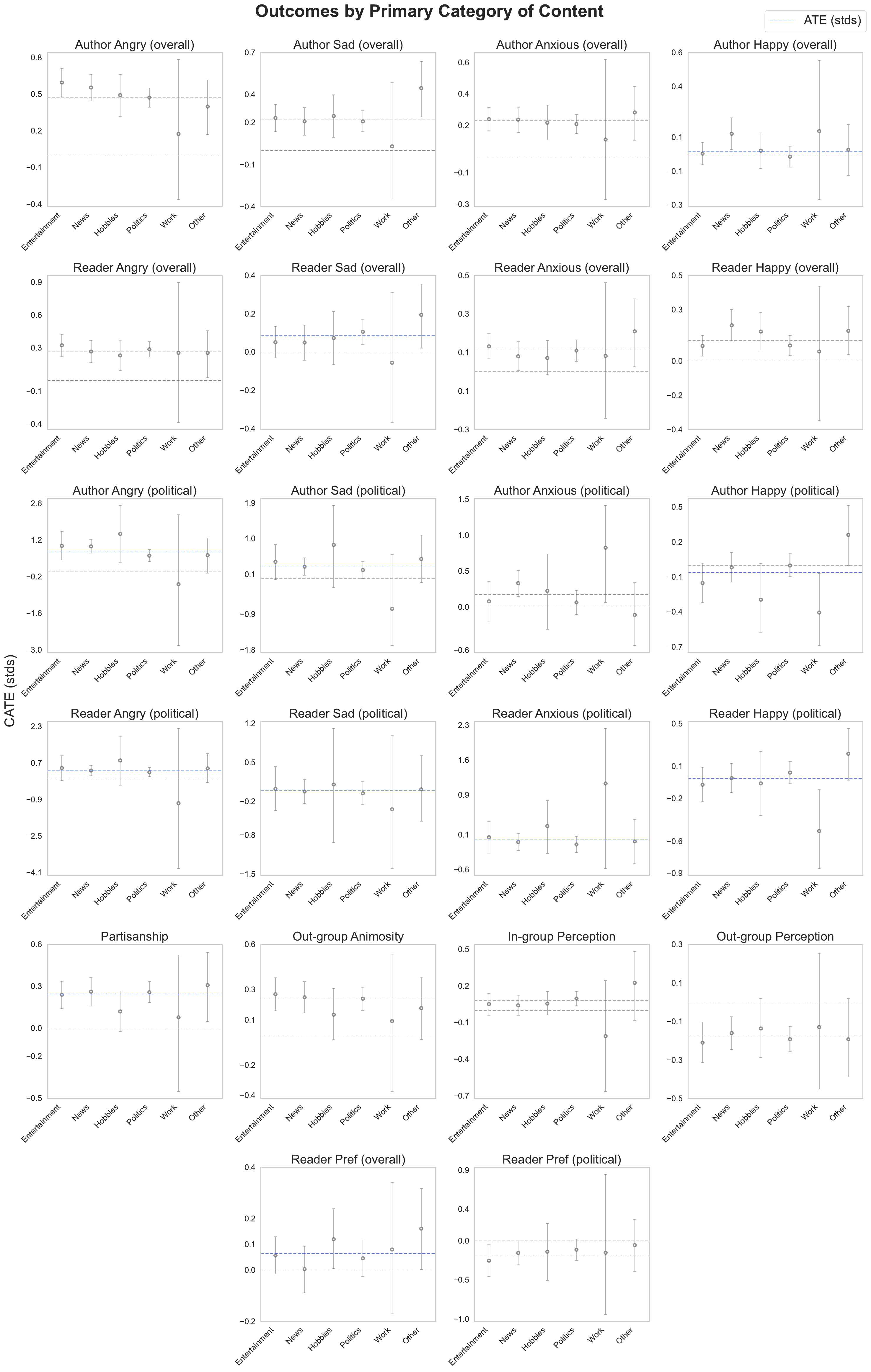}
    \vspace{-1em}
    \caption{\small Conditional average treatment effect (CATE) for all outcomes when conditioned on different types of primary content category shown. The blue line shows the average treatment effect (ATE). We ask participants about the content shown to them as follows: ``What were the tweets we showed you today predominantly about? Select a maximum of two'' (\Cref{app:survey}). The options to select from are ``News,'' ``Politics,'' ``Work,'' ``Entertainment,'' ``Hobbies.''}
\end{figure}

\newpage
\subsection{Effects of stated preference timeline}

\begin{table}[!h]
\centering
\begin{tabularx}{\textwidth}{@{} lccccc @{}}
\hline
\multirow{2}{*}{{\textbf{Outcome}}} & \textbf{Standardized} & \textbf{Unstandardized} & \textbf{Chron.} & \textbf{Eng.} & \multirow{2}{*}{\textbf{$p$-value}} \\
& \textbf{Effect} & \textbf{Effect} & \textbf{Mean} & \textbf{Mean} &\textbf{} \\ \hline

\hline
\multicolumn{6}{c}{\textbf{Emotional effects (all tweets)}} \\
\hline

Author Angry  &                0.110 &                  0.044 &       0.352 &      0.402 &   0.0002 \\
  Author Sad  &                0.070 &                  0.024 &       0.293 &      0.326  &   0.0008 \\
Author Anxious  &                0.082 &                  0.035 &       0.391 &      0.427 &   0.0002 \\
Author Happy  &                0.189 &                  0.150 &       1.307 &      1.455 &   0.0002 \\
Reader Angry  &                0.095 &                  0.038 &       0.306 &      0.350 &   0.0002 \\
  Reader Sad  &                0.056 &                  0.021 &       0.316 &      0.344 & 0.0018 \\
Reader Anxious  &                0.082 &                  0.035 &       0.346 &      0.384 &   0.0002 \\
Reader Happy  &                0.439 &                  0.314 &       0.941 &      1.249 &   0.0002 \\

\hline
\multicolumn{6}{c}{\textbf{Emotional effects (political tweets only)}} \\
\hline

  Author Angry  &                0.205 &                  0.081 &       1.084 &      1.173 &   0.0040 \\
    Author Sad  &                0.170 &                  0.059 &       0.675 &      0.731 &   0.0090 \\
Author Anxious  &                0.043 &                  0.018 &       0.804 &      0.821 &   0.3762 \\
  Author Happy  &                0.062 &                  0.049 &       0.523 &      0.589 &   0.0336 \\
  Reader Angry  &                0.142 &                  0.057 &       1.014 &      1.087 &   0.0240 \\
    Reader Sad  &                0.088 &                  0.033 &       0.814 &      0.858 &   0.1718 \\
Reader Anxious  &                0.058 &                  0.025 &       0.839 &      0.872 &   0.2546 \\
  Reader Happy  &                0.111 &                  0.079 &       0.450 &      0.548  &   0.0002 \\

\hline
\multicolumn{6}{c}{\textbf{Political effects}} \\
\hline
Partisanship &                0.042 &                  0.009 &       0.151 &      0.163 &   0.0342 \\
Out-group Animosity &                0.045 &                  0.006 &       0.085 &      0.093 &   0.0242 \\
In-group Perception (all users) &                0.094 &                  0.017 &       0.060 &      0.077 &   0.0002 \\
Out-group Perception (all users) &               -0.099 &                 -0.021 &      -0.108 &     -0.132 &   0.0002 \\
  In-group Perception (left users) &                0.083 &                  0.015 &       0.056 &      0.071 &   0.0002 \\
 Out-group Perception (left users)  &               -0.100 &                 -0.020 &      -0.097 &     -0.120 &   0.0002 \\
 In-group Perception (right users)  &                0.111 &                  0.023 &       0.072 &      0.100 &   0.0002 \\
Out-group Perception (right users) &               -0.093 &                 -0.024 &      -0.145 &     -0.175 &   0.0018 \\

\hline
\multicolumn{6}{c}{\textbf{Reader Preference}} \\
\hline

Reader Pref (all tweets) &                1.105 &                  0.396 &       0.507 &      0.892 & 0.0002 \\
Reader Pref (political tweets) &                0.788 &                  0.283 &       0.581 &      0.854 & 0.0002 \\

\hline
\end{tabularx}
\caption{The effects of our alternative, exploratory ranking of content based on users' stated preferences (SP). The table shows the average treatment effects (standardized and unstandardized), $p$-values, and $p$-values for all outcomes. All statistics are calculated as described in \Cref{app:ate-estimation} except we replace the engagement-based timeline with the SP timeline as the treatment of interest.}
\label{table:sp-effects}
\end{table}

\begin{table}[!bth]
\centering
\begin{tabularx}{\textwidth}{@{} lccccc @{}}
\hline
\multirow{2}{*}{{\textbf{Outcome}}} & \textbf{Standardized} & \textbf{Unstandardized} & \textbf{Chron.} & \textbf{Eng.} & \multirow{2}{*}{\textbf{$p$-value}} \\
& \textbf{Effect} & \textbf{Effect} & \textbf{Mean} & \textbf{Mean} &\textbf{} \\ \hline

\hline
\multicolumn{6}{c}{\textbf{Emotional effects (all tweets)}} \\
\hline

Author Angry  &               -0.058 &                 -0.023 &       0.352 &      0.338 &   0.0138 \\
  Author Sad  &               -0.011 &                 -0.004 &       0.293 &      0.299 &   0.6243 \\
Author Anxious  &                0.022 &                  0.009 &       0.391 &      0.402  &   0.2250 \\
Author Happy  &                0.242 &                  0.192 &       1.307 &      1.497  &   0.0002 \\
Reader Angry  &               -0.036 &                 -0.015 &       0.306 &      0.298 &   0.0750 \\
  Reader Sad  &               -0.015 &                 -0.006 &       0.316 &      0.319 &   0.4522 \\
Reader Anxious  &                0.025 &                  0.011 &       0.346 &      0.360 &   0.1466 \\
Reader Happy  &                0.490 &                  0.350 &       0.941 &      1.283 &   0.0002 \\

\hline
\multicolumn{6}{c}{\textbf{Emotional effects (political tweets only)}} \\
\hline

Author Angry  &               -0.275 &                 -0.109 &       1.036 &      0.938 &   0.0010 \\
Author Sad  &                0.015 &                  0.005 &       0.653 &      0.657 &   0.8191 \\
Author Anxious  &               -0.052 &                 -0.022 &       0.790 &      0.761 &   0.3370 \\
Author Happy  &                0.145 &                  0.115 &       0.536 &      0.668 &   0.0002 \\
Reader Angry  &               -0.158 &                 -0.063 &       0.991 &      0.942 &   0.0300 \\
Reader Sad  &               -0.002 &                 -0.001 &       0.800 &      0.810 &   0.9695 \\
Reader Anxious  &                0.011 &                  0.005 &       0.837 &      0.841 &   0.8471 \\
Reader Happy  &                0.195 &                  0.139 &       0.471 &      0.628 &   0.0002 \\

\hline
\multicolumn{6}{c}{\textbf{Political effects}} \\
\hline

Partisanship &               -0.166 &                 -0.035 &       0.151 &      0.119 &   0.0002 \\
Out-group Animosity &               -0.252 &                 -0.033 &       0.085 &      0.054 &   0.0002 \\
In-group Perception (all users) &               -0.004 &                 -0.001 &       0.060 &      0.059 &   0.8345 \\
Out-group Perception (all users) &                0.046 &                  0.010 &      -0.108 &     -0.101 &   0.0180 \\
In-group Perception (left users) &               -0.021 &                 -0.004 &       0.056 &      0.053 &   0.3364 \\
Out-group Perception (left users) &                0.053 &                  0.010 &      -0.097 &     -0.091 &   0.0170 \\
In-group Perception (right users) &                0.035 &                  0.007 &       0.072 &      0.077 &   0.2524 \\
Out-group Perception (right users) &                0.034 &                  0.009 &      -0.145 &     -0.137 &   0.3616 \\

\hline
\multicolumn{6}{c}{\textbf{Reader Preference}} \\
\hline

 Reader Pref (all tweets) &                1.117 &                  0.400 &       0.507 &      0.896 &   0.0002 \\
Reader Pref (political tweets) &                0.811 &                  0.291 &       0.585 &      0.870 &   0.0002 \\

\hline
\end{tabularx}
\caption{The effects of the SP-OA timeline, the timeline that ranks by users' stated preference and uses the presence of out-group animosity to break ties. The table shows the average treatment effects (standardized and unstandardized), $p$-values, and $p$-values for all outcomes. All statistics are calculated as described in \Cref{app:ate-estimation} except we replace the engagement-based timeline with the SP-OA timeline as the treatment of interest.}
\label{table:sp-oa-effects}
\end{table}

\newpage 
\subsection{Effects of SP-OA timeline}
\label{app:sp-oa}
Ranking by stated preferences (SP) reduced the amount of partisan animosity, relative to the engagement-based timeline. However, this primarily occurred through a reduction in animosity towards the reader's in-group, rather than animosity towards the reader's out-group. In other words, there was an asymmetry in which readers tended to be tolerant of animosity towards their out-group but not their own in-group. The SP timeline had the highest proportion of in-group content and the lowest proportion of out-group content (relative to both the engagement and chronological timeline). Is it possible to satisfy users' stated preferences without inducing in-group bias? To investigate this, we considered a variant of the SP timeline that used the presence of out-group animosity to break ties between tweets. We call this new timeline the \textbf{SP-OA timeline}. 

We constructed the SP-OA timeline by adding down-ranking for out-group animosity to the SP timeline. In particular, in the SP timeline, we scored the approximately twenty unique tweets for each user by the users' stated preference for the tweet: 1 = ``Yes,'' 0 = ``Indifferent,'' -1 = ``No.'' In  the SP-OA timeline, if a tweet was labeled as having out-group animosity, then we bumped its score down by $0.5$ points. The SP-OA timeline for each user consists of their top ten tweets as ranked by this modified score. Note that even if a tweet has out-group animosity, if the user stated that they valued the tweet, it will always be ranked higher than a tweet that they are indifferent to or do not value. Since the presence of out-group animosity is effectively only used to break ties among tweets with the same stated preference, the SP-OA timeline will satisfy users' preferences to the same extent that the regular SP timeline does.

\begin{figure}[t]
    \centering
    \includegraphics[width=0.98\columnwidth]{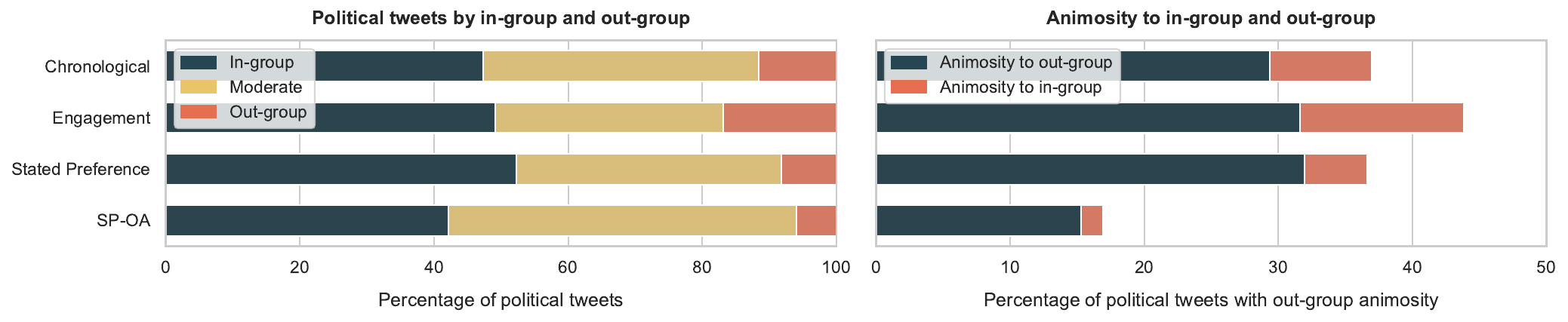}
    \caption{\small On the left, the graph illustrates the distribution of political tweets in each timeline, categorized by whether they align with the reader's in-group, out-group, or are moderate. Meanwhile, the right graph delineates the proportion of political tweets that express out-group animosity, broken down by whether they target the reader's in-group or out-group. Compared to the engagement and chronological timeline, the stated preference (SP) timeline reduces animosity, but only by reducing animosity towards the reader's \emph{in-group}. In contrast, the SP-OA timeline reduces animosity towards both the reader's in-group and out-group.}
    \label{fig:oga-all}
\end{figure}

\begin{figure}[hbtp]
    \centering
    \includegraphics[width=0.9\columnwidth]{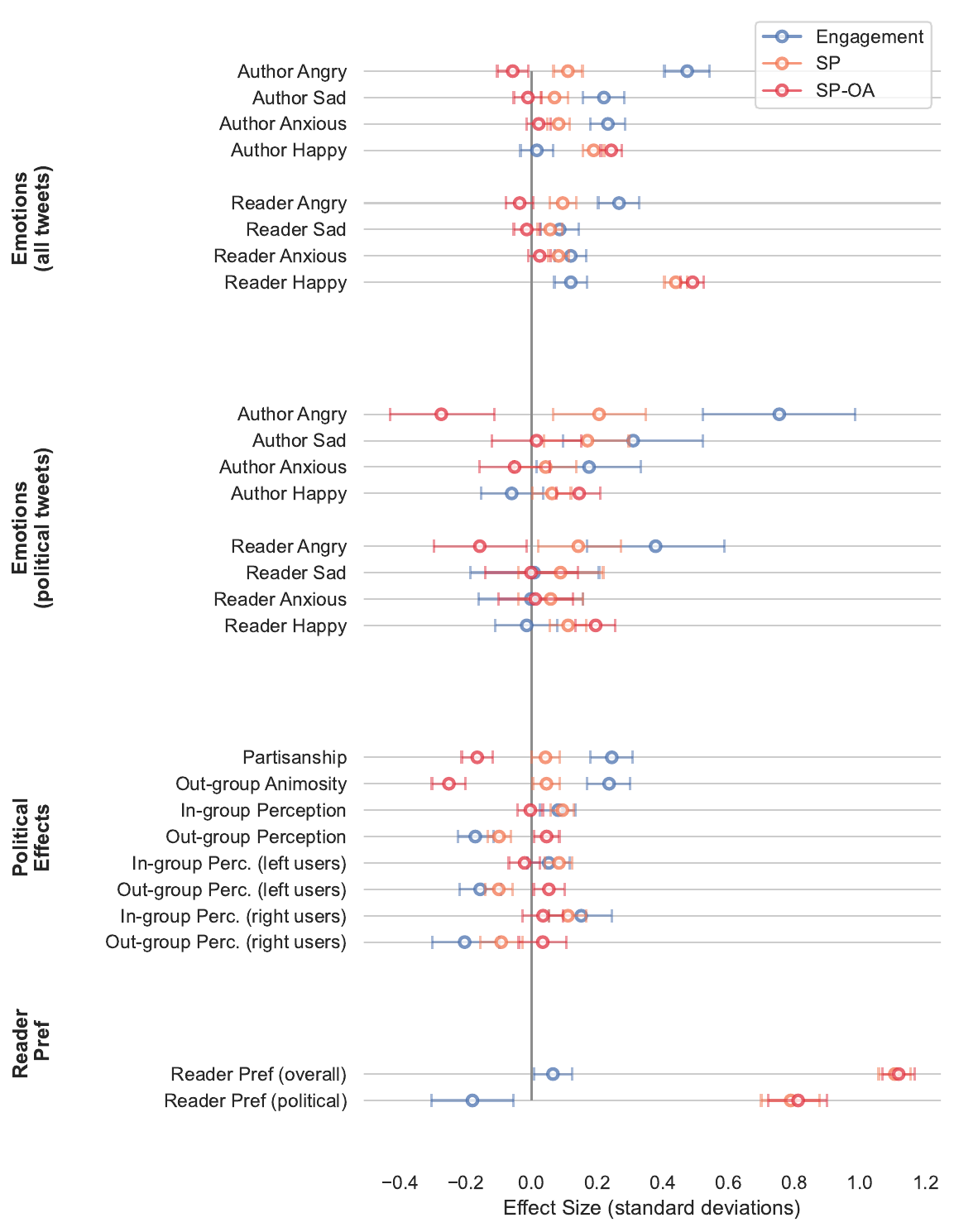}
    \caption{\small The average treatment effect for all pre-registered outcomes along with their 95\% Bootstrap confidence intervals (unadjusted for multiple testing). The effects of three different timelines are shown: (1) Twitter's own engagement-based timeline, (2) our exploratory timeline that ranks based on users' stated preferences (SP), (3) a variant that tie-breaks based on the presence of out-group animosity. The effect sizes for both timelines are relative to the reverse-chronological timeline (the zero line). Average treatment effects are standardized using the standard deviation of outcomes in the chronological timeline (see \Cref{app:ate-estimation} for details).}
    \label{fig:effects-all}
\end{figure}

\Cref{table:sp-oa-effects} shows results for the SP-OA timeline, and \Cref{fig:effects-all} compares the effects of all three timelines: the engagement-based timeline, the SP timeline, and the SP-OA timeline.  The SP-OA yielded the lowest instances of angry, partisan, and out-group hostile content when stacked against the chronological, engagement, and SP timelines. Furthermore, the SP-OA timeline was the most effective at reducing animosity towards the reader's \emph{out-group} (Figure \ref{fig:oga-all}). In the engagement timeline and SP timeline, respectively 34 percent and 33 percent of political tweets contained animosity towards the users' out-group. In contrast, the SP-OA timeline halved this proportion: only 17 percent contained animosity towards the users' out-group. 

In conclusion, the SP-OA timeline has high satisfaction of users' stated preferences, mitigates the amplification of divisive content, and does so without reinforcing in-group bias.

\newpage
\section{Survey questionnaires} \label{app:survey}


\section*{Supplementary material}

**Not Using Chrome**

You must be using Google Chrome on your laptop or desktop (not on your phone) in order to complete this survey. Please try again with Google Chrome.

**Screen**

**Thanks for considering participating in our study!**

Our study aims to investigate the effects of Twitter's personalization algorithm.

As part of our study, we will ask you to download a Chrome extension to collect **public** tweets from your own Twitter timeline (no other data will be collected from the Chrome extension). Data collection will only take a couple of minutes and the Chrome extension will automatically uninstall afterwards.

Afterwards, we will show you the tweets and ask you some questions about how you feel about them.

First, we will ask you a few questions to determine your eligibility for the study.

Are you located in the United States?

- Yes
- No

Are you 18 or older?

- Yes
- No

About how often do you use or visit Twitter?

> Several times a day
> About once a day
> A few times a week
> Once a week or less

How many people do you follow on Twitter?

> Less than 50 users
> 50 users or more

**Failed screen**

Thank you for considering participating in our study, however based on your answers to the questions we asked, you are not eligible. The survey will end now.

**Consent**

## Consent Form for Investigating Twitter experience
### Introduction
My name is Professor Anca Dragan. I am a faculty member at the University of California, Berkeley, in the Electrical Engineering and Computer Science (EECS) Department. I am planning to conduct a research study, which I invite you to take part in.

### Purpose
The purpose of this study is to collect data in order to assess the impact of social media algorithms on users.

### Procedures
If you agree to be in this study, you will be asked to download a Chrome extension that will collect public tweets from your Twitter timeline. The only data we will collect using the Chrome extension are public tweets, no other data (e.g. account credentials, messages, likes, bookmarks, etc) will be collected. Data collection will only take a few seconds and the Chrome extension will automatically uninstall after collecting data. Afterwards, we will survey you about the tweets collected from your timeline.

**Study time:** The whole study should take about 30 minutes.

**Study location:** You will participate online, from the comfort of your current location.

## Benefits

There is no direct benefit to you (other than compensation) from participating in this study. We hope that the information gained from the study will help us design better social media algorithms.

## Risks/Discomforts

This study represents minimal risk to you. As with all research, there is the risk of an unintended breach of confidentiality. However, we are taking precautions to minimize this risk (see below).

## Confidentiality

The data we collect will be stored on password-protected servers. Once the research is complete, we intend to scrub the data of all identifiable information. We will keep only the recorded survey responses, as well as a freshly generated identifier for each subject. The de-identified data will be retained indefinitely for possible use in future research done by ourselves or others. This cleaned dataset may be made public as part of the publishing process. No guarantees can be made regarding the interception of data sent via the Internet by any third parties.

## Compensation

We compensate workers based on the estimated duration of completing the study. The study will be prorated to $20/hour for the anticipated duration of completing the study, which is posted for your job on the CloudResearch interface you used to view the job (duration includes reviewing instructions, completing the task, and filling an exit survey). The payment is arranged by CloudResearch via credit to subjects' accounts.

## Rights

Participation in research is completely voluntary. You have the right to decline to participate or to withdraw at any point in this study without penalty or loss of benefits to which you are otherwise entitled.

## Questions

If you have any questions or concerns about this study, or in case anything goes wrong with the online interface, you can contact Anca Dragan at effects.of.twitter.algorithm@gmail.com.

If you have any questions or concerns about your rights and treatment as a research subject, you may contact the office of UC Berkeley's Committee for the Protection of Human Subjects, at 510-642-7461 or subjects@berkeley.edu.

**IRB review:**

This study was approved by an IRB review under the CPHS protocol ID number 2021-09-14618

**You should save a copy of this consent form for your records.**

**If you wish to participate in this study, please click the "I consent" button below**

I consent

**Connect ID**

What is your **Cloudresearch Connect ID**?

**Chrome extension instructions**

We will now have you install a Chrome extension to collect public tweets from your timeline so that we can ask you about them. No other information, including your username or other personally identifying information, will be collected. The data collection will only take a couple of minutes.

**Data collection will not work for everyone. As long as you have tried it, we will still pay you 0.75. If it does work for you, and you complete the rest of the survey, you will receive $10.**

Follow the steps below:

1. Download the Chrome extension from this link
2. The Chrome extension will open a new tab with instructions. Follow the instructions on screen.
3. When you are done, you will see either a "Success!" or "Error!" page and the Chrome extension will uninstall. You may now move on to the next part of the survey.

**Do not proceed until data collection is complete or you will not get the full $10.**

**If the extension has succeeded,** you will see a "Success!" message like the one shown below. You can skip to the bottom of this page and proceed to the next page.

## Success!

Data collection is complete. The extension will now automatically uninstall. You may now return to the survey.

**If you accidentally exited out of the tab or an error occurred,** you can restart the process through the following steps:

1. Open the extensions menu in the top right of your browser by clicking the puzzle piece icon.

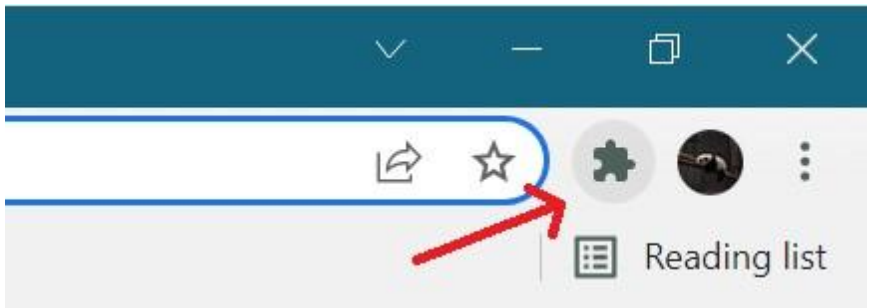

2. Click the bird icon to restart the authorization and data collection process.

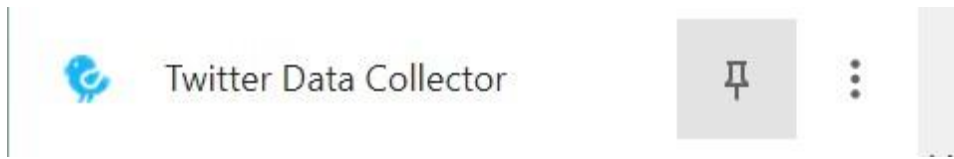

**If repeating the steps above still leads you to errors,** you can proceed to the next page, and you will still receive the base pay of 0.75$ for having attempted the survey.

I have followed the instructions on the Chrome extension

**Outcome success**

The previous step completed successfully. For the next part of the study you will be answering questions about individual Tweets we'll show you.

**Aim to spend no more than 1-2 minutes per tweet.**

The following choices represent the rank of the tweet. This is a hidden question that is used for the Qualtric loop / merge randomization that is used to for randomizing the order of tweets in our outcome survey.

[ dropdown ]

**Outcome questions**

You have ${e://Field/tweetsLeft} tweets left.

You will be answering questions about the following tweet(s):

Note that there are tweets by both ${lm://Field/4} and ${lm://Field/2}. We will first be asking you questions about **${lm://Field/4}**'s tweet. You can use for ${lm://Field/2}'s tweet for context, but answer the following questions while focusing on ${lm://Field/4}'s tweet.

How is **${lm://Field/4}** feeling in their tweet?

|         | Not at all | Slightly | Somewhat | Moderately | Extremely |
|---------|------------|----------|----------|------------|-----------|
| Angry   | O          | O        | O        | O          | O         |
| Anxious | O          | O        | O        | O          | O         |
| Happy   | O          | O        | O        | O          | O         |
| Sad     | O          | O        | O        | O          | O         |

How did **${lm://Field/4}**'s tweet make *you* feel?

|         | Not at all | Slightly | Somewhat | Moderately | Extremely |
|---------|------------|----------|----------|------------|-----------|
| Angry   | O          | O        | O        | O          | O         |
| Anxious | O          | O        | O        | O          | O         |
| Happy   | O          | O        | O        | O          | O         |
| Sad     | O          | O        | O        | O          | O         |

Is **${lm://Field/4}**'s tweet about a political or social issue?

Yes

No

How does **${lm://Field/4}**'s tweet lean politically?

| Far left | Left | Moderate | Right | Far right |
|---|---|---|---|---|
| -2 | -1 | 0 | 1 | 2 |

0

How does **${lm://Field/4}**'s tweet make you feel about people or groups on the Left?

| Much worse | Worse | The same as before | Better | Much better |
|---|---|---|---|---|
| -2 | -1 | 0 | 1 | 2 |

0

How does **${lm://Field/4}**'s tweet make you feel about people or groups on the Right?

| Much worse | Worse | The same as before | Better | Much better |
|---|---|---|---|---|
| -2 | -1 | 0 | 1 | 2 |

0

Is **${lm://Field/4}**'s tweet expressing anger, frustration, or hostility towards a person or group on the Left?

Yes

No

Is **${lm://Field/4}**'s tweet expressing anger, frustration, or hostility towards a person or group on the Right?

Yes

No

When you use Twitter, do you want to be shown tweets like **${lm://Field/4}**'s tweet?

Yes

No

Indifferent

Next, we will be asking you questions about **${lm://Field/2}**'s tweet.

How is **${lm://Field/2}** feeling in their tweet?

|  | Not at all | Slightly | Somewhat | Moderately | Extremely |
|---|---|---|---|---|---|
| Angry | ○ | ○ | ○ | ○ | ○ |
| Anxious | ○ | ○ | ○ | ○ | ○ |
| Happy | ○ | ○ | ○ | ○ | ○ |
| Sad | ○ | ○ | ○ | ○ | ○ |

How did **${lm://Field/2}**'s tweet make *you* feel?

|  | Not at all | Slightly | Somewhat | Moderately | Extremely |
|---|---|---|---|---|---|
| Angry | ○ | ○ | ○ | ○ | ○ |
| Anxious | ○ | ○ | ○ | ○ | ○ |
| Happy | ○ | ○ | ○ | ○ | ○ |
| Sad | ○ | ○ | ○ | ○ | ○ |

Is **${lm://Field/2}**'s tweet about a political or social issue?

Yes

No

How does **${lm://Field/2}**'s tweet lean politically?

| Far left | Left | Moderate | Right | Far right |
|---|---|---|---|---|
| -2 | -1 | 0 | 1 | 2 |

0

How does **${lm://Field/2}**'s tweet make you feel about people or groups on the Left?

|   | Much worse | Worse | The same as before | Better | Much better |
|---|---|---|---|---|---|
|   | -2 | -1 | 0 | 1 | 2 |
|   |   |   |   |   | 0 |

How does **${lm://Field/2}**'s tweet make you feel about people or groups on the Right?

|   | Much worse | Worse | The same as before | Better | Much better |
|---|---|---|---|---|---|
|   | -2 | -1 | 0 | 1 | 2 |
|   |   |   |   |   | 0 |

Is **${lm://Field/2}**'s tweet expressing anger, frustration, or hostility towards a person or group on the Left?

- Not at all
- Slightly
- Somewhat
- Moderately
- Extremely

Is **${lm://Field/2}**'s tweet expressing anger, frustration, or hostility towards a person or group on the Right?

- Not at all
- Slightly
- Somewhat
- Moderately
- Extremely

When you use Twitter, do you want to be shown tweets like **${lm://Field/2}**'s tweet?

- Yes
- No
- Indifferent

**Overall Content Questions**

**What were the tweets we showed you today predominantly about? Select a maximum of two.**

News
Politics
Work
Entertainment
Hobbies
Other

**Data collection failed**

Sorry, an error occurred and we were unable to collect your data. This is a known issue that happens with a small portion of the participants.

**You should fill out the form at the link below to be paid** *a base rate of compensation of* **$0.75:** https://forms.gle/CJNtmYVbMhHaeNTw7

**Please advance to the next page to complete this survey and then "Return"/"Withdraw from" the study in CloudResearch Connect. If you submit a completion code, your HIT will be rejected.**

If you have any feedback, share it with us on CloudResearch Connect.

**Additional background information**

To end, we will ask some basic questions about your background.

In which state do you currently reside?

What gender do you identify as?

Woman

Non-binary

Man

Other

How old are you?

18-24 years old

25-34 years old

35-44 years old

45-54 years old

55-64 years old

65-74 years old

75 years or older

Are you Spanish, Hispanic, or Latino or none of these (select all that apply).

Spanish

Hispanic

Latino

None of these

Choose one or more races that you consider yourself to be:

White

Black or African American

American Indian or Alaska Native

Asian

Native Hawaiian or Pacific Islander

Other

What is the highest level of schooling that you have completed?

Some high school

High school graduate

Associate degree

Bachelor's degree

- Master's degree or above
- Prefer not to answer
- Other: _____

### In politics today, do you consider yourself a:

- Republican
- Democrat
- Independent
- Something else

### As of today do you lean more to...

- The Republican Party
- The Democratic Party

### How do you identify politically?

- Far left
- Left
- Moderate
- Right
- Far right
- Other: _____

### Which way do you lean politically?

- Towards the Left
- Towards the Right

### What is your annual household income?

- Less than $25,000
- $25,000 - $50,000
- $50,000 - $100,000
- $100,000 - $200,000

More than $200,000

Prefer not to say

Why do you use Twitter? **Select all that apply**.

Entertainment

A way to stay informed

A way to express my opinions

Keeping me connected to other people

Lets me see different points of view

It's useful for my job or school

What would you say is the *main* reason you use Twitter?

Entertainment

A way to stay informed

A way to express my opinions

Keeping me connected to other people

Lets me see different points of view

It's useful for my job or school

Thank you! **To complete the survey, click next and you will be redirected back to Cloudresearch.**



\end{appendix}
\end{document}